%% file: BH_v04.tex
\documentclass[onecolumn,amsmath,amssymb,nofootinbib,12pt]{article}
\pdfoutput=1
\usepackage{jheppub}
\usepackage{ifpdf}

\usepackage[bb=boondox]{mathalfa}

\usepackage{graphicx,subcaption}
\graphicspath{ {figures/} }
\usepackage{amsfonts}
\usepackage{amsmath}
\usepackage{amssymb}
\usepackage{epsfig}
\usepackage{pdfpages}
\usepackage{bbm}
\usepackage{graphicx,epstopdf}
\usepackage[numbers]{natbib}
\usepackage[makeroom]{cancel}
\usepackage{hyperref}
\usepackage{xcolor}
\hypersetup{pdftex,colorlinks=true,allcolors=blue}
\usepackage{array}
\usepackage[export]{adjustbox}
\usepackage{mathrsfs}
\usepackage{booktabs}

\usepackage[normalem]{ulem}

\newcommand{\eg}{{\it e.g.,}\ }
\newcommand{\ie}{{\it i.e.,}\ }
\newcommand{\reef}[1]{(\ref{#1})}

\newcommand{\mt}[1]{\textrm{\tiny #1}}

\newcommand{\qml}{QM$_{\rm L}$}
\newcommand{\qmr}{QM$_{\rm R}$}
\newcommand{\aims}{AEM$^4$Z}
\newcommand{\qes}{\mt{QES}}
\newcommand{\QES}{\mt{QES}}
\newcommand{\yp}{\y_\mt{Page}}
\newcommand{\ys}{\y_\mt{shock}}
\newcommand{\y}{\sigma}
\newcommand{\GN}{\ensuremath{G_\textrm{N}}}
\newcommand{\uP}{u_\mt{Page}}
\newcommand{\upp}[1]{u_{\mt{Page},{#1}}}
\newcommand{\yb}{\y_\mt{Turn}}
\newcommand{\sigmab}{\sigma_\mt{Turn}}
\newcommand{\q}{\gamma}

\newcommand{\beq}{\begin{equation}}
\newcommand{\eeq}{\end{equation}}
\newcommand{\beqs}{\begin{equation}\begin{aligned}}
\newcommand{\eeqs}{\end{aligned}\end{equation}}
\newcommand{\beqa}{\begin{eqnarray}}
\newcommand{\eeqa}{\end{eqnarray}}
\newcommand{\beqar}{\begin{eqnarray*}}
\newcommand{\eeqar}{\end{eqnarray*}}

\newcommand{\Bcal}{\mathcal{B}}
\newcommand{\Hcal}{\mathcal{H}}
\newcommand{\Ocal}{\mathcal{O}}
\newcommand{\PP}{\mathrm{PP}}

\newcommand{\Page}{\mathrm{Page}}

\renewcommand{\(}{\left(}
\renewcommand{\)}{\right)}
\renewcommand{\[}{\left[}
\renewcommand{\]}{\right]}
\def\tr{{\text{Tr}}}

\definecolor{holo1}{HTML}{800080}
\definecolor{holo2}{HTML}{0000fe}
\definecolor{holo3}{HTML}{f60000}

\usepackage[thinlines]{easytable}
\usepackage{textcomp}

\title{Information Flow in Black Hole Evaporation}

\author[a,b]{Hong Zhe Chen,}
\author[a]{Zachary Fisher,}
\author[a,b]{Juan Hernandez,}
\author[a]{Robert C. Myers}
\author[a,b]{and Shan-Ming Ruan}
\affiliation[a]{Perimeter Institute for Theoretical Physics, Waterloo, ON N2L 2Y5, Canada}
\affiliation[b]{Dept.~of Physics $\&$ Astronomy, University of Waterloo, Waterloo, ON N2L 3G1, Canada}

\emailAdd{hchen2@pitp.ca}
\emailAdd{me@zachfisher.com}
\emailAdd{jhernandez@pitp.ca}
\emailAdd{rmyers@pitp.ca}
\emailAdd{sruan@pitp.ca}

\date{\today}

\abstract{Recently, new holographic models of black hole evaporation have given fresh insights into the information paradox~\cite{Penington:2019npb,Almheiri:2019psf,Almheiri:2019hni}. In these models, the black hole evaporates into an auxiliary bath space after a quantum quench, wherein the holographic theory and the bath are joined. One particularly exciting development is the appearance of `ER=EPR'-like wormholes in the (doubly) holographic model of~\cite{Almheiri:2019hni}.  At late times, the entanglement wedge of the bath includes the interior of the black hole. In this paper, we employ both numerical and analytic methods to study how information about the black hole interior is encoded in the Hawking radiation. In particular, we systematically excise intervals from the bath from the system and study the corresponding Page transition. Repeating this process ad infinitum, we end up with a fractal structure on which the black hole interior is encoded, implementing the \"uberholography protocol of~\cite{Pastawski:2016qrs}.}

\begin{document}

\maketitle

\section{Introduction}\label{sec:intro}
\input{sections/intro}

\section{Review of the {\aims} Model}\label{sec:review}
\input{sections/preamble}
\subsection{Setup}\label{sec:setup}
\input{sections/setup}

\subsection{Recovering the Page Curve}\label{sec:QES}
\input{sections/QES}

%\section{Phase transitions}
%\input{sections/Phase_transitions}

\section{Entanglement of Hawking Radiation}\label{sec:excise}
\input{sections/minbath}
\input{sections/HolyInfo}
\input{sections/Later}

\section{Discussion}\label{sec:discuss}
\input{sections/Discussion}

\section*{Acknowledgments}
We would like to thank Ahmed Almheiri, Raphael Bousso, Hugo Marrochio, Ignacio Reyes, Song He, Joshua Sandor, Antony Speranza, Zixia Wei, and especially Beni Yoshida  for useful comments and discussions. Research at Perimeter Institute is supported in part by the Government of Canada through the Department of Innovation, Science and Economic Development Canada and by the Province of Ontario through the Ministry of Economic Development, Job Creation and Trade. HZC is supported by the Province of Ontario and the University of Waterloo through an Ontario Graduate Scholarship. RCM is also supported in part by a Discovery Grant from the Natural Sciences and Engineering Research Council of Canada. JH is also supported by the Natural Sciences and Engineering Research Council of Canada through a Postgraduate Doctoral Scholarship. RCM also received funding from the BMO Financial Group. RCM and ZF also received funding from the Simons Foundation through the ``It from Qubit'' collaboration.

\appendix
%\section{Copy section 2}

%\input{sections/sec2copy}

\bibliography{references}
\bibliographystyle{utphys}

\end{document}

%% file: sections/intro.tex
% !TEX root = ../BH_v01.tex

More than four decades after its introduction, the information paradox~\cite{Haw76a} still looms large over the field of quantum gravity. Although a full solution remains elusive, investigations of the information paradox have led to some breakthroughs about the nature of spacetime in quantum gravity. Much of this research can be summarized with the slogan ``entanglement builds spacetime''~\cite{Van09}. Most famously, the ER=EPR connection~\cite{MalSus13} argues that entangled states in certain quantum systems have a dual interpretation as quantum gravitational wormholes.

The ER=EPR connection was developed to provide a resolution to the firewall paradox~\cite{AMPS, AMPSS, Mathur:2009hf,Mat10,MatPlu11,Mathur:2011uj}, a sharp version of the information paradox that concerned the entanglement between modes inside the black hole horizon and early-time Hawking modes. According to ER=EPR, the Hilbert spaces corresponding to the early radiation and the interior of the black hole are not independent because a wormhole connects those regions. This hypothesis resolves some of the confusion about black hole evaporation but also suggests many fascinating new questions. Yet historically, it has been difficult to study the ER=EPR connection in this context, for lack of a tractable model of an evaporating black hole where the quantum effects are under control.

\begin{figure}[t]
	\centering\includegraphics[width=3.0in]{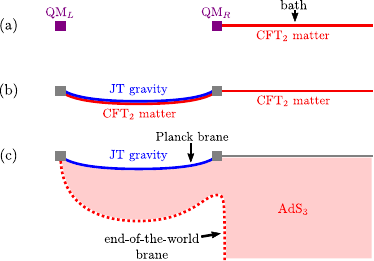}
	\caption{In the {\aims} model, the holographic principle is invoked twice, resulting in three different pictures of the same physical system. In the top picture, there are two quantum mechanics systems (\textcolor{holo1}{\qml} and \textcolor{holo1}{\qmr}) as well as a field theory (\textcolor{holo3}{CFT$_2$}) vacuum state prepared on the half-line. The middle picture includes the \textcolor{holo2}{2D holographic geometry (JT gravity)} dual to the entangled state of {\qml} and \qmr. The last picture contains the doubly-holographic description, with a \textcolor{holo3}{bulk AdS$_3$} dual to the matter in the middle picture.}\label{trio}
\end{figure}

Two recent papers~\cite{Penington:2019npb,Almheiri:2019psf} made remarkable progress by constructing holographic models of evaporating black holes.\footnote{See also the important follow-up discussions of~\cite{Almheiri:2019hni,Akers:2019nfi, Almheiri:2019yqk, Rozali:2019day}.} Here we will focus on the second of these, which considers a two-dimensional model of Jackiw-Teitelboim (JT) gravity~\cite{Jackiw:1984je,Teitelboim:1983ux,Maldacena:2016upp} coupled to a conformal field theory (CFT). In this model, we begin with an eternal black hole, which has a holographic description in terms of a thermofield double state of two entangled quantum mechanical systems~\cite{Mal01}. We denote the latter as {\qml} and {\qmr} -- see the top illustrations in figure~\ref{trio}. At some finite time, we couple the right boundary system {\qmr} to a (zero temperature) bath, which consists of a copy of the same two-dimensional CFT prepared its vacuum state on a half-line. The quench joining the two states creates two shockwaves, one of which propagates into the black hole and the other into the bath. Following \aims, we will not worry too much about how the quench is regularized\footnote{It remains an  open problem to apply a more rigorous analysis of the quench, \` a la \cite{Calabrese:2007mtj,Shimaji:2018czt,Asplund:2013zba}, to the {\aims} model.}. With this new connection, the  Hawking radiation from the bulk black hole can then escape into the  bath, allowing the black hole to evaporate.

For this simple two-dimensional model, the backreaction can be explicitly calculated because of the topological nature of JT gravity.\footnote{Recall that for JT model, the geometry is fixed to be at constant curvature, \ie it is always locally AdS$_2$, and the backreaction only involves the evolution of the scalar dilaton on this background, and the subsequent motion of the asymptotic boundary~\cite{Maldacena:2016upp}.} Additionally, the von~Neumann entropy of CFT$_2$, defined by the analog of Shannon entropy for quantum states $\rho$,
\begin{equation}
	S_\text{bulk}(\rho) = - \tr \rho \log \rho
\end{equation} is also computationally tractable~\cite{CalCar09}. The situation is fortuitous because both the geometry and the quantum entanglement in the bulk play a role in determining the entanglement wedge, \ie the region of a bulk which can be reconstructed from a subset of the boundary theory~\cite{Czech:2012bh,Headrick:2014cta,Wall:2012uf,Jafferis:2015del,Dong:2016eik,Cotler:2017erl,Penington:2019npb}. At leading order in $1/N$, the edge of this region is determined by the Ryu-Takayanagi (stationary area) surface; but at next-to-leading order, corrections arise from the entanglement of bulk fields, specifically as computed by the von~Neumann entropy~\cite{FauLeu13, EngWal14}. The quantum prescription is to instead minimize the generalized entropy, defined by
\begin{equation}\label{genX}
	S_{\text{gen}}[C] = \frac{A[\chi]}{4\GN } +  S_\text{bulk}(\tr_C \rho)
\end{equation}
%where we understand the latter term as a renormalized von~Neumann entropy in which the bare $\GN $ appears as the leading counterterm.
% \rcm{Zach, okay with changes??}
for a region $C$ of a Cauchy surface divided into an interior and exterior by a codimension-2 surface $\chi$. Fixing a region $B$ on the boundary, we scan over all $\chi$ homologous to $B$ (that is to say, satisfying $\partial C=\chi\cup B$) to find the surface which minimizes $S_\textrm{gen}$. The bulk surface $\chi$ is called a quantum extremal surface (QES) for $B$~\cite{EngWal14}.

The two-dimensional model of~\cite{Almheiri:2019psf} reproduces many expected features of semiclassical black hole evaporation. In particular, the model reproduces the information paradox for the Hawking radiation, \ie the entropy of the Hawking radiation absorbed by the bath continues to grow without end. However, the entropy of the black hole, \ie of \qmr,\footnote{Of course, the entanglement entropy of {\qml} remains fixed at the Bekenstein-Hawking entropy of the initial eternal black hole.} undergoes a Page transition. That is, the {\qmr} entropy initially rises to track the increasing entropy of the bath, but then there is a sharp transition to a phase where it decreases again. This rise and fall of the black hole entropy are characteristic of the behaviour exhibited by the classic Page curve~\cite{Page:1993wv,Page:2013dx}.  This novel transition occurs in this holographic model (and in the model described by~\cite{Penington:2019npb}) as  a result of the existence of a new class of QESs just inside the event horizon of the evaporating black hole. These surfaces are in fact the minimal solutions at late times, and thus delineate the true boundary of the entanglement wedge of the dual {\qmr} theory.

This two-dimensional model~\cite{Almheiri:2019psf} was then extended with an extra layer of holography by~\cite{Almheiri:2019hni}. In this variant, the matter theory in the bulk and bath is chosen to be itself a holographic CFT coupled to JT gravity. This theory is itself the boundary theory of a dual AdS$_3$ bulk -- see the third illustration in figure~\ref{trio}.  The JT gravity theory resides on a Planck brane suspended in an asymptotically AdS$_3$ bulk. The latter can be thought of as a Randall-Sundrum brane~\cite{RS,RS1}, which cuts off the asymptotic AdS$_3$ geometry at a finite radius, but it is also engineered as a Dvali-Gabadadze-Porrati brane~\cite{Dvali:2000hr}, in that the brane carries an intrinsic gravity action (confined to one lower dimension), \ie the JT action. Since the CFT is defined on manifolds with boundary (a boundary conformal field theory, or BCFT), the bulk also contains a second class of branes on which the AdS space ends: end-of-the-world (ETW) branes ~\cite{Tak11}. This doubly holographic model, which we refer to as the {\aims} model\footnote{The suggested pronunciation is `aims'.} from the combined authors' initials of~\cite{Almheiri:2019psf,Almheiri:2019hni} will be central to our considerations.

In this setup, the contribution of the CFT to the generalized entropy is calculated by finding extremal HRT surfaces, \ie geodesics, in the AdS$_3$ bulk, in accord with the usual prescription for holographic entanglement entropy~\cite{RyuTak06,HubRan07}. In general, these geodesics may connect the endpoints of the relevant intervals in the boundary theory, however, they may also end on the ETW brane~\cite{Tak11,FujTak11}  or the Planck brane. In the latter case, the gravitational entropy associated with the end-point must also be included as part of the generalized entropy. The doubly holographic {\aims} model yields much the same behaviour as found with the two-dimensional model~\cite{Almheiri:2019psf} described above. In particular, a Page-like curve is recovered for the entropy of \qmr. In the three-dimensional bulk, the corresponding HRT surface undergoes a phase transition at the Page time, where the endpoint on the Planck brane jumps to the new QES described above.  However, since the total system, \ie \qml, {\qmr} and bath, is in a pure state,  the information paradox is resolved and a Page curve is \textit{also} recovered for the Hawking radiation absorbed by the bath. That is, the same HRT surface in the bulk describes the entanglement entropy for {\qmr} and for the complementary system, \qml+bath. A remarkable feature of this doubly holographic description is that after the Page time, the new HRT surfaces delineate an entanglement wedge which includes (a portion of) the black hole interior. Invoking entanglement wedge reconstruction, the bath (plus \qml) is in principle able to reconstruct the black hole interior. Hence the {\aims} model provides an explicit manifestation of ER=EPR.

It is natural to ask how the black hole interior is encoded in the bath. In this paper, we begin to investigate this question. Our approach is straightforward: we start by considering the entire bath (plus \qml) as our entangled subsystem. We then systematically excise various subregions of the bath from our entangling region, each time studying the corresponding entanglement wedge in the three-dimensional dual. We perform the excisions such that the system always sits at the transition where the entanglement wedge of the remaining bath in combination with \qml begins to include the interior of the black hole. By identifying the Page transition for these various `hole-y' subregions of the bath, we can find which regions of the bath are important for encoding the black hole interior. In this simple case, we find that the late radiation contains somewhat redundant information to reconstruct the black hole interior, and the early time radiation is more important; a similar effect was observed recently in~\cite{Rozali:2019day}.
We also study the limiting case where we excise a large number of subintervals in the bath.  By repeating this process ad infinitum, the remaining bath has a fractal structure. In this way, we implement the \"uberholography of~\cite{Pastawski:2016qrs}, and we can determine the support of the black hole interior encoding in the bath.

\paragraph{Outline.} In section~\ref{sec:review}, we review the {\aims} model. In particular, we show that there are three phases that the entanglement entropy evolves through after the quench. We study the entanglement properties of the holographic model in section~\ref{sec:excise}, removing increasingly large entangling segments from the bath. We explain how the information encoding the interior of the black hole is encoded in the CFT via an increasingly refined boundary-bath operator algebra. In section~\ref{sec:discuss}, we conclude with a discussion of our calculations and future directions.

%% file: sections/preamble.tex
% !TEX root = ../BH_v04.tex

In this section, we review the {\aims} model~\cite{Almheiri:2019psf,Almheiri:2019hni} and describe the salient quantitative results for the quantum extremal surfaces and generalized entropies. We also examine numerical solutions in certain instances to compare with our analytical approximations.

The process described in the introduction involves a quantum quench where the {\qmr} system is connected to the bath, as well as the subsequent evaporation of the black hole on the Planck brane. In the three-dimensional bulk description, the quench involves connecting the corresponding end-of-the-world (ETW) branes and letting them fall into the AdS$_3$ geometry. Similarly, the black hole evaporation is described by the dynamics of the joint between the Planck brane and the asymptotic AdS$_3$ boundary.

In principle, the problem of finding quantum extremal surfaces for the extremely dynamical bulk geometry described above seems an intimidating one. However, this difficulty is mitigated by several simplifying features in the {\aims} model.\footnote{Certainly, one of the simplifying features is that the evaporating black hole is constructed in the two-dimensional JT model, which means any candidate QES is simply a point and its extremity is easily tested by taking ordinary derivatives, \eg see eqs.~\reef{difconstraints}.} First, the theory in the first holographic description (panel (b) in figure~\ref{trio}) is a {\it two-dimensional} boundary CFT (BCFT). Hence in the dual description, after an analytic continuation, the entire evolution can be conformally mapped to the vacuum state in the upper half-plane (UHP), \ie with a simple boundary running along the real axis.
Given this configuration and turning to the second holographic description (panel (c) in figure~\ref{trio}), we exploit the fact that {\it holographic} BCFTs have relatively simple expressions for the entanglement entropy, \eg see eq.~\reef{Sbulk_hol}. Lifting this result back to the two-dimensional description (b), the remainder of the analysis involves undoing the previous conformal transformations. That is, we are essentially following the analysis of~\cite{Almheiri:2019psf}, but the key difference is that we have a specific formula for the entanglement entropy determined by the holographic BCFT. This also allows us to consider more complicated situations, \eg multiple intervals, in the two-dimensional description in a straightforward way.

\begin{figure}[t]
	\centering\includegraphics[width=0.7\textwidth]{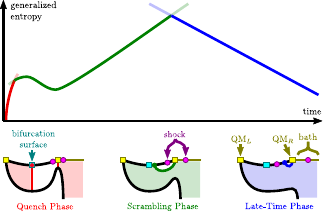}
	\caption{ A cartoon illustration of the three phases for the entanglement entropy of {\qmr} or \qml+bath, after the quench where {\qmr} is connected to the bath. The darker colors indicate the true generalized entropy, while the lighter colors indicate the general shape of each of the branches slightly beyond the regime where it provides the minimal value for the generalized entropy. Below the plot is a sketch of the shape of the extremal surfaces in AdS$_3$ which contribute to the generalized entropy in each phase.}\label{faze}
\end{figure}

\begin{figure}[t]
	\centering\includegraphics[width=3.50in]{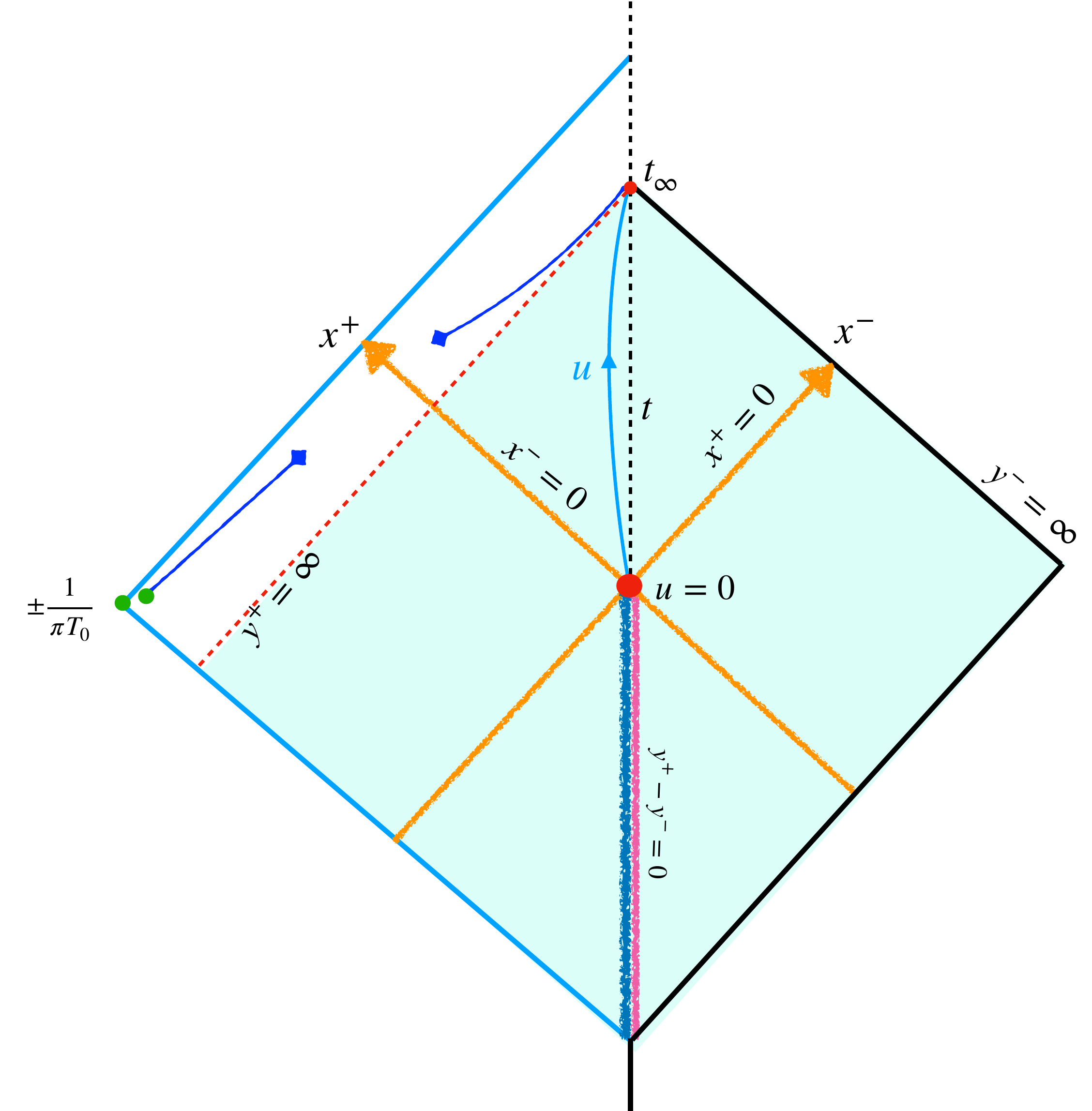}
	\caption{In the {\aims} model, the $\text{AdS}_2$ black hole is coupled to bath along the boundary $\sigma=0$ at time $\tau=0=t$. This results in the shock indicated by the yellow solid line. The evolution of quantum extremal surfaces is indicated by the solid blue curve. The first phase transition occurs when the QES jumps from the green point at $x^\pm = (\pi T_0)^{-1}$ to the other green point, and the second (Page) phase transition happens at the jump between the blue block. In this final phase, the QES tracks close to the new horizon.}\label{fig:ads}
\end{figure}

When, as in~\cite{Almheiri:2019psf}, we consider the entanglement entropy of \qmr, or alternatively its purification, the bath plus \qml, we find the entropy evolves through three phases, which are sketched in figure~\ref{faze} -- see also the spacetime diagram of the two-dimensional boundary in figure~\ref{fig:ads}. These three phases are as follows:

\paragraph{a) Quench Phase:} This is a short period after the bath and {\qmr} systems are joined, in which the entanglement entropy rapidly rises. The three-dimensional description involves the HRT surface having two separate components. The first is anchored to the bifurcation surface of the initial eternal black hole on the Planck brane and falls straight down into the AdS$_3$ bulk to terminate on the ETW brane (which is stationary at this point).
Similarly, the second connects {\qmr} to the ETW brane where the new connection was made and where it quickly falling into the bulk. Hence the rapid rise in the entanglement entropy is entirely due to the stretching of this second component of the HRT surface.
\paragraph{b) Scrambling Phase:} The transition to this phase occurs on a thermal time scale (see eq.~\reef{eq:scramble}).
The entanglement entropy shows some transient behaviour at the beginning of this phase, \eg depending on the precise choice of parameters, the entropy may initially decrease, as shown in figure~\ref{fig:ads}.  However, after roughly the scrambling time (see eq.~\reef{Constant}), the entanglement entropy begins to grow linearly as the bath steadily absorbs more and more Hawking radiation from the black hole (or from \qmr). The gradual increase in entropy is consistent with the heuristics from efficient scrambling systems where only a small but increasing amount of the radiation can be decoded before the Page transition~\cite{HayPre07}. During this phase, the corresponding HRT surface consists of a single geodesic which connects {\qmr} to a point very close to the bifurcation surface of the initial black hole (see figure~\ref{fig:ads}). In particular, it connects boundary points on opposite sides of the shock wave propagating into the Planck brane.
\paragraph{c) Late-Time Phase:} In this phase, the entanglement entropy decreases, as required by the late time behaviour of the Page curve. Of course, the bath continues to absorb Hawking radiation and so this decrease indicates there must be correlations between the Hawking quanta emitted at early and late times. In this phase, the corresponding HRT surface again consists of a single component, but now the geodesic connects {\qmr} to the new QES behind the event horizon of the evaporating black hole -- see figure~\ref{fig:ads}. Hence these geodesics are distinguished from the previous class since the two boundary points which they connect both lie to the future of the shock wave.

%% file: sections/setup.tex
% !TEX root = ../BH_v04.tex

The {\aims} model consists of a $\text{AdS}_2$ black hole in JT gravity, dual to a Hartle-Hawking state of two copies of a one-dimensional quantum mechanics theory~\cite{Almheiri:2019psf,Almheiri:2019hni}. At Lorentzian time $t = 0$, we perform a quantum quench on the CFT, joining it to a field theory vacuum state defined on the half-line $\sigma > 0$. In the bulk, Hawking radiation can now escape to the bath and land on $\mathscr{I}^+$, and the black hole thus evaporates. Additionally, the quench results in two shockwaves, one propagating into the black hole and one into the bath, corresponding to the propagation of a large amount of energy arising from the joining quench. The energy of these shockwaves $E_S$ should be thought of as one of the UV scales for the model. The spacetime diagram of the coupled system is shown in figure~\ref{fig:ads}.

The two-dimensional gravity solution is locally $\text{AdS}_2$, described by the Poincar\'e metric
\begin{equation}
ds^2_{\mt{AdS}} = -\frac{4\,L_{\mt{AdS}}^2 }{(x^+-x^-)^2}dx^+ dx^-\,\qquad (x^\pm= t\pm s)\,.
\end{equation}
Note that the Poincar\'e depth coordinate is denoted $s$, so that the (unregulated) asymptotic boundary is at $s = 0$. Further, we will generally set the AdS curvature scale $L_{\mt{AdS}}=1$ in the following. Meanwhile, the bath is represented by a flat Minkowski half-space:
\begin{equation}
ds^2_{\textrm{bath}}=  - \frac{dy^+ dy^-}{\epsilon^2} \qquad (y^\pm = u \mp \sigma)\,.
\end{equation}
where $\sigma$ denotes the spatial coordinate.\footnote{Our unconventional choice in defining $y^\pm$ ensures that moving further into the bath corresponds to moving towards larger positive $\y$. That is, $\sigma$ is positive in the bath, while $s$ is positive in $\textrm{AdS}$.} These two spaces are to be glued along their respective boundaries, \ie $\sigma=0$ in the bath region and $s \sim \epsilon \approx 0$ in the $\text{AdS}_2$ space, where $\epsilon$ is an UV cutoff. After this quench, energy can flow freely through the boundary from one space to the other. The $x^\pm$ coordinates can be extended to cover the bath, and the $y^\pm$ coordinates can be extended to cover a Rindler patch of the $\textrm{AdS}$.

To prepare the corresponding bulk quantum state, we Wick rotate to Euclidean signature. The Euclidean coordinates and Lorentzian coordinates are related by $x^- \to -x,\,x^+ \to \bar{x}$. This state can be mapped to the vacuum of the CFT in the upper half plane (UHP) $\operatorname{Im}\{{z}\} \geq 0$ by a local Weyl rescaling
\begin{equation}
\begin{split}
ds^2_{\mt{\text{AdS}}} &\longrightarrow \Omega(x^+,x^-)^{2} ds^2_{\mt{AdS}} = dzd\bar{z} \,, \\
ds^2_{\mt{\text{bath}}} &\longrightarrow \Omega'(y^+,y^-)^{2} ds^2_{\mt{bath}} = dzd\bar{z} \,. \\
\end{split}
\end{equation}
Explicitly,
\begin{equation}
\Omega = \frac{x^+-x^-}{2} \sqrt{z'(x)\bar{z}'(\bar{x})}\,, \quad \Omega' =  \epsilon\sqrt{z'(y)\bar{z}'(\bar{y})}\,.
\end{equation}
Before the quench, the reparameterization function $f(u)$ relating the $x$ and $y$ coordinates is given by the solution of a black hole with temperature $T_0$ in JT gravity, \ie
\begin{equation}
f(u) = \frac{1}{\pi T_0} \tanh \( \pi T_0 u \)\, \qquad (u<0),
\end{equation}
where we identify the physical time on the boundary with the coordinate $t$ via the inverse function $u = f^{-1}(t)$.
The quench occurs at $u=0$.  The quench introduces a localized positive energy shock followed by a flux of energy:\footnote{The Schwarzian is defined as $\{ f(u), u\}= \frac{-3f''^2+2f'f'''}{2f'^2}$.}
\begin{equation}
\langle T_{x^-x^-} \rangle = E_{\mt{S}} \delta (x^-) -\frac{c}{24 \pi }\{y^-,x^-\} \Theta(x^-)\,.
\end{equation}
Consistency of the change in black hole energy with this flow of energy between the AdS and bath systems demands that $f$ satisfies the following equation 
\begin{equation}\label{Schwarzian}
\{f(u), u\} = -2(\pi T_1)^2 e^{-ku} \,. 
\end{equation}
The solution was found in~\cite{Almheiri:2019psf} to be
\beq
f(u) = \frac{1}{\pi T_1} \frac{I_0\[\frac{2\pi T_1}{k}\] K_0 \[\frac{2\pi T_1}{k}e^{-ku/2}\]-I_0\[\frac{2\pi T_1}{k}e^{-ku/2}\] K_0 \[\frac{2\pi T_1}{k}\]}{I_1\[\frac{2\pi T_1}{k}\] K_0 \[\frac{2\pi T_1}{k}e^{-ku/2}\]+I_0\[\frac{2\pi T_1}{k}e^{-ku/2}\] K_1 \[\frac{2\pi T_1}{k}\]}\,.
\eeq
where $k \ll T_1$ is a constant that determines the relative strength of backreaction compared to the entropy: 
\begin{equation}
	\frac{c}{12} = k \frac{\bar{\phi}_r}{4\GN }~.
\end{equation}

After the quench, the horizon shifts, corresponding to the change in temperature. The new horizon corresponds to $x^+ = t_\infty$, where
\begin{equation}\label{tinfinity}
t_{\infty}=f(u=\infty)=\frac{1}{\pi T_{1}} \frac{I_{0}\left[\frac{2 \pi T_{1}}{k}\right]}{I_{1}\left[\frac{2 \pi T_{1}}{k}\right]}=\frac{1}{\pi T_{1}}+\frac{k}{4\left(\pi T_{1}\right)^{2}}+O\left(k^{2}\right)
\end{equation}
After taking the limit of very large $E_S \equiv \frac{\phi_r \pi }{4\GN }\( T_1^2 -T_0^2\)$, the map to the UHP is achieved by the piecewise-M\"obius map~\cite{Almheiri:2019psf}
\begin{equation}
z = \begin{cases}
\(\frac{12\pi}{c}E_S\)^{-2} \frac{i}{f(y)} & \textrm{for } y>0, \\
- i y & \textrm{for } y<0,       \\
\end{cases}
\end{equation}
or equivalently in terms of $x$ coordinates,
\beq
\label{eq:zmap}
z = \begin{cases}
	\(\frac{12\pi}{c}E_S\)^{-2} \frac{i}{x} & \textrm{for } x>0\,, \\
	i f^{-1}(-x) & \textrm{for } x<0\,.                 \\
\end{cases}
\eeq

 We are looking for the quantum corrections to the entanglement wedge of {\qmr}. This means we need to evaluate the generalized entropy \reef{genX}, which in JT gravity means the function
 \begin{equation}\label{Sgen}
 S_{\mt{gen}}(x^+, x^-) = \frac{\phi}{4\GN } + S_{\mt{bulk}}\,,
 \end{equation}
 where
 \beq
 \phi =\phi_0 + \frac{\phi_r(x^+, x^-)}{\epsilon}\,,
 \eeq
 is the value of the dilaton. The large constant contribution from $\phi_0$ is related to the divergences associated to the short range entanglement across the end points of an interval. The spacetime-dependent $\phi_r$ takes the value $\bar{\phi}_r$ at the boundary where $\rm{AdS}$ and the bath are joined.
 
 We solve the quantum extremal surfaces, \ie the codimension-2 surfaces (points) which minimize the generalized entropy. Before the quench, the dilaton takes the simple static solution
 \begin{equation}\label{dilaton01}
 \phi=2 \bar{\phi}_{r} \frac{1-\left(\pi T_{0}\right)^{2} x^{+} x^{-}}{x^{+}-x^{-}} = 2 \bar{\phi}_{r}  \pi T_0\coth\( \pi T_0 (y^+ -y^-)\) \,,
 \end{equation}
where we used the reparameterization function for static black hole with temperature $T_0$. After the quantum quench, the $\text{AdS}_2$ geometry is modified due to the backreaction. Since 2D gravity is topological, this corresponds to a modification of the boundary. Alternatively, we can consider the $\text{AdS}_2$ geometry as fixed and account for the backreaction by putting it in the dilaton. After the shock $x^- >0$, the new solution is
 \begin{equation}\label{dilaton02}
 \phi=2 \bar{\phi}_{r} \frac{1-\left(\pi T_{1}\right)^{2} x^{+} x^{-}  + \frac{k}{2} I( x^+,x^-)}{x^{+}-x^{-}}
 \end{equation}where
 \begin{equation}
 	I = \int_0^{x^-}(x^+-t)(x^--t)\{u,t\} dt \end{equation}
 	accounts for the presence of stress-energy exchange through the boundary~\cite{Maldacena:2016upp,Almheiri:2014cka}.

In the original iteration of the {\aims} model~\cite{Almheiri:2019psf}, no assumptions are made about the bulk BCFT. The calculation of the entanglement entropy can then be carried out using replica techniques~\cite{Cardy:1984bb,Calabrese:2007rg,Coser_2014}.
% On the other hand, the bulk entropy can be calculated by taking replica trick and considering the R\'enyi entropy. Thanks to the two-dimensional CFT, the question is transfered to the calculations of expectation value of twist operators at the endpoints of the intervals.
% We can first calculate it in the upper half plane (UHP)~\cite{Cardy:1984bb,Calabrese:2007rg,Coser_2014} and then map it to the physical coordinate system. The method is based on the results for two point function of twist operators on UHP
% \begin{equation}
% \langle \mathcal{T}_n \(z_1\)\mathcal{T}_{-n} \(z_0\) \rangle_{\text{UHP}} = \frac{\mathcal{F}_n\( \eta\)}{\left|(1-\eta) \cdot (z_1-\bar{z}_1) (z_0-\bar{z}_0)) \right|^{\Delta_n}}  = \frac{\mathcal{F}_n\( \eta\) }{\left|\eta\, (z_1-z_0) (\bar{z}_1-\bar{z}_0)) \right|^{\Delta_n}} \,,
% \end{equation}
% where $\Delta_n = \frac{c}{12}(n-1/n)$ is the scaling dimension of twist operator, the $\eta$ is the conformally invariant cross-ratio
 %\begin{equation}
 %\zeta = \frac{(z_1 -z_0)(\bar{z}_1-\bar{z}_0)}{(z_1-\bar{z}_0)(\bar{z}_1-z_0)} \equiv %\frac{z_{10}z_{\bar{1}\bar{0}}}{z_{1\bar{0}}z_{\bar{1}0}} \quad \in [0,1] \,, \(1-\zeta \equiv %(-)\frac{(z_1-\bar{z}_1)(z_0-\bar{z}_0)}{(z_1-\bar{z}_0)(\bar{z}_1-z_0)}\)
 %\end{equation}
In terms of the conformal cross-ratio
 \begin{equation}\label{cross}
 \eta\equiv \frac{(z_1-\bar{z}_1)(z_0-\bar{z}_0)}{(z_1-\bar{z}_0)({z}_0-\bar{z}_1)} ,
 \end{equation}
% and function $\mathcal{F}$ only depends on the invariant cross-ratio but is not universal which means it varies for different CFT theories. The above two point function of CFT on UHP can be derived from the four point function on the full plane according to the image method~\cite{Cardy:1984bb,DiFrancesco:1997nk}.
the entanglement entropy of an interval with endpoints $z_0$ and $z_1$ in a two-dimensional BCFT with boundary at $z-\bar{z}=0$ is \begin{equation}
S_{\mt{UHP}} = \frac{c}{6} \log \(\frac{|z_0-z_1|^2}{\delta^2} \eta\) + \log \mathcal{F}(\eta).
\end{equation}
Here, $\delta$ is a UV cutoff and $\mathcal{F}(\eta)$ is a function which depends on the theory living on the boundary defect. In the limit $\eta \rightarrow 1$, we are in the OPE limit, whence $\mathcal{F}(1) = 1$; in the opposite limit $\eta \rightarrow 0$, we instead have $\mathcal{F}(0) = g^2$, where $\log g$ is the Affleck-Ludwig boundary entropy~\cite{Affleck_1994}.

For our purposes, however, we wish to work with the holographic model described in~\cite{Almheiri:2019hni}. In this case, the matter theory is a holographic BCFT. Thus, we can imagine the JT gravity theory plus bath system as living on the boundary of a new, asymptotically-AdS$_3$ bulk. Because of the boundary defects, there is a dynamical ETW brane hanging into the space~\cite{Takayanagi:2011zk,Fujita:2011fp}. After the quench, the ETW brane detaches from the asymptotic boundary (where the JT gravity and bath are connected) and falls into the bulk.

A particularly convenient aspect of the holographic model is that the entanglement entropy is now determined simply using the Hubeny-Rangamani-Ryu-Takayanagi prescription~\cite{RyuTak06, Rangamani:2016dms}. In this setup, this simply means evaluating the length of the minimal geodesic homologous to the entangling region. In this case, the HRT surfaces are allowed to end on the ETW brane.

In this case, the entanglement entropy of one interval reduces to
\begin{equation}\label{Sbulk_hol}
S_{\mt{UHP}} = \begin{cases}
\frac{c}{6} \log \( \frac{|z_1-z_0|^2}{\delta^2}\) & \textrm{for } \eta >\eta_* \\
\frac{c}{6} \log  \( \frac{|z_1-\bar{z}_1|}{\delta}\cdot \frac{|z_0 -\bar{z}_0|}{\delta}\)+2\,\log g & \textrm{for } \eta<\eta_*
\end{cases}
\end{equation}
where $\eta_* = {(1+g^{12/c})}^{-1}$ is the value of the cross-ratio at which the transition between families of HRT surfaces occurs. Without loss of generality for our purposes, we take $g=1$ from now on, so that $\eta_* = 1/2$. We will discuss the role of $g$ in more detail in section~\ref{sec:discuss}.
%This phase transition corresponds to the transition from the quench phase to the scrambling phase, as described above.

\begin{figure}[t]
	\centering\includegraphics[width=3.5in]{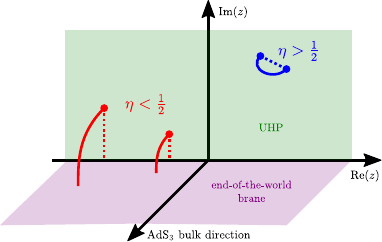}
	\caption{The entanglement entropy for an interval in a holographic BCFT on the upper half-plane has two branches. The dominant branch is determined by the cross ratio $\eta$ defined in eq.~\reef{cross}. The case illustrated here corresponds to a tensionless ETW brane in the bulk, or alternatively $\log g=0$ in the BCFT. For other choices of $\log g$, the ETW brane will be tensionful and intersect the UHP at some other angle.}\label{holoEE}
\end{figure}

Figure~\ref{holoEE} illustrates the two families of bulk geodesics in the two different branches contained in eq.~\reef{Sbulk_hol} (with $\log g=0$). The $\eta \ge \frac{1}{2}$ channel corresponds to a single geodesic stretching between the two endpoints, while the $\eta \le \frac{1}{2}$ channel corresponds to two geodesics (one from each endpoint) terminating on the ETW brane. This formula matches with CFT calculations in the large $c$ limit; the phase transition between the two channels follows from the universality of the four-point function on the full plane in a holographic theory~\cite{Hartman:2013mia,Leichenauer:2015xra}.

% conformal block decomposition always factorizes in either the s-channel or t-channel resulting in the factorized function $\mathcal{F}$ in large $c$ limit, \ie
%\begin{equation}
%\mathcal{F}_n \( \eta\) = \begin{cases}
%(1-\eta)^{\Delta_n}\,, \quad  \eta \in [0,\frac{1}{2}]   \\
%\eta^{\Delta_n}\,, \quad \qquad \eta \in [\frac{1}{2},1] \,,\\
%\end{cases}\,.
%\end{equation}
%where the sharp phase transition happens at $\eta= \frac{1}{2}$ \footnote{Here except for the large $c$ limit, we also need to assume the CFT only has a sparse spectrum, \ie the number of operator not scale with the $c$. And we should also expect the transition for general CFT should be very smooth due to the finite $c$ contributions from the two channels.}.

%\begin{figure}[t]
%	\centering\includegraphics[width=3.50in]{UHP_01.pdf}
%	\caption{The two point function for a holographic CFT ($c \gg 1$ and sparse spectrum) on the upper half-plane, with two different channels. The dominant channel is determined by the cross ratio between $z_0$ and $z_1$. }\label{UHP_plot}
%\end{figure}

Employing this holographic formula~\eqref{Sbulk_hol} for the entropy of an interval on the upper half-plane, we can find the bulk entropy we need by taking the conformal transformation to the physical coordinate system. Because of the conformal invariance, this reduces to the answer on the upper half-plane, except for the transformation of the cutoff at each endpoint:
 \begin{equation}
 S_{\mt{bulk}} = S_{\Omega^{-2}g} = S_{\mt{UHP}}- \frac{c}{6} \sum_{x_i \in \partial} \log \Omega(x_i)
 \end{equation}
 where the sum runs over all the endpoints of the intervals.

Note that all of the entanglement entropies which we calculate in the following are formally UV divergent, because of the UV cutoff $\delta$ appearing in eq.~\reef{Sbulk_hol}. However, in any of our analyses, we are also comparing different branches with the same number of endpoints in the bath and so these $\delta$ contributions do not play a role. Hence in any expressions which are explicitly shown in equations or plotted in the figures, we simply subtract $\frac{c}{6}\,\log(L_{\mt{AdS}}/\delta)$ for each of the endpoints. Of course, in the holographic description, these UV divergences appear because of the infinite length appearing when the HRT geodesics extend to the AdS$_3$ boundary. A similarly large length appears when these bulk geodesics terminate on the Planck brane. In this case, the divergences are absorbed by the gravitational contribution in the generalized entropy \reef{Sgen}. In particular, these divergences are associated with renormalizing the coupling to the topological Einstein term in the JT action, \ie $\phi_0/4\GN$~\cite{Almheiri:2019psf,Almheiri:2019hni}.\footnote{For details on how this occurs in general dimensions, refer to the appendix of~\cite{BouFis15a}.}

%% file: sections/QES.tex
% !TEX root = ../BH_v04.tex
We now review the results in~\cite{Almheiri:2019psf} for finding quantum extremal surfaces. Finding these surfaces requires computing the generalized entropy for an interval with one point in the $\text{AdS}_2$ and another point on the boundary. We assume the simple holographic results for bulk entropy, where we found a small change in the behavior of the quantum extremal surface before the shock relative to the results of~\cite{Almheiri:2019psf}. Unless otherwise specified, we will use the parameters in table~\ref{tab:baseline} as our baseline parameters in all of our numerics.

\begin{table}[t]
	\centering
	\begin{tabular}{l|c|c|c|c|c|c|c|c}
	\hline
		\hline
		& & &    & & & &\\[-2ex]
		\textbf{Parameter} & $L_{\mt{AdS}}$ &  $k$ & 	$T_1$ & $T_0$ &c &$\epsilon$ &	$\phi_0$ &	$\bar{\phi}_r$  \\ \hline
		& & &    & & & &\\[-2ex]
		\textbf{Value} & 1	& $\frac{1}{4096}$  & $\frac{1}{\pi}$ &$\frac{63}{64\pi}$ & 4096	& $\frac{1}{4096}$ & 0 & $\frac{1}{4096^2}$\\[0.5ex] \hline \hline
	\end{tabular}
	\caption{Baseline parameters for this work. Unless otherwise specified, all of our figures are generated using these values for the parameters.}
	\label{tab:baseline}
\end{table}

Before the shock, the possible contributions to generalized entropy is also divided into two different phases according to the position of endpoints. After the shock, the cross-ratio is fixed to be $1$ at leading order in $E_S^{-2}$, as in~\cite{Almheiri:2019psf}.

\subsubsection{Finding the phase transitions}
\begin{figure}[t]
	\centering
	\begin{subfigure}[t]{0.45\textwidth}
		\centering
		\includegraphics[width=0.8\textwidth]{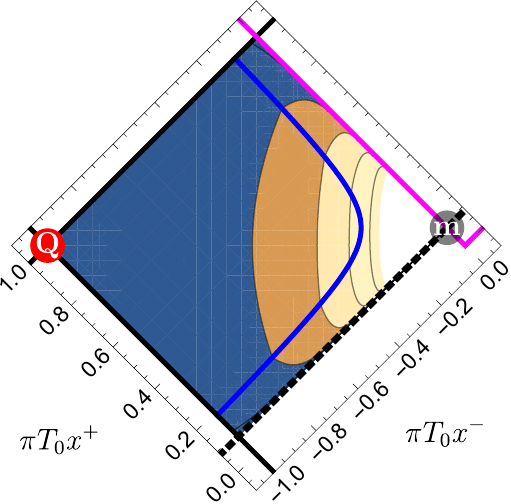}
		\caption{Quench Phase.}
	\end{subfigure}
	\hfill
	\begin{subfigure}[t]{0.45\textwidth}
		\centering
		\includegraphics[width=0.8\textwidth]{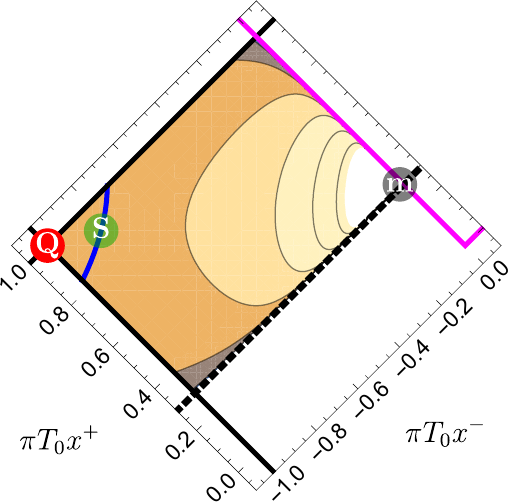}
		\caption{A new extremum (point S) emerges, but is non-minimal.}
	\end{subfigure}
	\\
	\begin{subfigure}[t]{0.45\textwidth}
		\centering
		\includegraphics[width=0.8\textwidth]{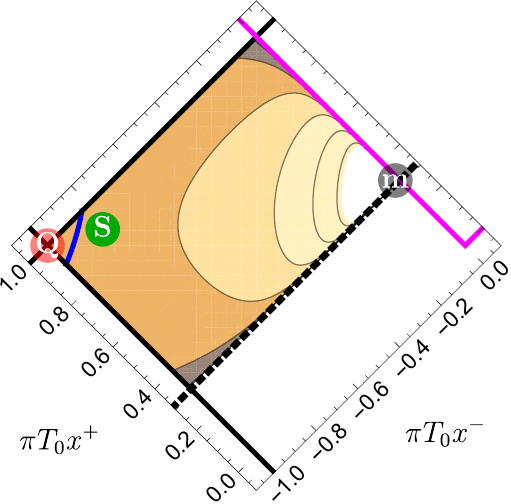}
		\caption{Transition to Scrambling Phase.}
	\end{subfigure}
	\hfill
	\begin{subfigure}[t]{0.45\textwidth}
		\centering
		\includegraphics[width=0.8\textwidth]{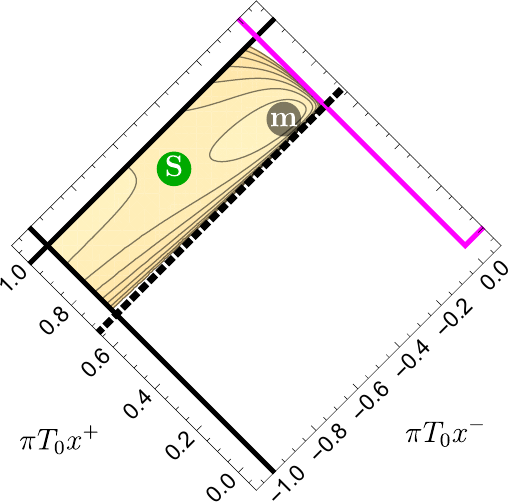}
		\caption{Instant before Page transition.}
	\end{subfigure}
	\caption{Motion of QES and other (non-minimal) extrema in the Quench and Scrambling Phases. The sub-figures show contour plots of generalized entropy as a function of $x_{\QES}$ in the region bounded by the initial black hole horizon (solid black lines), a past null ray (dotted black line) emanating from the point $x_1$ on the AdS-bath boundary, and the shock (magenta lines); dark blue and bright yellow shading indicate low and high generalized entropies respectively. The blue curve marks points for which $\eta=1/2$. Three extrema of generalized entropy are shown: the bifurcation point (Q), a saddle point (S), and a maximum point (m). The QES (opaque point) in the Quench and Scrambling Phases is given respectively by Q and S. In order to make various qualitative features visible in this figure, we have chosen parameters differing from the baselines listed in table~\ref{tab:baseline}; here, $\epsilon=\frac{1}{16}$, $c=16$, $k=\frac{1}{16}$, $T_0=\frac{2}{3\pi}$, $T_1=\frac{1}{\pi}$, $\phi_0=0$, and $\phi_r=\frac{1}{256}$.}
	\label{fig:nosleep}
\end{figure}

%We start from the simple interval with only two points, with one point in the bulk AdS$_2$ space and another on the AdS$_2$ boundary. The result is
%\begin{equation}
%\begin{split}
%\text{AdS}_2, \text{ before}\, \quad &x^{\pm}_0 :  x^+_0 >0, x^-_{0}<0 \,,  \text{ or} \quad \text{after}\quad   x^+_0 >0, x^-_{0}>0\,,\\
%\text{bounday},\text{after} \, \quad  &x_{1}^{\pm} : x^+_1>0, x^-_1>0 \,,
%\end{split}
%\end{equation}
Consider a bulk region defined by the interval between two points, $x_\QES^\pm$ and $x_1^\pm$. (More correctly, consider the domain of dependence of this interval.) As a warm-up, we take $x_0$ to lie in the bulk and $x_1$ to be near the boundary. In this case we can relabel the point $x_1$ in terms of the proper time $u$ along the boundary,
\begin{equation}
t= f(u)=\frac{x^+_1 + x^-_1}{2}\,,\quad  z=\frac{x^+_1 - x^-_1}{2}=\epsilon f'(u)\,.
\end{equation}
From the above holographic formula for entanglement entropy with two points, we can fix the choice of bulk entropy by taking account of the cross-ratio decided by the position of $\text{AdS}_2$ point, giving
\begin{equation}\label{Sbulk_hol_2}
S_{\rm bulk}=\begin{cases}
\frac{c}{6} \log \( \frac{|z_1-\bar{z}_1|\cdot|z_{\QES} -\bar{z}_\QES|}{\Omega_1 \Omega_\QES \delta^2} \) & \text{for } \eta \in [0,\frac{1}{2}) \\
\frac{c}{6} \log \( \frac{|z_1-z_\QES|\cdot|\bar{z}_1 -\bar{z}_\QES|}{\Omega_1 \Omega_\QES\delta^2} \) & \text{for } \eta \in [\frac{1}{2},1]
\end{cases}
\end{equation}
The first formula (where $\eta <  \frac{1}{2}$) is only applicable when the bulk endpoint lies before the shock with $x^-_\QES < 0$. In this formula, the entropy factorizes into contributions from the two endpoints. (For an idea of what the $\eta < \tfrac{1}{2}$ region looks like, consult figure~\ref{fig:nosleep}.)

The second formula (where $\eta \ge \tfrac{1}{2}$) holds both before and after the shock. However, because the map from the upper half plane to the physical coordinates depends on whether the interval straddles the shock or lies to its future, the formulas for the bulk entropy will still depend on this distinction.

\begin{table}[t]
  \centering
  \begin{tabular}{c|ccc}
  \hline
  \hline
  \textbf{Phase} & \textbf{Range of $\eta$} & \textbf{Position relative to shock} \\ \hline
  &&&\\[-2.2ex]
  Quench & $[0, \tfrac{1}{2})$ & Straddling ($x^-_\QES \le 0$) \\
  Scrambling & $[\tfrac{1}{2},1)$ & Straddling ($x^-_\QES \le 0$) \\
  Late-Time & $\approx 1$ & Above ($x^-_\QES \ge 0$) 
  \\[.3ex] \hline \hline
  \end{tabular}
  \caption{A summary of the range of parameters determining the phase of the von~Neumann entropy. In Lorentzian coordinates, $\eta = x^+_1(x^+_{\QES}-x^-_{\QES}) / [x_\QES^+(x_1^+-x_\QES^-)]$.}
    \label{tab:phases}
\end{table}

In total, we end up with the following bulk von~Neumann entropy formulas for the three phases defined in table~\ref{tab:phases}:
\begin{align}
  S_\text{bulk, quench} &= \frac{c}{6} \log \left(\frac{24\pi E_{S}}{\epsilon c} \frac{u t}{\sqrt{f^{\prime}(u)}}\right) \label{bulk_entropy}\,,\\
  S_\text{bulk, scrambling} &= \frac{c}{6} \log \left(\frac{24 \pi E_{S}}{\epsilon c} \frac{u x_\QES^-\(t-x_\QES^+\)}{(x_\QES^+-x_\QES^-)\sqrt{f^{\prime}(u)}}\right)\label{bulk_entropy2}\,, \\
  S_\text{bulk, late-time} &= {\frac{c}{6} \log \left[\frac{2\left(u-y^{-}_\QES\right)\left(x_\QES^{+}-t\right)}{\epsilon\left(x_\QES^{+}-x_\QES^{-}\right)} \sqrt{\frac{f^{\prime}(y^-_\QES)}{f^{\prime}(u)}}\right]}\,.
\end{align}

With these ingredients in place, we can find the generalized entropy in each of the three phases,
\begin{equation}\label{Sgen_two}
S_{\mt{gen}} = \frac{\phi }{4\GN } + S_{\mt{bulk}}\,,
\end{equation}
and find the quantum extremal surfaces which are stationary points of this equation, using
\begin{equation}
\partial_{+} S_{\text{gen}} =0 \,, \qquad \partial_{-} S_{\text{gen}} =0.
\end{equation}
where we abbreviate $\partial_\pm$ to mean $\partial_{x_{QES}^\pm}$ to simplify the notation.

% \begin{equation}
% S_{\mt{bulk}}=\left\{\begin{array}{ll}
% \frac{c}{6} \log \left(\frac{24\pi E_{S}}{\epsilon c} \frac{u t}{\sqrt{f^{\prime}(u)}}\right)\,,(\text{A})                                                                        & \eta \in [0,\frac{1}{2}]\,, {x^{-}<0<t<x^{+}} \\
% \,                                                                                                                                                                                                                                              \\
% \frac{c}{6} \log \left(\frac{24 \pi E_{S}}{\epsilon c} \frac{u x^-\(t-x^+\)}{(x^+-x^-)\sqrt{f^{\prime}(u)}}\right)\,,(\text{B})                                                    & \eta \in [\frac{1}{2},1]\,, {x^{-}<0<t<x^{+}}  \\
% \,                                                                                                                                                                                                                                              \\
% {\frac{c}{6} \log \left[\frac{2\left(u-y^{-}\right)\left(x^{+}-t\right)}{\epsilon\left(x^{+}-x^{-}\right)} \sqrt{\frac{f^{\prime}(y^-)}{f^{\prime}(u)}}\right]} \,, (\text{u_HP}) &\quad  \eta=1\,,{0<x^{-}<t<x^{+}}\,,
% \end{array}\right.
% \end{equation}
% {Even though we find the holographic CFT has a sharp transition at $\eta = \frac{1}{2}$, later we will see the transition between two phases for generalized entropy actually happens at other point because the QES is defined at the branch with minimal entropy. }
\subsubsection{Quantum extremal surfaces at early times}
\begin{figure}[t]
	\centering\includegraphics[width=3.00in]{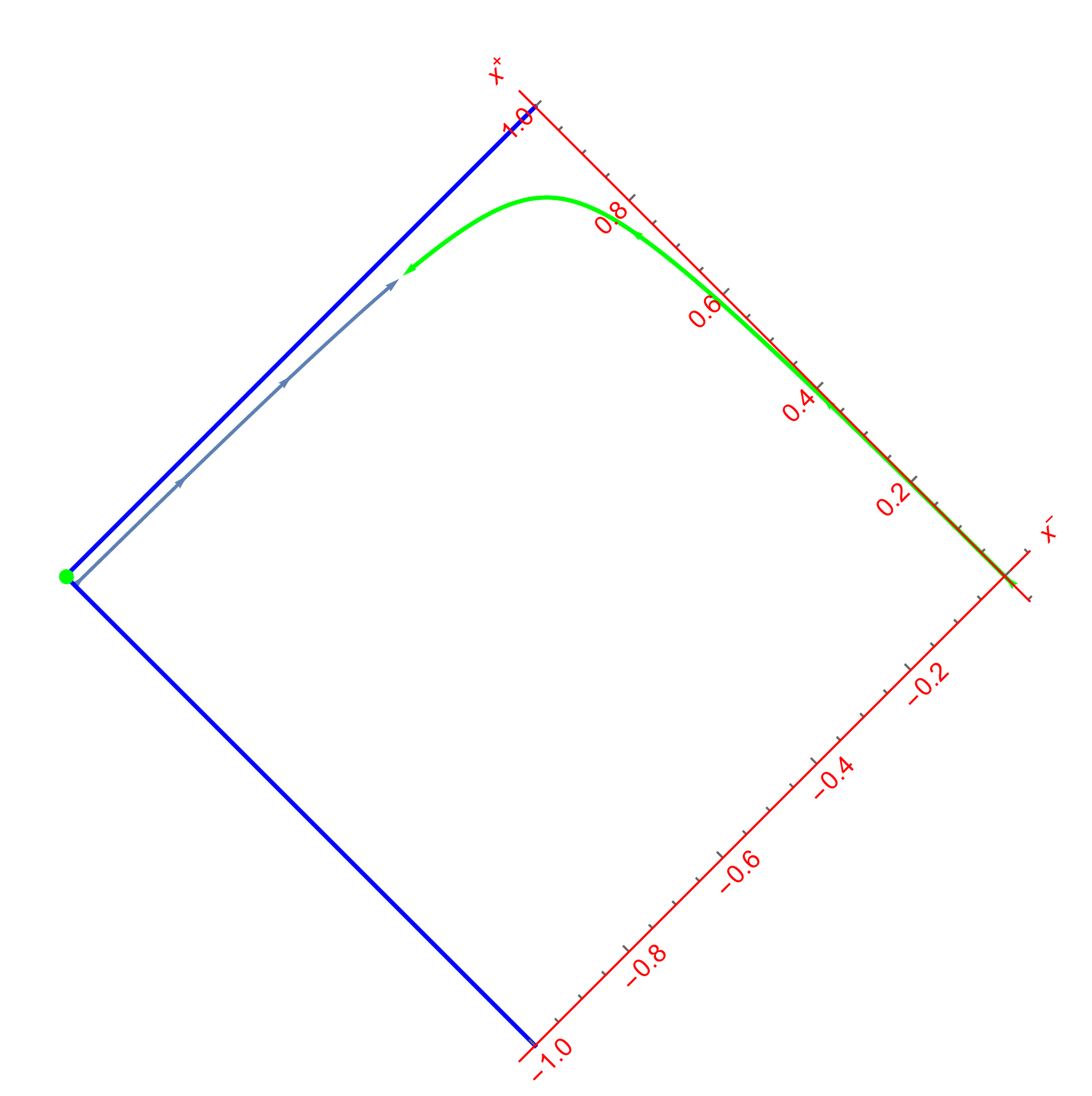}
	\caption{The time evolution of quantum extremal surfaces. The arrow indicates the direction of the flow. The blue line is the physical solution we considered in the paper. It starts at the bifurcation point and ends at a point away from shock. The green one is another branch of the solution with larger entropy. Here, we choose a large $k$ to make the deviation from the horizon more obvious when plotted.}\label{QES_plot}
\end{figure}
It is easy to minimize the generalized entropy in the quench phase, because the bulk von~Neumann entropy in this phase is independent of $x_\QES$. The problem reduces to finding the saddle point of the dilaton, which is of course the bifurcation surface of the original (temperature $T_0$) black hole at
\begin{equation}
x^{\pm}_\QES= \pm \frac{1}{\pi T_0}\,.
\end{equation}
Even though the quantum extremal surface is fixed to the bifurcation surface, the generalized entropy still evolves with time, and is given by 
\begin{equation}
S_{\text{gen, quench}} = \frac{\bar{\phi}_r}{4 \GN } \(  2\pi T_0    + 2k \log \left(\frac{24 E_{S}}{\epsilon c} \frac{u t}{\sqrt{f^{\prime}(u)}}\right)\) \qquad (\eta \le  \tfrac{1}{2}).
\end{equation}
This solution is relevant only when
\begin{equation}
t\le \frac{1}{3\pi T_0}.
\end{equation}

\begin{figure}[t]
	\includegraphics[width=0.56\textwidth]{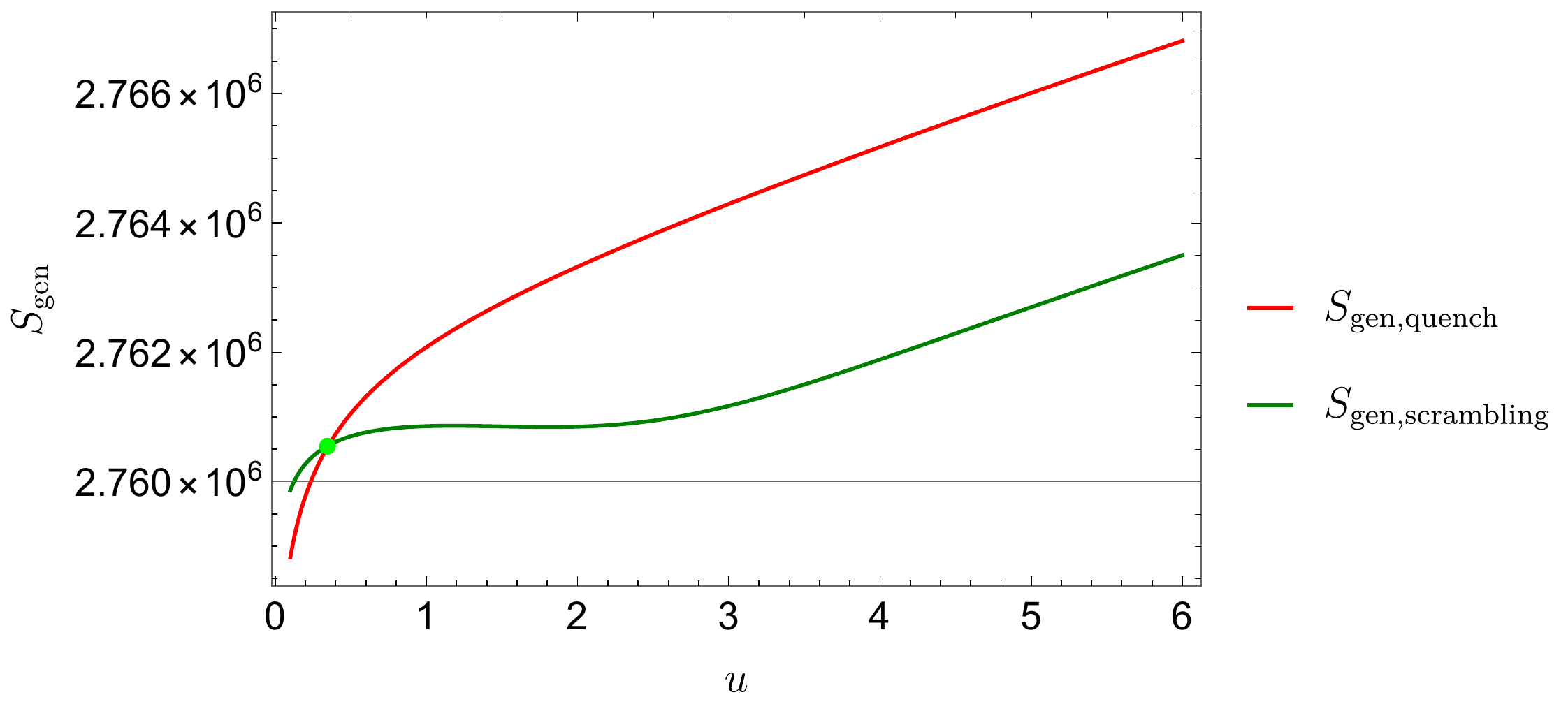} \hspace{0.001\textwidth}
	\includegraphics[width=0.47\textwidth]{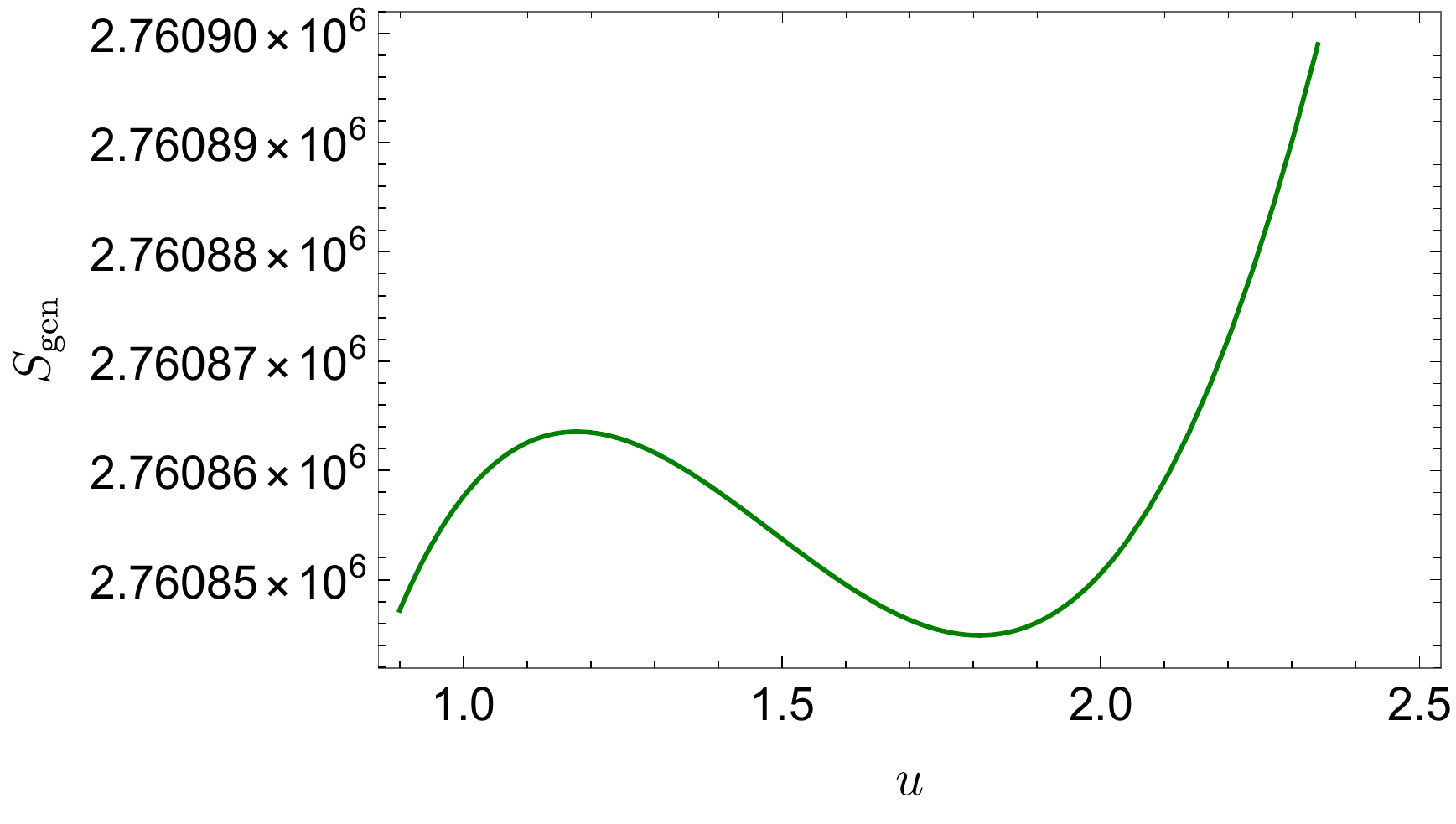}
	\caption{The generalized entropy from full solutions. The green curve is derived from~\eqref{bulk_entropy2} with exact solutions of~\eqref{difconstraints}. The red one represents the generalized entropy with endpoint at bifurcation point. The green point in the right plot indicates the point $u_{\mt{QS}}$ where $S_{\mt{gen,scrambling}} =S_{\mt{gen,quench}}$.}
	\label{fig:Sgen01}
\end{figure}

Now we consider the scrambling phase. The quantum extremal surfaces in this phase are found from solving the equations
\begin{align}\label{difconstraints}
0 &= \frac{4\GN }{\bar{\phi_r}}\partial_{+} S_{\text{gen}}  =  \frac{2((\pi T_0 x_\QES^-)^2-1)}{ (x_\QES^+-x_\QES^-)^2}  +  2k \frac{x_\QES^- - t}{ (t-x_\QES^+)(x_\QES^+-x_\QES^-)}\,,
\\
0 &= \frac{4\GN }{\bar{\phi_r}}\partial_{-} S_{\text{gen}}  =  \frac{2(1-(\pi T_0 x_\QES^+)^2)}{ (x_\QES^+-x_\QES^-)^2}  +   \frac{2k x_\QES^+}{ x_\QES^-(x_\QES^+-x_\QES^-)}\,.
\end{align}
An exact solution\footnote{The above equations actually have several branches of solutions. Here we only take the solutions satisfying the constraints. Even still there is another solution, shown in figure~\ref{QES_plot} in green, which satisfies the constraints but with larger generalized entropy. This occurs because of the factor of $\log x_\QES^-$ in the entropy of the scrambling phase; the solution lies close to the shock located at $x_\QES^-=0$.} for $x_\QES^{\pm}$ can easily be found. Using these exact solutions, we plot the generalized entropy in this phase in figure~\ref{fig:Sgen01}. An approximate solution (using a small-$k$ expansion) is
\begin{align}\label{k_QES_02}
x_\QES^+ (t) &= \frac{1}{\pi  T_0}-\frac{k}{\pi ^2 T_0^2}+\frac{k^2 \left(3 \pi  T_0 t-1\right)}{2 \pi ^3 T_0^3 \left(\pi  T_0 t-1\right)}+\mathcal{O}(k^3) < \frac{1}{\pi T_0}\,,\\
x_\QES^-(t) &= -\frac{1}{\pi  T_0}-\frac{k \left(\pi  T_0 t+1\right)}{\pi ^2 T_0^2 \left(\pi  T_0 t-1\right)}+\frac{k^2 \left(\pi  T_0 t+1\right)}{2 \pi ^3 T_0^3 \left(\pi  T_0 t-1\right)} + \mathcal{O}(k^3)\,,
\end{align}
The small $k$ expansion is a good approximation for this early-time regime and we need to consider more and more orders of $k$ when we move to later time region. From the $k$ expansion, we can also derive the leading contributions to generalized entropy for $u T_1 = {\cal O}(1)$ as
\begin{equation}\label{Sgen_before_k}
	\begin{split}
S_{\text{gen, scrambling}}\approx \frac{\bar{\phi}_r}{4 \GN } \bigg[&  2\pi T_0 + 2k\pi T_1u + 2k \log \left(\frac{24 \pi E_{S}}{\epsilon c}\frac{u}{2 \pi T_0 }  \right) \\
& \qquad + 2k \log \left(\frac{  T_0 \left(2 e^{-2 \pi  T_1 u}-1\right)+T_1}{2 T_1}\right)\bigg] + \mathcal{O}(k^2) \,,
	\end{split}
\end{equation}
keeping the first two orders in $k$. The first line is the dominant term in this approximation, because the second line is order $\mathcal{O}\left(k \log\left(\frac{T_1-T_0}{T_1} \right)\right)$.  The above approximation captures the behavior of generalized entropy at early time in the scrambling phase.  We also note the contribution from dilaton is almost constant up to a linear increase of order $k^2$:
\begin{equation}
 \frac{\phi}{\bar{\phi}_r} \approx 2 \pi  T_0+\frac{k^2 \(\pi T_0t+1\)}{\pi  T_0\(1-\pi T_0t\)}+\mathcal{O}\left(k^3\right) \,,
\end{equation}
which is  negligible at early times ($\pi T_0 t \ll 1 $).

For later times, closer to the Page time, we need to push the above approximation to next order. The next order correction to the dilaton takes the form
\begin{equation}
\begin{split}
\frac{\phi}{\bar{\phi}_r} &\approx   2 \pi  T_0+\frac{k^2 }{\pi  T_0}  \frac{2T_1}{T_1- T_0} +\mathcal{O}\left(k^3\right) \,.
\end{split}
\end{equation}
which is not negligible when $T_1-T_0$ is order $k$. Similarly, the bulk terms are corrected at order ${k^n/(T_1-T_0)^{n-1}}$. For example, we can find the linearized generalized entropy as
\begin{equation}\label{Sgen_before_k2}
\begin{split}
S_{\text{gen, scrambling}}\approx \frac{\bar{\phi}_r}{4 \GN } \bigg[  2\pi T_0 &+ 2k\pi T_1u + 2k \log \left(\frac{24 \pi E_{S}}{\epsilon c}\frac{u}{2 \pi T_0 } \frac{  T_1 -T_0}{2 T_1}\right)\\
&+ \frac{1}{2} k^2 \left(- \pi  T_1 u^2+\frac{5}{\pi  \left(T_0-T_1\right)}+ u\right)\bigg] +\mathcal{O}(k^2) \,.
\end{split}
\end{equation}

From this approximation, it is easy to find the almost linear growth of generalized entropy as a function of proper time $u$ in the regime $ 1/\pi T_1 \ll u < k^{-1}$, as shown in figure~\ref{fig:Sgen01}. This linear growth is dominated by the second term tracing back to the bulk entropy term. However, at later times, the terms which we dropped at small $k$ become important. Even still, it is easy to find that the evolution in $x^+$ direction at late time is very small. So we can take the approximation
\begin{equation}
 x_\QES^+(t) \approx  x_\QES^+(t_{\infty}) >  t_{\infty}\,, \qquad t \approx t_{\infty}\,.
\end{equation}
and most parts in generalized entropy will be around a constant decided by its value at $t_{\infty}$. For example,
\begin{equation}
\log \( \frac{1}{x_\QES^+-x_\QES^-} \) \approx  - \log \( \frac{2}{\pi T_0} + \frac{2k}{(\pi T_0)^2(\pi T_0 t_{\infty}-1)} \) \,.
\end{equation}
In order to get a simple expression for generalized entropy, we define a constant to approach parts of generalized entropy
\begin{equation}\label{chi88}
\kappa = \frac{2(1- \pi T_0 x_\QES^+)(\pi T_0 x_\QES^- +1)}{x_\QES^+ -x_\QES^-} + 2k \log \( \frac{\pi T_1 x_\QES^- (t-x_\QES^+)}{x_\QES^+-x_\QES^-} \)  \bigg|_{t \to t_{\infty}} \,,
\end{equation}
and rewrite the entropy for very late $u$ as
\begin{equation}\label{Sgen_B_late}
\begin{split}
S_{\text{gen, scrambling}}&\approx   \frac{\bar{\phi}_r}{4 \GN } \bigg[  2\pi T_0 + 2k \log \left(\frac{24\pi E_{S}}{\epsilon c}\frac{u}{\pi T_1\sqrt{f^{\prime}(u)}}  \right)  + \kappa\bigg]  \,,
\end{split}
\end{equation}
The evolution of generalized entropy is dominated by the derivative term whose approximation is derived as
\begin{equation}
 \log \frac{1}{\sqrt{f'(u)}} \approx  \frac{2\pi T_1}{k} \( 1- e^{-ku/2} \) -\frac{1}{2}\log \(4\pi T_1 t_{\infty} \)  +\frac{k u}{4} + \mathcal{O}(k e^{ku})
\end{equation}
For the late-time region $ku <1$, the above term leads us to a linear increasing entropy
\begin{align}
S_{\text{gen, scrambling}} \approx   \frac{\bar{\phi}_r}{4 \GN } \bigg[  2\pi (T_0 + T_1 k u) &+2k \log \left(\frac{24\pi E_{S}}{\epsilon c}\frac{u}{2\pi T_1}   \right)\nonumber \\
&-\frac{\pi T_1}{2}k^2 u^2+\frac{k^2}{2}u+  \kappa\bigg]  \,,
\end{align}
as show in figure~\ref{fig:Sgen01}. Physically, we can understand this linear increase of entropy as the increase of entanglement between the Hawking radiations and their partners left behind the event horizon. For very late times $u > k^{-1}$, one can see from the above formula ~\eqref{Sgen_B_late} that the linear dependence on time $u$ breaks down and the entropy is dominated by the logarithmic term.

Having located the candidates for the quantum extremal surfaces in each phase using eq.~\eqref{bulk_entropy}, we need to compare their generalized entropies and pick the minimal solution. Using the approximate formulae, we can find the transition occurs at
\begin{equation}
\log t_{\mt{QS}} \approx \log \( \frac{x_\QES^- (t_{\mt{QS}}-x_\QES^+)}{x_\QES^+-x_\QES^-} \)  +\frac{k \(\pi T_0t_{\mt{QS}}+1\)}{2\pi  T_0\(1-\pi T_0t_{\mt{QS}}\)} + \dots \,,
\end{equation}
which gives the approximate solution
 \begin{equation}
 \label{eq:scramble}
  t_{\mt{QS}}\equiv f(u_{\mt{QS}}) \approx \frac{1}{3 \pi  T_0}-\frac{4 k}{9 \pi ^2 T_0^2}+\frac{7 k^2}{27 \pi ^3 T_0^3} + \dots \,.
\end{equation}

\subsubsection{Quantum extremal surfaces at later times}
\begin{figure}[t]
	\includegraphics[width=0.515\textwidth]{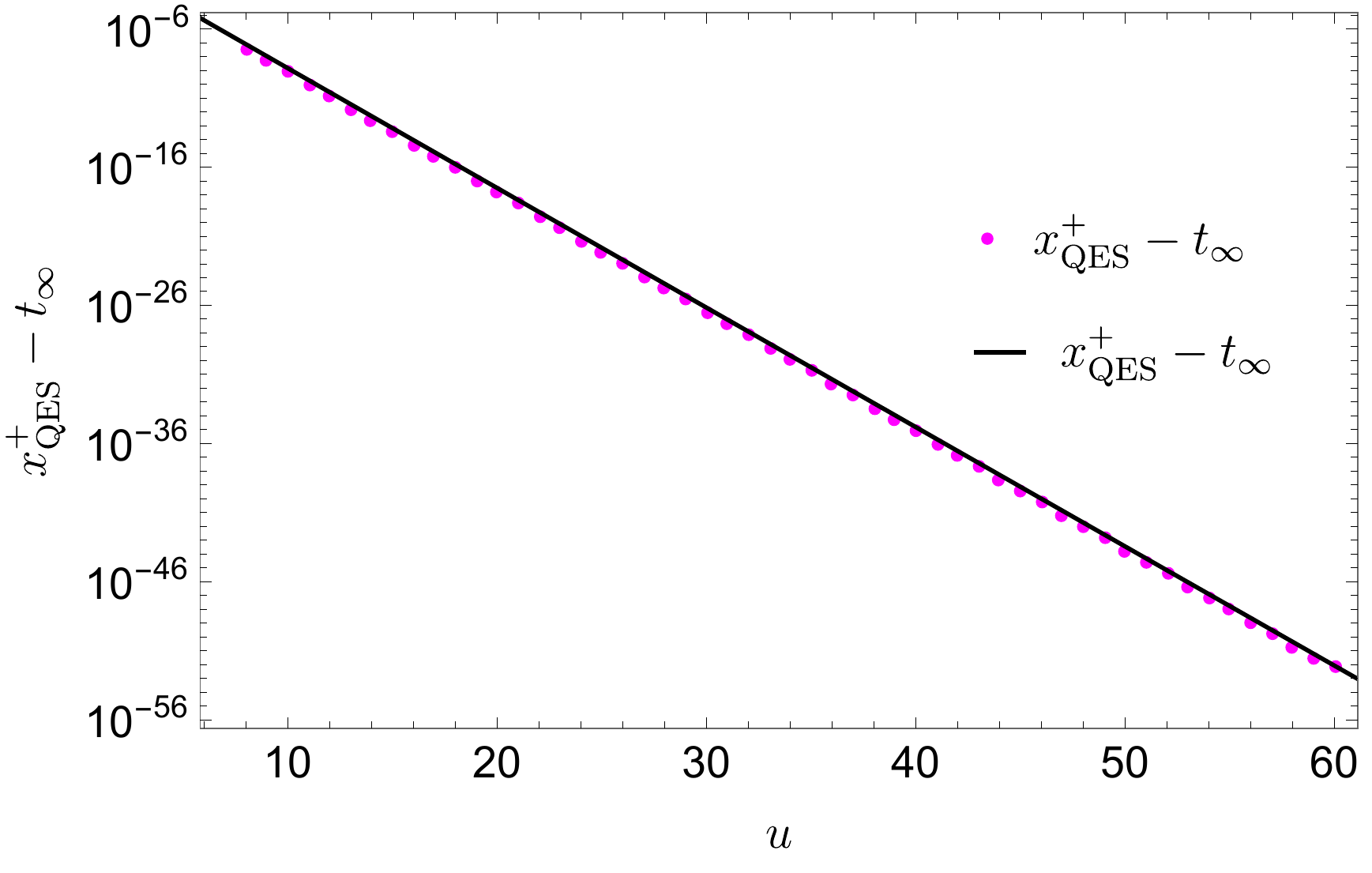} \hspace{0.01\textwidth}
	\includegraphics[width=0.49\textwidth]{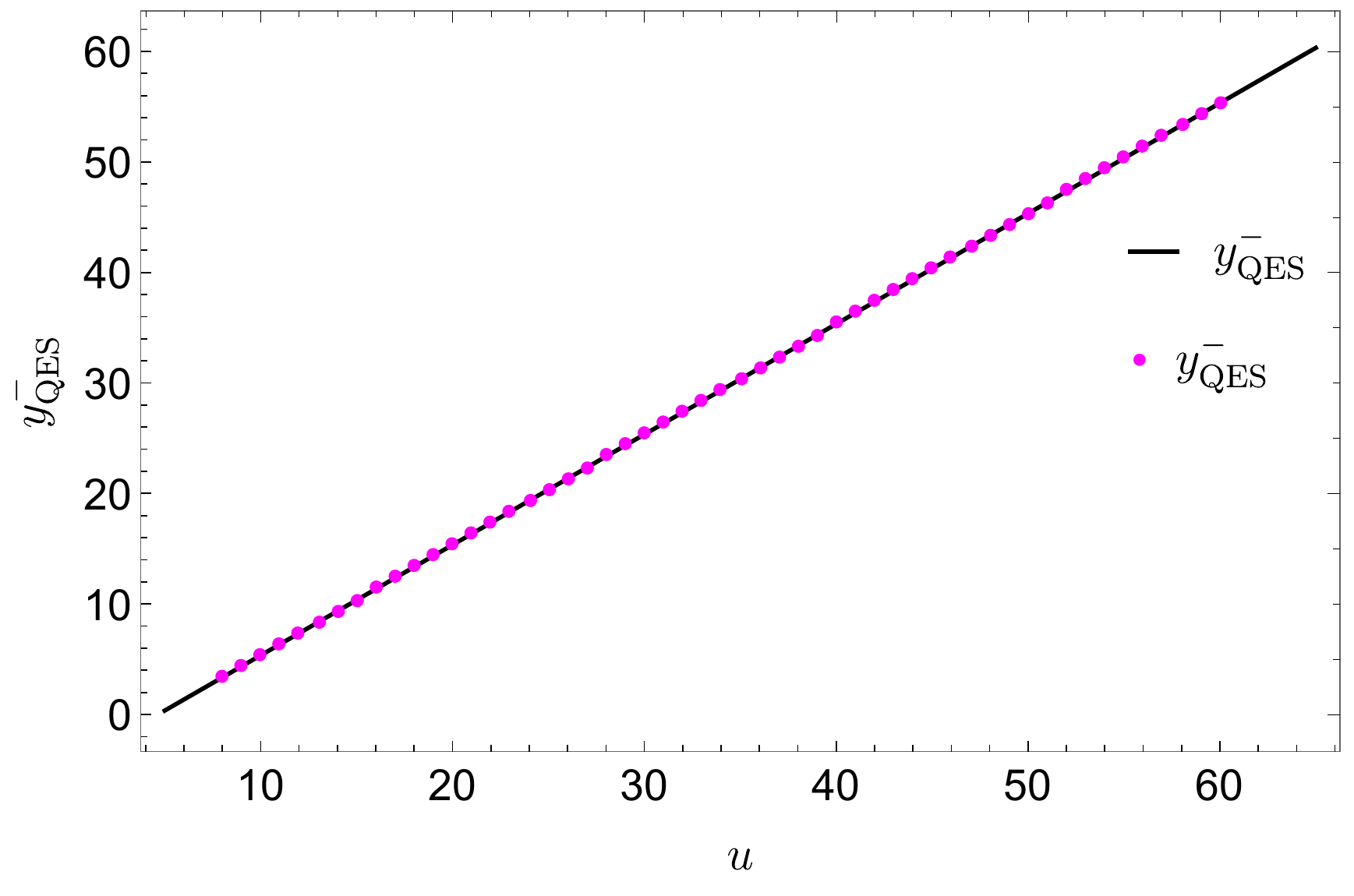}
	\caption{The numerical solutions $x_\QES^{+}, y_\QES^-$ from eqs.~\eqref{QES_equations} is presented by the dotted lines. Note that the left plot is a log plot. The solid line is the linear approximation from~\eqref{QES_linear}. }
	\label{fig:QESxpym}
\end{figure}
\begin{figure}[t]
	\centering\includegraphics[width=5.80in]{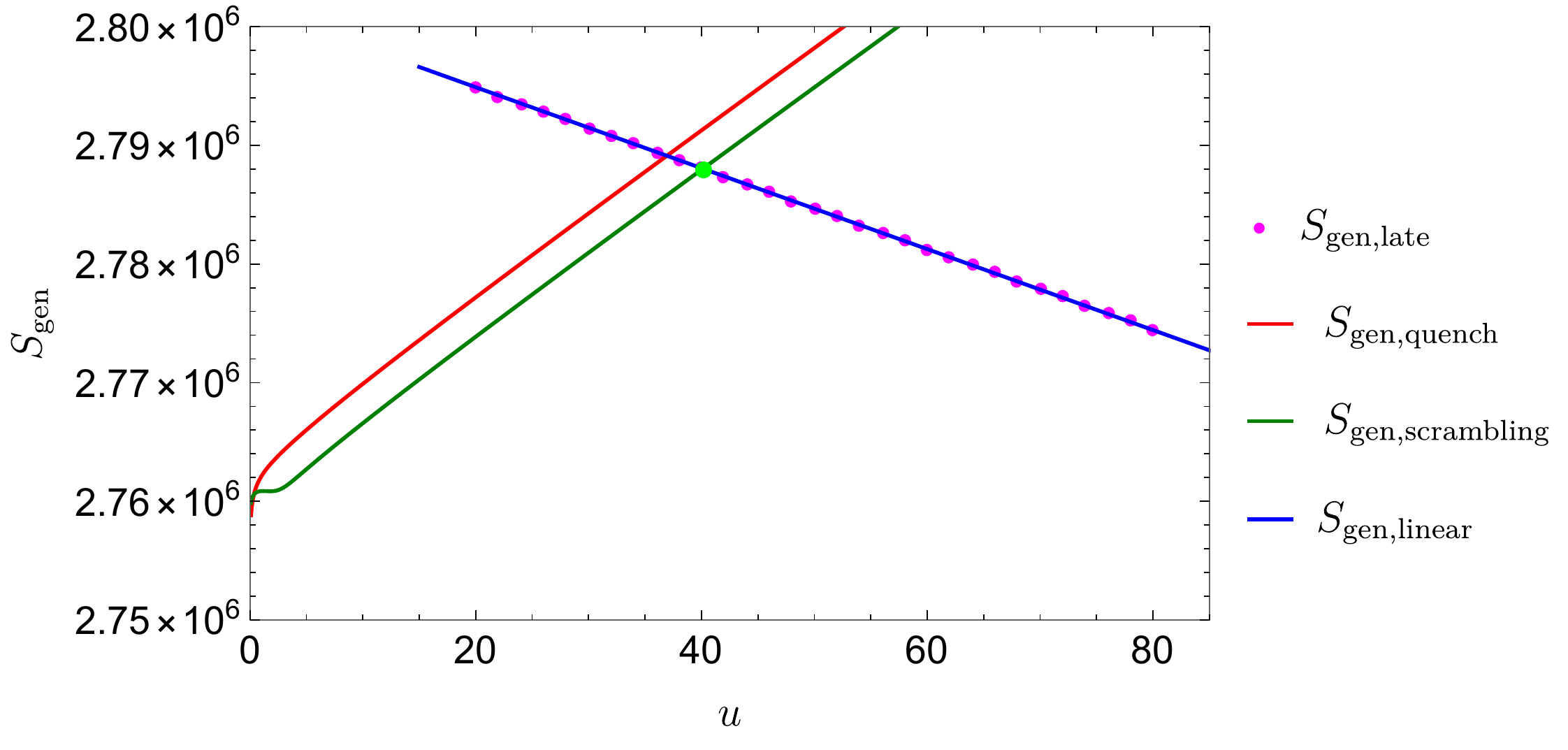}
	\caption{The dotted pink line shows the numerical results for generalized entropy with endpoint after the shock. The Page time and the first transition at the early time are both indicated by the green point in this plot. The solid red line is derived from the linear approximation, \ie eq.~\eqref{Slinear}. The difference between analytical and numerical results is approximately constant, due to the constant error from the approximation of the dilaton term.  }\label{fig:Pagecurve}
\end{figure}

If the bulk endpoint is located in the region after the shock (\ie if $x^+_\QES \ge 0$), then the bulk entropy is in its late-time phase. This case is exactly the same as that analyzed in~\cite{Almheiri:2019psf}. The QES is derived from the solutions of the following  two equations
\begin{align}\label{QES_equations}
0 = \frac{4\GN }{\bar{\phi_r}}\partial_{+} S_{\text{gen}}  &=  \frac{2(\pi T_1 x_\QES^-)^2-2-k \int_{0}^{x_\QES^-} (x_\QES^- -t)^2\{u,t\}\, dt  }{ (x_\QES^+-x_\QES^-)^2}  \\
&\qquad+  2k \(\frac{1}{ x_\QES^+-t} - \frac{1}{x_\QES^+-x_\QES^-} \)\,, \notag\\
0 = \frac{4\GN }{\bar{\phi_r}}\partial_{-} S_{\text{gen}}  &=  \frac{2-2(\pi T_1 x_\QES^+)^2+k \int_{0}^{x_\QES^-} (x_\QES^+ -t)^2\{u,t\}\, dt   }{ (x_\QES^+-x_\QES^-)^2}  \\
& \qquad +2k \( \frac{1}{x_\QES^+-x_\QES^-}  -\frac{1}{(u-y_\QES^-)f^{\prime}(y_\QES^-)} + \frac{f^{\prime\prime}(y_\QES^-)}{2(f^{\prime}(y_\QES^-))^2} \)\,.\notag
\end{align}
Because of the integral in the dilaton term, it is not easy fo find the analytical solutions for these equations. Therefore we first turn to numerics. The numerical answer is presented in figure~\ref{fig:QESxpym}, and the corresponding generalized entropy is shown in figure~\ref{fig:Pagecurve}.

From the numerical plot, it is interesting to find that around the Page time, the two branches both display linear behavior. For the solution before the Page time, the linearity can be seen in the small $k$ expansion in~\eqref{Sgen_before_k}. The post-Page time analysis is performed in the original paper~\cite{Almheiri:2019psf} by carefully dealing with the integral with Schwarzian. The key idea is we can use the approximation for $f$ around small $k$ for fixed $ku$, specifically
\begin{equation}
\label{logapprox}
\log \left(\frac{t_{\infty}-f(u)}{2 t_{\infty}}\right) \sim-\frac{4 \pi T_{1}}{k}\left(1-e^{-\frac{k}{2} u}\right)+O\left(k e^{\frac{k}{2} u}\right) \,,
\end{equation}
% and notice the fact $x^+ \sim t \sim x^-$ but they satisfy  the following conditions  \footnote{We can easily see that from our numerical results.}
% \begin{equation}
% x^+ \approx  t_{\infty}\,, \quad x^+ -t \ll t-x^- \approx x^+-x^-  \approx t_{\infty}  -x^- \,.
% \end{equation}
Keeping only the leading terms in the QES equations ($\partial_\pm S_\text{gen} = 0$), we arrive at these extremely simple equations 
% \begin{align}
% 0&=\frac{4\GN }{\bar{\phi_r}}\partial_{+} S_{\text{gen}}  \approx 4\pi T_1 \frac{e^{-\frac{k}{2} y_\qes^-}}{t_{\infty}-x_\QES^-} - \frac{2k}{ x_\QES^+ -t} \,, \\
% 0&=\frac{4\GN }{\bar{\phi_r}}\partial_{-} S_{\text{gen}}  \approx  \frac{1}{(x_\QES^+-x_\QES^-)^2} \( 4\pi T_1 (t_{\infty}-x_\QES^+) e^{-\frac{k}{2}y_\QES^-}   + \frac{k}{2} (t_{\infty}-x_\QES^-) \) \,,
% \end{align}
\begin{align}
0&\approx 4\pi T_1 \frac{e^{-\frac{k}{2} y_\qes^-}}{t_{\infty}-x_\QES^-} - \frac{2k}{ x_\QES^+ -t}  \\
0&\approx 4\pi T_1 (t_{\infty}-x_\QES^+) e^{-\frac{k}{2}y_\QES^-}   + \frac{k}{2} (t_{\infty}-x_\QES^-) \,,
\end{align}
or, solving at the same order,
\begin{align}\label{QES_linear}
x_\QES^+  &= \frac{4}{3} t_{\infty} -\frac{1}{3} t+ \mathcal{O}(k ( t_{\infty} - t))\\
y_\QES^- &= u -u_{\mt{HP}}+ \frac{k}{2} \(u_{\mt{HP}}-\frac{1}{2\pi T_1}\) \( u_{\mt{HP}}-u \)+ \mathcal{O}(k^2)\,,
\end{align}
where we define the delay of $y^-$ in time direction as
\begin{equation}
\label{Constant}
u_{\mt{HP}} = \frac{1}{2\pi T_1}  \log \( \frac{8\pi T_1}{3k} \) \,,
\end{equation}
which is (to leading order) the Hayden-Preskill scrambling time~\cite{HayPre07}, as explained in~\cite{Almheiri:2019psf}. Note that the quantum extremal surface after the shock $(x_\QES^+,y_\QES^-)$ lies close to but behind the new horizon located at $x^+=t_{\infty}$.

The above linear solution captures the leading-order behavior of QES and also the generalized entropy. In figure~\ref{fig:QESxpym}, we compare this analytic approximation to the numerical solution. We can find an approximation for generalized entropy\footnote{Compared to~\cite{Almheiri:2019psf}, here we added the contributions from bulk terms and also two sub-leading corrections for dilaton which are ignored in~\cite{Almheiri:2019psf}.}
\begin{equation}\label{bulkentropy}
	S_{\mt{bulk}} \approx  \frac{c}{6} \( \log \( \frac{8 u_{\mt{HP}}}{3\epsilon} \) -  \pi T_1u_{\mt{HP}} + \frac{k}{4}u_{\mt{HP}}  \) +\mathcal{O}(k^2)
\end{equation}
\begin{align}
\phi & \approx  2\bar{\phi}_r \(    \frac{1-\left(\pi T_{1}\right)^{2} x_\QES^{+} x_\QES^{-}  + \frac{k}{2} I\( x_\QES^+,x_\QES^-\)}{t_{\infty}-x_\QES^{-}}    \)  \(  1- \frac{x_\QES^+ -t_{\infty} }{t_{\infty} - x_\QES^-}  \) +  \mathcal{O}(k^2 \log k)\,,
\end{align}
% and
% \begin{equation}
% \begin{split}
% & \frac{1-\left(\pi T_{1}\right)^{2} x^{+} x^{-}  + \frac{k}{2} I\( x^+,x^-\)}{t_{\infty}-x^{-}}    \\ % \(  \frac{1-\left(\pi T_{1}\right)^{2} x^{+} x^{-}  + \frac{k}{2} I\( t_{\infty},x^-\)}{t_{\infty}-x^{-}}    \) \\
% &\qquad\approx (\pi T_1)^2 t_{\infty}  \(1- \frac{x^+ -t_{\infty} }{t_{\infty} - x^-} \) + \frac{k}{4} \(\log  \(\frac{t_{\infty} -x^-}{ t_{\infty}}\) - 1 \)   +\mathcal{O}(k^2\log k) \\
% \end{split}
% \end{equation}
% where the integral is approximated by
\begin{equation}
I\(t_{\infty},x^-\) \approx  \frac{2}{k}\( (\pi T_1 t_{\infty})^2-1 \) +\frac{t_{\infty} -x^-}{2} \(  \log \(\frac{t_{\infty}-x^-}{t_{\infty}} \) -1  \) \,.
\end{equation}
For times much smaller than $k^{-1}$ we can further simplify these expressions  by taking the limit
\begin{equation}
\begin{split}
\log \left(\frac{t_{\infty}-f(u)}{t_{\infty}-f(y^-)}\right) &\sim \frac{4\pi T_1}{k}\left(e^{-\frac{k}{2} u}-e^{-\frac{k}{2} y^-}\right)+O\left(k e^{\frac{k}{2} u}\right) \approx  2 \pi T_1(y^--u) \,,\\
\log \(  \frac{f^{\prime}(y^-)}{f^{\prime}(u)}\frac{t_{\infty}-f(u)}{t_{\infty}-f(y^-)}\) &\approx \log \( e^{\frac{k}{2} (u-y^-)} \) =\frac{k}{2}(u-y^-)\,.
\end{split}
\end{equation}
We find the entropy from the dilaton contribution decreases linearly:
\begin{equation}\label{dilatexpasion}
\begin{split}
\phi
% &\approx \phi_r \[ 2(\pi T_1)^2 t_{\infty} + \frac{k}{2} \( \log \(\frac{t_{\infty}-x^-}{t_{\infty}} \) -1   \) - 4(\pi T_1)^2 t_{\infty} \frac{x^+ -t_{\infty} }{t_{\infty} - x^-} \]\\
% &\approx \phi_r \[  2\pi T_1 - 2 \pi T_1 (1-e^{-ky^-/2}) -\frac{k}{2} \log 2 -\frac{4\pi T_1}{3} \frac{t_{\infty} -t}{ t_{\infty} -x^-} \]\,,\\
&\approx \bar{\phi}_r \(  2\pi T_1 - k \pi T_1 (u-u_{\mt{HP}}) -\frac{k}{2} \log 2e \) \,.
\end{split}
\end{equation}
We arrive at an equation displaying linear decrease of the generalized entropy
\begin{equation}\label{Slinear}
\begin{split}
S_{\mt{linear}}
&\approx   \frac{\bar{\phi}_r}{4 \GN } \[  2\pi T_1 - k \pi T_1 (u-u_{\mt{HP}}) + k  \log \( \frac{8 k u_{\mt{HP}}^2}{3\sqrt{2e}\epsilon^2 \pi T_1}    \) +\mathcal{O}(k^2 \log (k)) \]
\end{split}
\end{equation}
where the first two terms are derived from the dilaton term which lead to the linear decrease of the entropy around the Page time, and the extra constant terms are contributions from the bulk entropy.

The linear formula given above matches the numerical results shown in figure~\ref{fig:Pagecurve}. As shown in this figure, when the time is large than the Page time $u_{\text{Page}}$, the endpoint of QES jumps from the point before the shock to that after the shock.

\begin{figure}[t]
	\centering\includegraphics[width=3.80in]{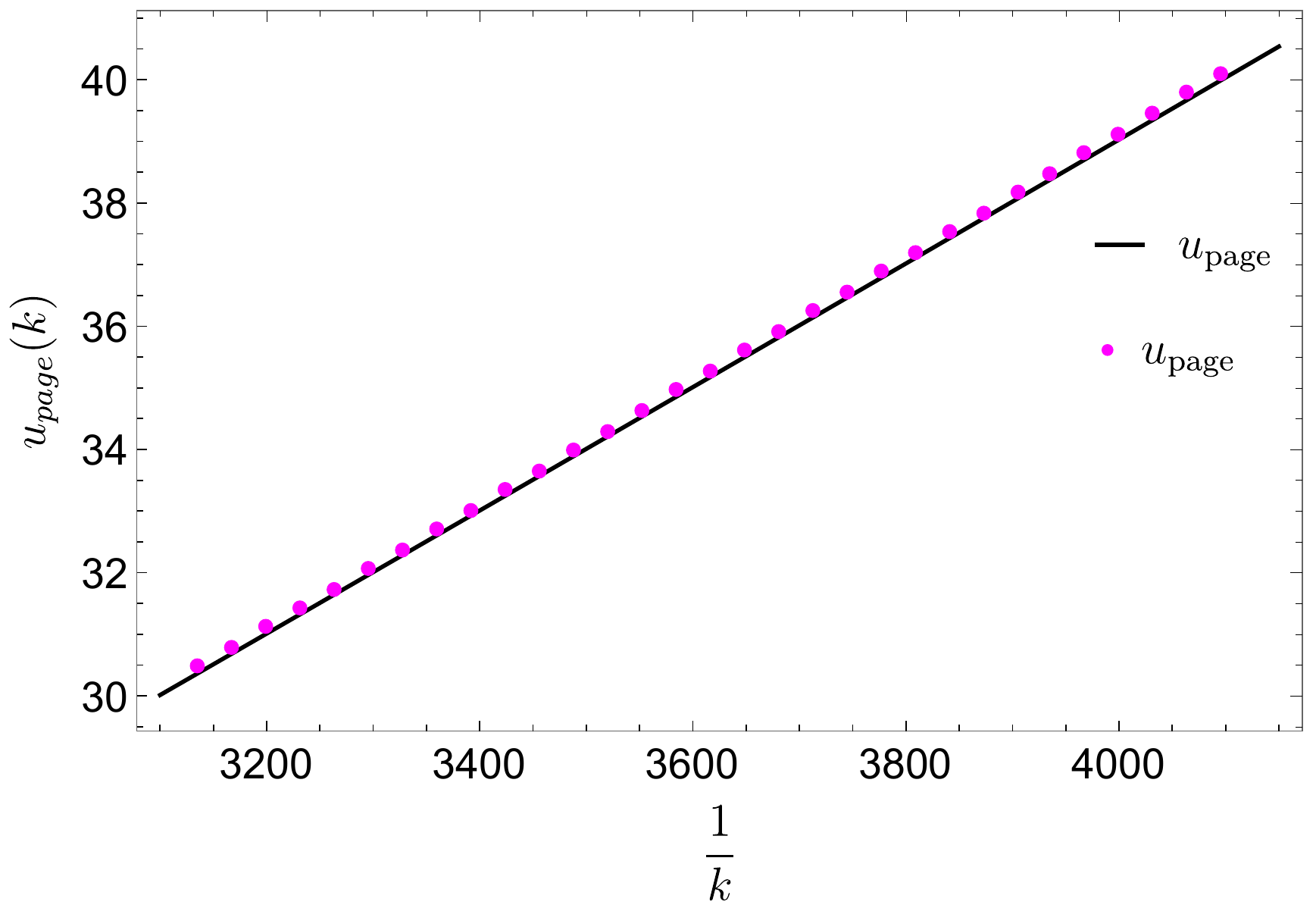}
	\caption{The Page time for fixed temperatures $T_0$ and $T_1$, as a function of $k^{-1}$. The dots are derived from numerical results without any approximation and the solid line is the approximate Page time defined in~\eqref{uPage}.}\label{fig:Pagetime}
\end{figure}

From the approximations in eqs.~\eqref{Slinear} and~\eqref{Sgen_before_k2}, we can find the approximate Page time
\begin{equation}\label{uPage}
\begin{split}
u_{\text{Page}}&\approx \frac23 \frac{T_1 -T_0}{T_1k} +\frac{u_{\mt{HP}}}{3}+\frac{k}{6 \pi T_1} \frac{5}{(T_1 - T_0)\pi} \\
&\qquad+ \frac{2}{3 \pi  T_1} \log\(\sqrt{\frac{8k\pi T_1}{3\sqrt{2e}}} \frac{u_{\mt{HP}}}{u_{\mt{P}}^0} \frac{c}{6 \pi E_{\mt{S}}} \frac{T_0 }{T_1-T_0}\) +\mathcal{O}(T_1-T_0) \,,
\end{split}
\end{equation}
where we have defined\footnote{Here we have kept the $\frac{k}{T_1-T_0}$ term, which may be order one for some choices for parameters.} \begin{equation}\label{eq:u0}
	u_{\mt{P}}^0=\frac{2}{3}\frac{T_1 -T_0}{T_1k} +\frac{u_{\mt{HP}}}{3}
\end{equation} as the leading-order approximation to $u_\textrm{Page}$. A comparison with numerical results is given in figure~\ref{fig:Pagetime}.
%
% Finally, before we close the warm-up section, we can summarize the results described in this section. First, the QES from the generalized entropy  in AEMM model is derived as
% \begin{equation}
% (x^+,x^-) _{\mt{QES}}=
% \begin{cases}
%     x^{\pm} = \pm \frac{1}{\pi T_0}\,, \qquad    u\le u_{\mt{AB}}\,,                                                                                 \\
%     \,\\
%    x^{\pm}(t) \quad \text{in}~\eqref{full_QES01} \,, \quad  u_{\mt{AB}}\le u \le u_{\mt{Page}}\\
%     \,\\
%     x^+ \approx \frac{4}{3} t_{\infty} -\frac{1}{3} t  \,, \qquad u_{\mt{Page}}\le u \,.\\
% \end{cases}
% \end{equation}
% With the above QES, we can find the generalized entropy for the two points, \ie $\text{AdS}$-bath, will be exactly the Page curve as shown in figure~\ref{fig:Pagecurve}, which implies the evaporation of black hole in AEMM model is unitary.

%% file: sections/minbath.tex
\begin{figure}[t]
	\centering
	\includegraphics[width=\textwidth]{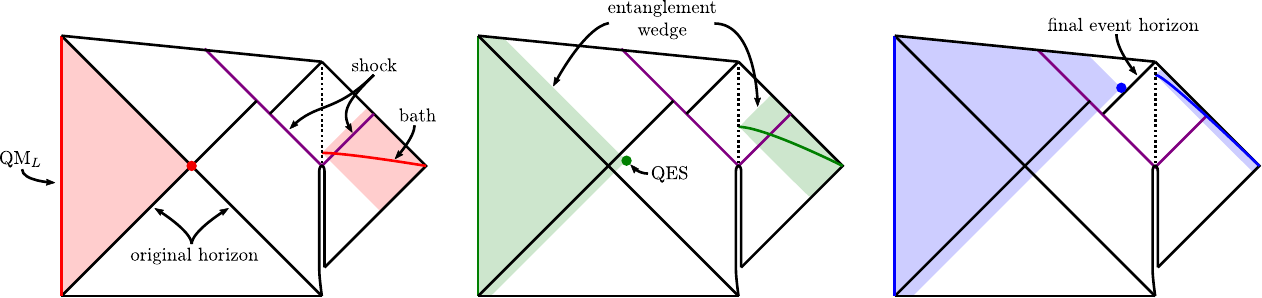}
	\caption{Entanglement wedges for the three phases of evolution. The quench phase is in red, the scrambling phase in green and the late-time phase in blue. }
	\label{fig:entwedges}
\end{figure}

% !TEX root = ../BH_v04.tex
As was shown in section~\ref{sec:QES}, the evaporating model we are considering exhibits two phase transitions. Each phase corresponds to a different location for the quantum extremal surface inside the new horizon. An important consequence of these transitions is that the entanglement wedge of {\qml}+bath contains a bigger region of the bulk geometry after each phase transition. In particular, there is an area inside the black hole that is contained in the entanglement wedge after the transitions, but not before. This is illustrated in figure~\ref{fig:entwedges}. By entanglement wedge reconstruction~\cite{Czech:2012bh,Headrick:2014cta,Wall:2012uf,Jafferis:2015del,Dong:2016eik,Cotler:2017erl,Penington:2019npb}, this implies that after some time, {\qml} plus the bath contain information about the interior of the black hole. In this section, we investigate how much of the bath is essential to keep in order to still reconstruct the black hole interior. For concreteness we will focus on the Page transition in which the quantum extremal surface jumps across the infalling shock, because this transition allows much more of the interior to be reconstructed; but a qualitatively similar story occurs for the first transition, in which the extremal surface jumps from the bifurcation point to a point perturbatively close from it.

Before the evaporation begins, the black hole interior cannot be reconstructed from only the {\qml} or the {\qmr}  system since it is not contained in the entanglement wedge of either. On the other hand, the combination of the {\qml} and {\qmr} is enough to reconstruct the entire bulk geometry. This implies that the information required to reconstruct the interior of the black hole is shard between the two sides of the black hole. One can also ask what is the entanglement wedge of {\qmr} (or \qml)+bath before evaporation begins, but the answer is trivial because there is no entanglement between the two: it is the entanglement wedge of {\qmr} (\qml) plus the empty set. After the Page time, enough evaporation has taken place and the quantum extremal surface of {\qmr} is located after the shock perturbatively close to the apparent horizon. The entanglement wedge of {\qmr} is \emph{smaller} than it was before evaporation began: {\qmr} has \emph{lost} part of the information required to reconstruct some of the bulk geometry it was originally encoding before the evaporation. On the other hand, the entanglement wedge of the complement, the {\qml} plus the bath, \emph{gained} information encoding part of the interior of the black hole. This reflects the fact that some of the initial entanglement between {\qml} and {\qmr} has been transferred to the bath by the Hawking radiation so that {\qml}+bath can reconstruct a portion of the black hole interior.

\subsection{Early-time protocol: forgetting the late-time radiation} \label{two}
Our first modification of the {\aims} model as described in section~\ref{sec:review} is to move the endpoint $y^{\pm}_1$ of the bulk interval into the bath region. This corresponds to omitting the late-time Hawking radiation from the entanglement wedge of {\qml}+bath. In this regime, we will have $ y^\pm_1>0$.
% \beqs
% &\text{AdS}_2, \text{before}\, \quad x^{\pm}_\QES :  x^+_\QES >0\,,\ x^-_{\QES}<0 \,, \ \text{or} \quad \text{after}\quad   x^+_\QES >0, x^-_{\QES}>0\,,\\
% &\text{bath},\ \text{after} \, \quad  y_{1}^{\pm} : y^+_1>0\,, \ y^-_1>0  \,.
% \eeqs

We parameterize the distance from a bath point to the AdS boundary by specifying the coordinate distance $\sigma_1$ from the boundary to the bath, \ie we set $y^{\pm}_1=u\mp \sigma_1$. 
%rcm added subscript 1 to the \sigma
Similar to eq.~\eqref{bulk_entropy}, we can identify three phases of the von~Neumann entropy of the interval in the bulk, which we label the same way: the quench phase, the scrambling phase, and the late-time phase. The most important difference is in the Lorentzian cross-ratio, where we must now account for the fact that the endpoints are not fixed on the boundary
\begin{equation}
	\eta = \frac{{f(y_{1}^+ ) (x_\QES^+-x_\QES^+))}}{{x_\QES^+ (f(y_{1}^+ )- x_\QES^-)}}
\end{equation}The phase boundary between the quench and scrambling phases still lies at $\eta = \tfrac{1}{2}$. The entropy functions in each phase now read
% \begin{table}[t]
%   \centering
%   \begin{tabular}{c|ccc}
%   \hline
%   \hline
%   \textbf{Phase} & \textbf{Range of $\eta$} & \textbf{Bulk Region}  \\ \hline
%     &&&\\[-2.2ex]
%   Quench & $[0, \tfrac{1}{2})$ & $x_\QES^{-}<0<t<x_\QES^{+}$ \\
%   Scrambling & $[\tfrac{1}{2},1)$ & $x_\QES^{-}<0<t<x_\QES^{+}$ \\
%   Late-Time & $\approx 1$ & $0<x_\QES^{-}<t<x_\QES^{+}$ 
%   \\[.3ex] \hline \hline
%   \end{tabular}
%   \caption{Removing the late-time radiation is equivalent to excising a Minkowski coordinate amount $\sigma_1$ of the bath. The phases of the von~Neumann entropy are qualitatively similar but shifted. In Lorentzian coordinates, $\eta = {f(y_{1}^+ ) (x_\QES^+-x_\QES^+))}/[{x_\QES^+ (f(y_{1}^+ )- x_\QES^-)}]$. \rcm{Zach will fix this, as discussed.}}
%     \label{tab:phases}
% \end{table}
\begin{align}\label{bulk_entropy_bath_quench}
S_{\textrm{bulk, quench}}&=
\frac{c}{6} \log \left(\frac{24\pi E_{S}}{\epsilon c} \frac{y_1^- f(y_1^+)}{\sqrt{f^{\prime}(y_1^+)}}\right)   \\
\label{bulk_entropy_bath_scrambling}
S_{\textrm{bulk, scrambling}}&=
\frac{c}{6} \log \left(\frac{24 \pi E_{S}}{\epsilon c} \frac{y_1^- x_\QES^-\(f(y_1^+)-x_\QES^+\)}{(x_\QES^+-x_\QES^-)\sqrt{f^{\prime}(y_1^+)}}\right) \\
\label{bulk_entropy_bath_latetime}
S_{\textrm{bulk, late-time}}&=
{\frac{c}{6} \log \left[\frac{2\left(y_1^--y_\QES^{-}\right)\left(x_\QES^{+}-f(y_1^+)\right)}{\epsilon\left(x_\QES^{+}-x_\QES^{-}\right)} \sqrt{\frac{f^{\prime}(y^-_\QES)}{f^{\prime}(y_1^+)}}\right]}
\end{align}
% with the cross ratio defined as
% \begin{equation}
% \eta = \frac{f(y_{1}^+ ) (x_\QES^+-x_\QES^+))}{x_\QES^+ (f(y_{1}^+ )- x_\QES^-)} \,,
% \end{equation}
Again we need to find the location of the new quantum extremal surface $x_{\QES}^{\pm}$ by minimizing the generalized entropy ($\partial_\pm S_\text{gen} = 0$). Before we move to finding the solutions, let's first comment on the effect of taking the point $x_1$ into the bath region, \ie
\begin{equation}
y^\pm_1=u\mp\sigma_1\,.
\end{equation}
It is obvious this operation has nontrivial effect on the location of the quantum extremal surface and bulk entropy because the entropy in the three cases all depend on both $y^{\pm}_1$. However, the effect from $y_1^-$ only appears in $S_\text{gen}$ as a term like
\begin{equation}
\begin{cases}
\frac{c}{6} \log \( y_1^-\)  &\text{for}\quad  {x_{\QES}^{-}<0<t<x_{\QES}^{+}}\,, \\
\frac{c}{6} \log \( y_1^--y_{\QES}^-\)  &\text{for} \quad {0<x_{\QES}^{-}<t<x_{\QES}^{+}}\,.
\end{cases}
\end{equation}
If the above contribution is negligible, it is easy to claim that the effect from moving the endpoint to the bath corresponds to a reparameterization, changing $u$ to $u-y_0$. At leading order, this is what happens, as we will now explain.

As before, the bulk entropy in the quench phase is independent of the location of the quantum extremal surface. Again the dilaton term is minimized at the bifurcation point $x^{\pm}= \pm \frac{1}{\pi T_0}$, so this is the location of the quantum extremal surface in the quench phase. The generalized entropy in this quench ($\eta \le  \frac{1}{2}$) phase reads
\begin{equation}
S_{\mt{gen, quench}}= \frac{\bar{\phi}_r}{4 \GN} \(  2\pi T_0    + 2k \log \left(\frac{24 \pi E_{S}}{\epsilon c} \frac{y_1^- f(y_1^+)}{\sqrt{f^{\prime}(y_1^+)}}\right)\) \,.
\end{equation}
which reduces to the $\text{AdS}$-boundary case when we take the limit $y^{\pm} \sim u$ or $\sigma_1 \to 0$, as expected.
The cross-ratio region  $\eta \le \frac{1}{2}$ that defines the quench phase is equivalent to
\begin{equation}
f(y_1^+) \le \frac{1}{3\pi T_0}\,, \qquad y^+_1 = u-\sigma_1 \le f^{-1}(\frac{1}{3\pi T_0}) \,.
\end{equation}
The location of the quantum extremal surface in the scrambling phase and with $\sigma_1 > 0$ is delayed with respect to the $\sigma_1=0$ solution, because the solutions to the extrema equations ($0 = \partial_\pm S_\text{gen}$), which read
% \begin{equation}
% \begin{split}
% 0 &= \frac{4\GN}{\bar{\phi_r}}\partial_{+} S_{\text{gen}}  =  \frac{2((\pi T_0 x_\QES^-)^2-1)}{ (x_\QES^+-x_\QES^-)^2}  +  2k \frac{x_\QES^- - f(y^+_1)}{ (f(y_1^+)-x_\QES^+)(x_\QES^+-x_\QES^-)}\,, \\
% 0 &= \frac{4\GN}{\bar{\phi_r}}\partial_{-} S_{\text{gen}}  =  \frac{2(1-(\pi T_0 x_\QES^+)^2)}{ (x_\QES^+-x_\QES^-)^2}  +   \frac{2k x_\QES^+}{ x_\QES^-(x_\QES^+-x_\QES^-)}
% \end{split}
% \end{equation}
\begin{align}
0 &=  \frac{(\pi T_0 x_\QES^-)^2-1}{ x_\QES^+-x_\QES^-}  +  k \frac{x_\QES^- - f(y^+_1)}{ f(y_1^+)-x_\QES^+}\,, \\
0 &= \frac{1-(\pi T_0 x_\QES^+)^2}{x_\QES^+-x_\QES^-}  +  k \frac{x_\QES^+}{ x_\QES^-}
\end{align}
only depend on $f(y_1^+)$. The location is similar to eq.~\eqref{k_QES_02} after making the replacement $u \to y_1^+$, $t \to f(y_1^+)$, \ie
\begin{equation}
x_\QES^\pm  = x_\QES^{\pm} (f (y_1^+)) \,.
\end{equation}
Although the location of the quantum extremal surface is simply delayed by $\sigma_1$, the generalized entropy
still has the non-trivial term from $\log y^-_1$ as we claimed before:
\begin{equation}\label{Sgen_before_k_bath}
\begin{split}
S_{\text{gen, scrambling}}&\approx   \frac{\bar{\phi}_r}{4 \GN} \left[  2\pi T_0 + 2k \log \left(\frac{24\pi E_{S}}{\epsilon c}\frac{u+\sigma_1}{\pi T_1\sqrt{f^{\prime}(u-\sigma_1)}}  \right)  + \kappa\right]  \,,
\end{split}
\end{equation}
where $\kappa$ is defined in eq.~\reef{chi88} and we have assumed $\eta \ge  \frac{1}{2}$ and $u\gg 1$. This extra term is still sub-leading, with the full generalized entropy dominated by the linear growth at early times $u \ll \frac{1}{k}$.

Similar to the $\sigma_1 =0$ case, the transition between the quench and scrambling phases happens at the point where
\begin{equation}\label{simple}
\begin{split}
S_{\text{gen, scrambling}} &=S_{\text{gen, quench}}\qquad \, \longleftrightarrow \qquad  y_{\mt{QS}}^+= u_{\mt{QS}}\,.\\
\end{split}
\end{equation}
where the equivalence is exact because of the cancellation of $\log y^-$ in \sloppy{${S_{\mt{gen,scrambling}}-S_{\mt{gen,quench}}}$}. Just like the $\sigma_1 =0$ case considered in section~\ref{sec:review}, the quantum extremal surface is at the bifurcation point until $u_{\mt{AB}}$ and then jumps to $x^\pm \( y^{+}\)$ in eq.~\eqref{k_QES_02}. This marks the transition between the quench phase and the scrambling phase.

\begin{figure}[t]
	\includegraphics[width=0.49\textwidth]{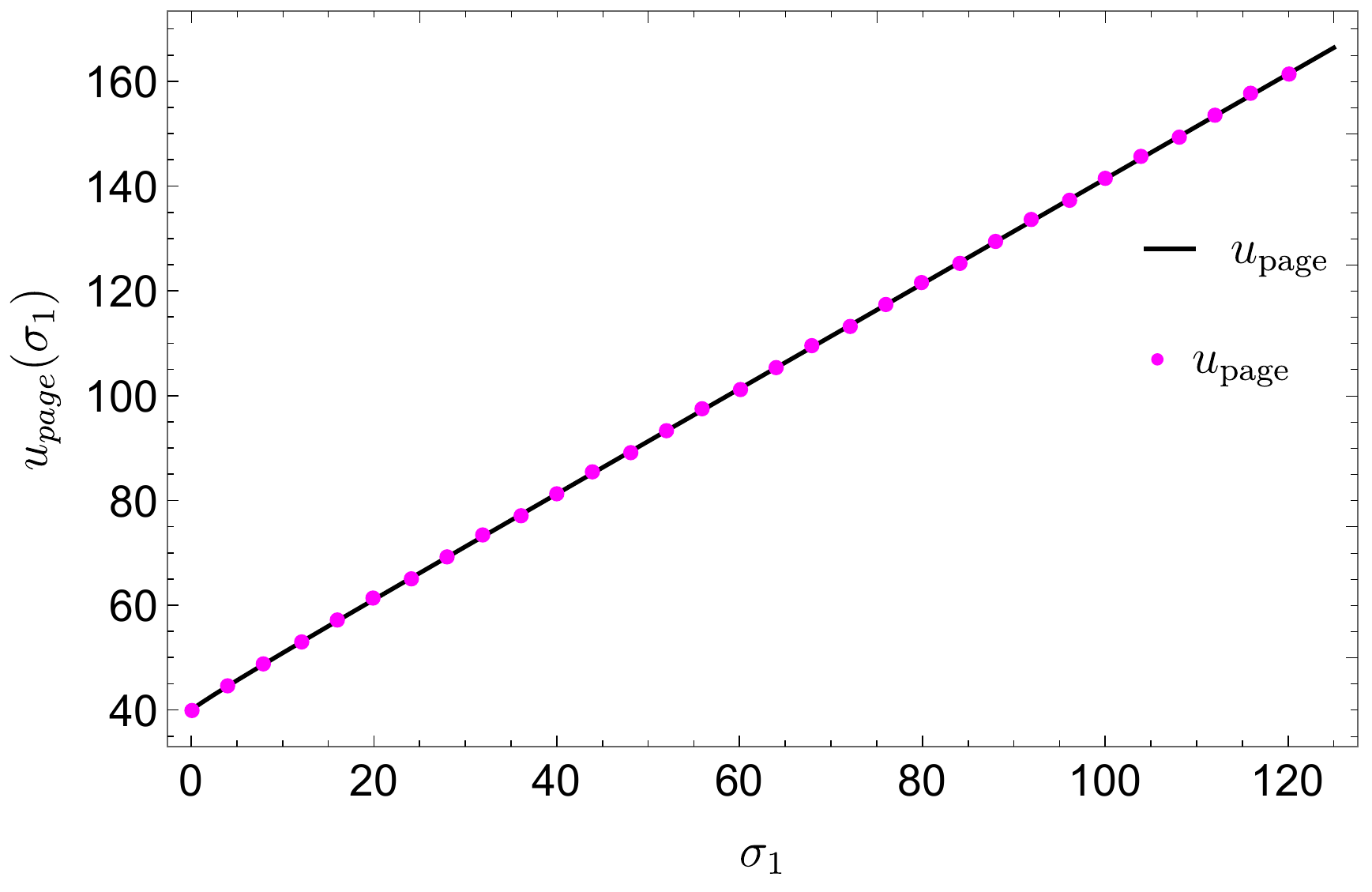} \hspace{0.01\textwidth}
	\includegraphics[width=0.49\textwidth]{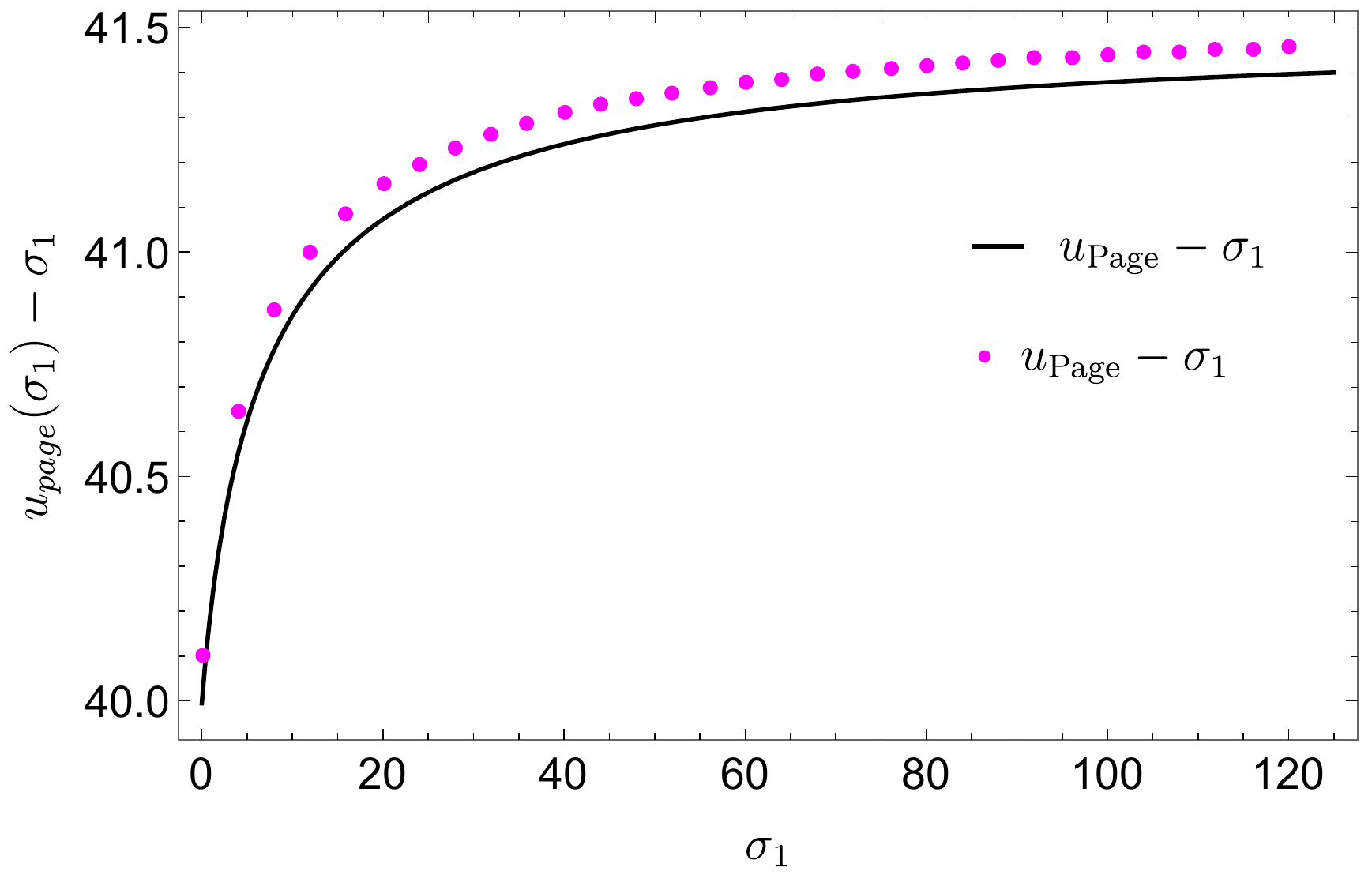}
	\caption{Left: The numerical results for $u_{\text{Page}}$ on the dependence on $\sigma_1$ and the comparison with analytical result defined in~\eqref{uPage_bath}. Right: $u_{\text{Page}} - \sigma_1$}
	\label{fig:uPagetwo}
\end{figure}

In the late time phase, the quantum extremal surface is located after the shock, and the extremum equations $0 = \partial_\pm S_{\text{gen}}$
% \jh{Maybe remove these?}
% \begin{equation}\label{QES_equations_bath}
% \begin{split}
% \frac{4\GN}{\bar{\phi_r}}\partial_{+} S_{\text{gen}}  &=  \frac{2(\pi T_0 x^-)^2-2-k \int_{0}^{x^-} (x^- -t)^2\{u,t\}\, dt  }{ (x^+-x^-)^2}  +  2k \(\frac{1}{ x^+-f(y_1^+)} - \frac{1}{x^+-x^-} \) =0\,, \\
% \frac{4\GN}{\bar{\phi_r}}\partial_{-} S_{\text{gen}}  &=  \frac{2-2(\pi T_0 x^+)^2+k \int_{0}^{x^-} (x^+ -t)^2\{u,t\}\, dt   }{ (x^+-x^-)^2}  \\
% &\qquad +2k \( \frac{1}{x^+-x^-}  -\frac{1}{(y_1^--y^-)f^{\prime}(y^-)} + \frac{f^{\prime\prime}(y^-)}{2(f^{\prime}(y^-))^2} \)=0\,.
% \end{split}
% \end{equation}
can be expanded into first order in $k$ to read
\begin{align}
0&\approx 2\pi T_1 \frac{e^{-\frac{k}{2} y_{\QES}^-}}{t_{\infty}-x_{\QES}^-} - \frac{k}{ x_{\QES}^+ -f(y_1^+)} \,, \\
0&\approx  4\pi T_1 (t_{\infty}-x_{\QES}^+) e^{-\frac{k}{2}y_{\QES}^-}   + \frac{k}{2} (t_{\infty}-x_{\QES}^-) \,.
\end{align}
This leads to the  linear solution
\begin{align}\label{QES_linear_bath}
x_{\QES}^+  &= \frac{4}{3} t_{\infty} -\frac{1}{3} f(y^+_1)+ \mathcal{O}(k (t_{\infty} - f(y_1^+)))\,,\\
y_{\QES}^- &= y^+_1 -u_{\mt{HP}}+ \frac{k}{2} \(u_{\mt{HP}}-\frac{1}{2\pi T_1}\) \( u_{\mt{HP}}-y^+_1 \)+ \mathcal{O}(k^2)\,.
\end{align}
The generalized entropy in the late time phase is given by eq.~\eqref{Sgen_before_k2}
\begin{equation}\label{Slinear_bath}
\begin{split}
S_{\text{linear}}
&\approx   \frac{\bar{\phi}_r}{4 \GN} \[  2\pi T_1 - k \pi T_1 (u-\sigma_1-u_{\mt{HP}}) + k  \log \( \frac{8 k (u_{\mt{HP}}+2\sigma_1)^2}{\sqrt{2e}3\epsilon^2 (\pi T_1)}     \) +\mathcal{O}(k^2 \log (k)) \] \,.
\end{split}
\end{equation}
With the new approximations~\eqref{Slinear_bath}, we can also define the Page time for this late-radiation-excised bath with fixed $\delta$ as
\begin{equation}\label{uPage_bath}
\begin{split}
u_{\text{Page}}(\sigma)&\approx \frac23\frac{T_1 -T_0}{ T_1k} +\frac{u_{\mt{HP}}}{3} +\sigma+\frac{2}{3 \pi  T_1} \log\(\sqrt{\frac{8k\pi T_1}{3\sqrt{2e}}} \frac{(u_{\mt{HP}}+2\sigma)}{(u_{\mt{P}}^0+2\sigma)} \frac{c}{6 \pi E_S} \frac{T_0 }{T_1-T_0}\)  \\
 &\quad +\frac{k}{6 \pi T_1} \frac{5}{(T_1 - T_0)\pi}
+\mathcal{O}(T_1 -T_0) \,,
\end{split}
\end{equation}
with the crossing condition $S_{\mt{linear}}=S_{\mt{gen, scrambling}}$. It is clear that it is just the $u_{\mt{Page}} +\sigma$ with corrections from one log term which is decreasing with the increase of $\sigma$. As a final check, in figure~\ref{fig:uPagetwo} we compare the numerical results for the Page time with the analytical approximation.

\subsubsection{Time evolution of $\sigma_\text{Page}$}\label{sec:pepsi}
\begin{figure}[t]
	\centering\includegraphics[width=5.80in]{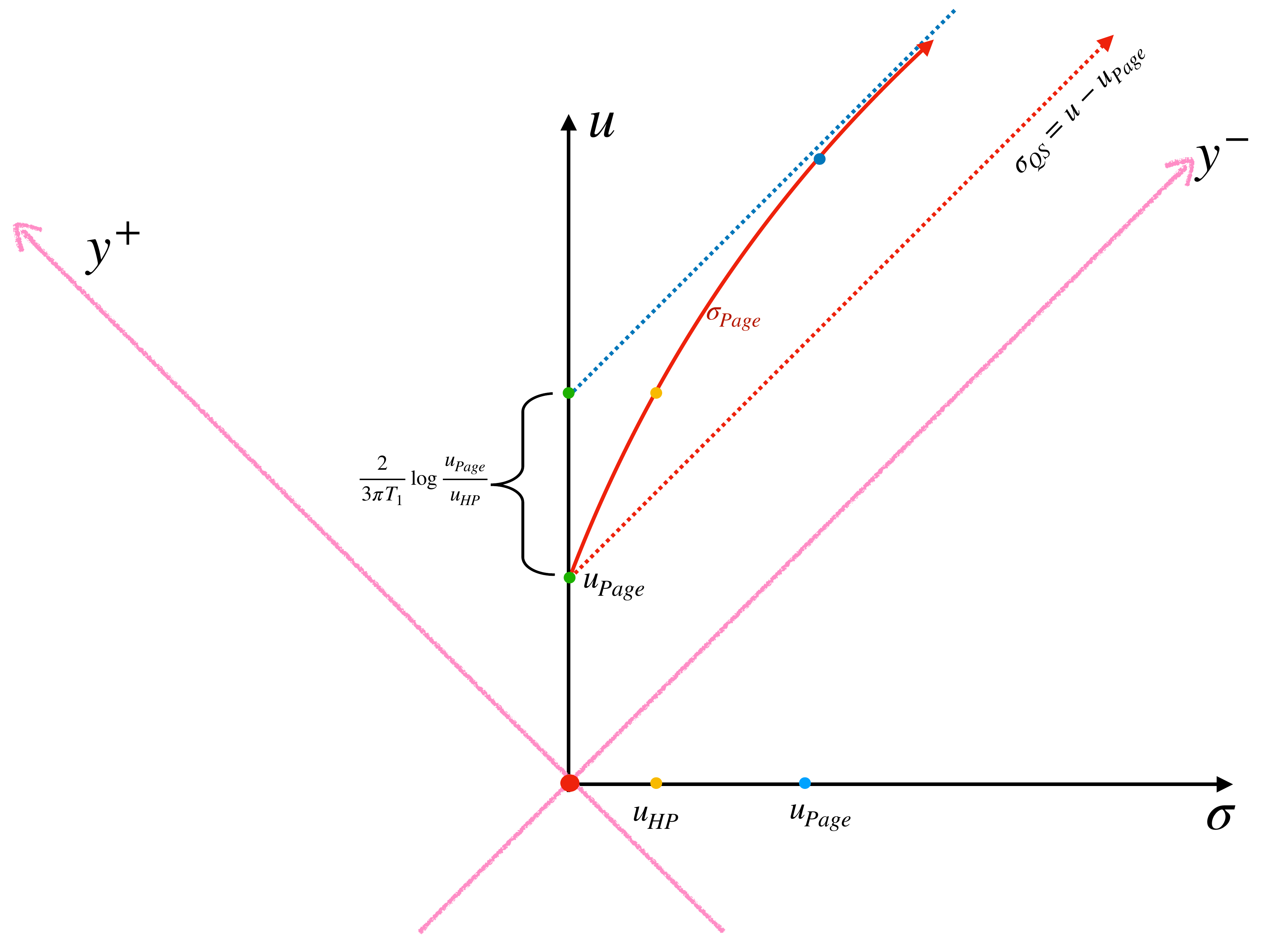}
	\caption{The red line indicates the evolution of $\sigma_{\mt{Page}}$. It starts from the boundary point at $u=u_{\mt{Page}}$ and evolve with time $u$.  Finally approach the another null surface with shift $\frac{2}{3\pi T_1} \log \( \frac{u_{\mt{Page}}}{u_{\mt{HP}}}\)$.}\label{fig:bath_point}
\end{figure}
Armed with the approximate solution~\eqref{uPage_bath}, we can fix a time slice $u$ after the Page time and ask how far into the bath we need to move to arrive at the Page transition. To fix the notation, we will say that this happens at
\begin{equation}
y^+_{\mt{Page}} \equiv u - \sigma_{\mt{Page}}\,,
\end{equation}
We can thus consider the evolution of the distance of the second endpoint to AdS boundary $ \sigma_{\mt{Page}} $ such that
\begin{equation}
u - \sigma_{\mt{Page}} - u_{\mt{Page}} = \frac{2}{3\pi T_1}\(\log \(  \frac{u_{\mt{HP}} +2\sigma_{\mt{Page}}}{u_{\mt{HP}}} \) - \log \(  \frac{u_{\mt{Page}} +2\sigma_{\mt{Page}}}{u_{\mt{Page}}}\)  \) \,,
\end{equation}
which is derived from the approximation of $y^+_{\mt{Page}}$ and $u_{\mt{Page}}$. It is still hard to solve the above equation for $\sigma_\text{Page}$. However, let's first comment on its speed with respect to $u$, \ie
\begin{equation}
\partial_{u} \sigma_{\mt{Page}} = \( 1+ \frac{4}{3\pi T_1} \frac{u_{\mt{Page}}-u_{\mt{HP}}}{(u_{\mt{HP}}+ 2\sigma_{\mt{Page}}) (u_{\mt{Page}}+2\sigma_{\mt{Page}}) }     \)^{-1}  < 1\,,
\end{equation}
which approaches 1 when $\sigma_{\mt{Page}}\to\infty$. In order to get insight on the simple form of $\sigma_{\mt{Page}}$, we consider three different regions for $\sigma_{\mt{Page}}$ using the separation of scales: $\frac{1}{\pi T_1} \ll u_{\mt{HP}}\ll u_{\mt{Page}}$. 

First of all, if we start from a small $\sigma_{\mt{Page}}$, it is easy to find for $\sigma < u_{\mt{HP}}$
\begin{equation}\label{sigmapage01}
\sigma_{\mt{Page}} ( \sigma) \simeq \( 1 +\frac{4}{3\pi T_1} \frac{u_{\mt{Page}}-u_{\mt{HP}}}{u_{\mt{Page}}u_{\mt{HP}}} \)^{-1} \( u -u_{\mt{Page}}\)\,,
\end{equation}
where the coefficient is a little bit smaller than one. Then we can move to the middle region with the approximate solution for $u_{\mt{HP}} < \sigma  < u_{\mt{Page}} $:
\begin{equation}\label{sigmapage02}
\sigma_{\mt{Page}} \simeq
\(
	1 -\frac{4}{3\pi T_1 u_{\mt{Page}}}
\)^{-1}
\(
u- u_{\mt{Page}} - \frac{2}{3\pi T_1}
\log \frac{u_{\mt{HP}}+2(u - u_{\mt{Page}})}{u_{\mt{HP}}}
\) .
\end{equation}
Note that although the coefficient looks larger than 1, it is easy to check that with the logarithmic correction, the velocity in this region still satisfies $\partial_{u}\sigma_{\mt{Page}} < 1$.

Finally we arrive at the region with $\sigma_{\mt{Page}} > u_{\mt{Page}}$, one can still find a linear result when $ u_{\mt{Page}} < \sigma$:
\begin{equation}\label{slackr1}
\sigma_{\mt{Page}} \simeq u- u_{\mt{Page}} -\frac{2}{3\pi T_1} \log \( \frac{u_{\mt{Page}}}{u_{\mt{HP}}}\) +\frac{u_{\mt{Page}}-u_{\mt{HP}}}{3\pi T_1(u - u_{\mt{Page}})} \,.
\end{equation}
So we can find that the evolution of $\sigma_{\mt{Page}}$ is time-like, however, in this regime, it quickly approaches a null line as the last term decays as $u-\uP$ grows. We note that the third term above represents a (small) finite shift of the asymptotic line above the simple leading approximation $\yp\simeq u-\uP$. We show a sketch of the evolution of $\yp$ in figure~\ref{fig:bath_point}, summarizing our results here.

In closing here, we comment that a similar but even simpler conclusion applies to the transition between the quench and scrambling phases. Recall that this transition occurs at $u=u_{\mt{QS}}$ defined in eq.~\eqref{eq:scramble}. Then on later time slices, we push $\y_1$ into the bath and define $\sigma_{\mt{QS}}$ in analogy with $\yp$, \ie $\sigma_{\mt{QS}}$ is the value of $\sigma$ on a constant $u$ slice where the transition between the quench and scrambling branches occurs. From eq.~\eqref{simple}, it is straightforward to show that $\sigma_{\mt{QS}}$  {\it exactly} satisfies the simple relation
\begin{equation}\label{sigmaQS}
 \sigma_{\mt{QS}}  =  u- u_{\mt{QS}}\,.
\end{equation}

\subsubsection{Importance of the early radiation} \label{sec:cocacola}

So far we have seen how much of the later radiation can be discarded while still being able to reconstruct the interior of the black hole with the remaining radiation + \qml. This was done by starting at some time slice after the Page time and removing an interval of the bath starting from the AdS-bath juncture until the generalized entropy of the late-time branch matches the entropy of the scrambling branch. That is, we found the point $x^+_1= f(y^+_1)$ in the bath such that
\beq
S_{\QES'-1}=S_{\QES-1}\,,
\eeq
where $x^+_{\QES'}$ is at the extrema of the generalized entropy with one endpoint before the shock, and $x^+_{\QES}$ is at the extrema of the generalized entropy with both endpoints after the shock. This allows us to remove part of the bath close to AdS that is not essential for black hole interior reconstruction. We can now ask the question of how much of the early-radiation regime of the bath we can remove while still keeping information about the black hole interior. That is, we consider a bath interval ${\cal B}_0=[\y_1=\yp(u), \y_2]$ on a constant time slice $u$, and ask how close can we move $\y_2$ to the initial endpoint while still being able to reconstruct the black hole interior. Unsurprisingly, we must place $\y_2$ near the shockwave falling into the bath, since more distant points are out of causal contact with the quench point. However, we will also find that $\sigma_2$ must be positioned slightly to the right of the shock, \ie we need to keep all of the early radiation. 

\begin{figure}[t]
	\centering
	\includegraphics[width=0.7\textwidth]{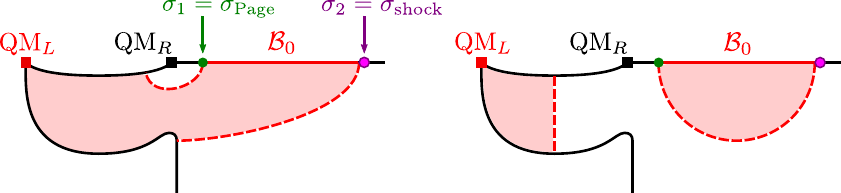}
	\caption{The smallest connected bath interval ${\cal B}_0$ that, together with the \qml, still has enough information to reconstruct the black hole interior with is the one in which the generalized entropy in the two channels depicted are equal.}
	\label{fig:3pts}
\end{figure}

As above, consider an interval of the bath ${\cal B}_0=[\y_1=\yp(u), \y_2]$, and then in terms of the null coordinates, the endpoints are positioned at $y_1^\pm = u\mp \sigma_{\mt{Page}}$ and $y_2^\pm= u \mp \sigma_2$. Now we ask for the smallest of $\sigma_2$ such that
\beq\label{barcode}
S^\textrm{gen}_{\QES''} + S_{1-2} = S^\textrm{gen}_{\QES-1}+S_2\,
\eeq
where $x_{\QES''}$ is at the bifurcation point and $x_{\QES}$ is at the extrema of the late time generalized entropy. This is illustrated in figure~\ref{fig:3pts}

We begin by assuming that $y_2$ is close to spacelike infinity of the bath, so that $y_2^->0$, $y_2^+<0$, and see how much closer to AdS we can bring it without losing the information required to reconstruct the interior of the black hole. After the coordinate transformation~\eqref{eq:zmap} and the Weyl rescaling required to bring to the evaporating black hole model,  we find the generalized entropy for these two channels read
\begin{align}
\label{eq:twochannels}
S^\textrm{gen}_{\QES''} + S_{1-2} &= \frac{c}{6} \log \(\frac{24 \pi E_S }{c \epsilon^2} \frac{-y_2^+ x_1^+(y_2^--y_1^-)}{\sqrt{f'(y_1^+)}}\)+ \frac{\phi(x_{\QES''})}{4G_N}\,,\\
\label{eq:twochannels-other}
	S^\textrm{gen}_{\QES-1}+S_2 & = \frac{c}{6} \log \(\frac{ 2}{ \epsilon^2}
	\frac{ (y_1^--y_\QES^-)  (x_\QES^+-x_1^+)(y_2^--y_2^+)\sqrt{f'(y_\QES^-)}}{(x_\QES^+-x_\QES^-)\sqrt{f'(y_1^+)}} \) + \frac{\phi(x_\QES)}{4G_N}\,.
\end{align}
The value of the dilaton at the bifurcation $x_{\QES''}$ is
\beq
\phi(x_{\QES''}) = \phi_0+2\pi T_0\bar{ \phi}_r\,.
\eeq
The dilaton at the extremal point $x_\QES$ is given by eq.~\eqref{dilatexpasion} to first order in $k$. The position $x_0$ of the extremal surface in eq.~\eqref{QES_linear} to leading order in $k$ is
\beq
x_\QES^+ \approx t_\infty\,, \quad y_\QES^- \approx y_1^+-u_{HP}\,.
\eeq
Using the leading order in eq.~\eqref{logapprox} and its derivative
\beq
\frac{f'(u)}{t_\infty-f(u)} \approx 2\pi T_1\,,
\eeq
we find $\Delta S^\textrm{gen} = S^\textrm{gen}_{\QES''}+S_{1-2}-S^\textrm{gen}_{\QES-1}-S_2$ is
\beqs
\frac{4G_N}{\bar{\phi}_r} \Delta S^\textrm{gen} =&\(2\pi (T_0-T_1) + k\pi T_1(3y_1^++u_{HP})+\frac{k}{2}\log 2e\)\\
&+2k \log \( \frac{6 \pi E_S}{c} \frac{-y_2^+ x_1^+ (y_2^--y_1^-)}{(y_1^--y_1^++u_{HP}) (y_2^--y_2^+)}\)+ {\cal O}(k^2)\,,
\eeqs
where we have used $\frac{\bar{\phi}_r}{4G_N} = \frac{c}{12k}$.

The very large negative $\frac{c \pi}{6 k}(T_0-T_1)$ term is offset by the $\frac{c \pi }{4k}T_1 y_1^+$ term because we are choosing $y_1^+ = u_\mt{Page}(\sigma_{\mt{Page}})-\sigma_{\mt{Page}} $ where $u_{\mt{Page}}$ can be read off eq.~\eqref{uPage_bath}. Plugging the value of $y^\pm_1$ and $y_2^\pm$, we find
\beq
\frac{4G_N}{\bar{\phi}_r} \Delta S^\textrm{gen} = 2k \log \( \frac{ 8 }{3\sqrt{\pi T_1 t_\infty}}  \frac{T_0}{T_1-T_0} \frac{(\sigma_2-u_\mt{Page}) (\sigma_2-\sigma_\mt{Page})}{(2\sigma_\mt{Page} +u_P^0) (2\sigma_2)}\)+ {\cal O}(k^2)\,,
\eeq
where we have used $u_\mt{Page}=u_\mt{Page}(\sigma_\mt{Page})$ to simplify the equation and once again $u_P^0$ in eq.~\eqref{eq:u0} is the leading order approximation to $u_\mt{Page}$.
The $ (T_1-T_0) (u_P^0+2\sigma_\mt{Page})$ term in the denominator is small, and the only term that can offset this to bring the argument of the logarithm close to one is $y_2^+ = \sigma_2-u_\mt{Page}$. But this requires us to anchor the end of the bath interval a distance $\sim (T_1-T_0)u/T_1$ to the right of the shock. The takeaway from this calculation is that we can remove most of the bath behind the shock. This should be expected because these intervals do not capture any of the radiation of the evaporating black hole, so they should not be essential for interior reconstruction.

\begin{figure}[t]
	\centering
	\includegraphics[width=0.7\textwidth]{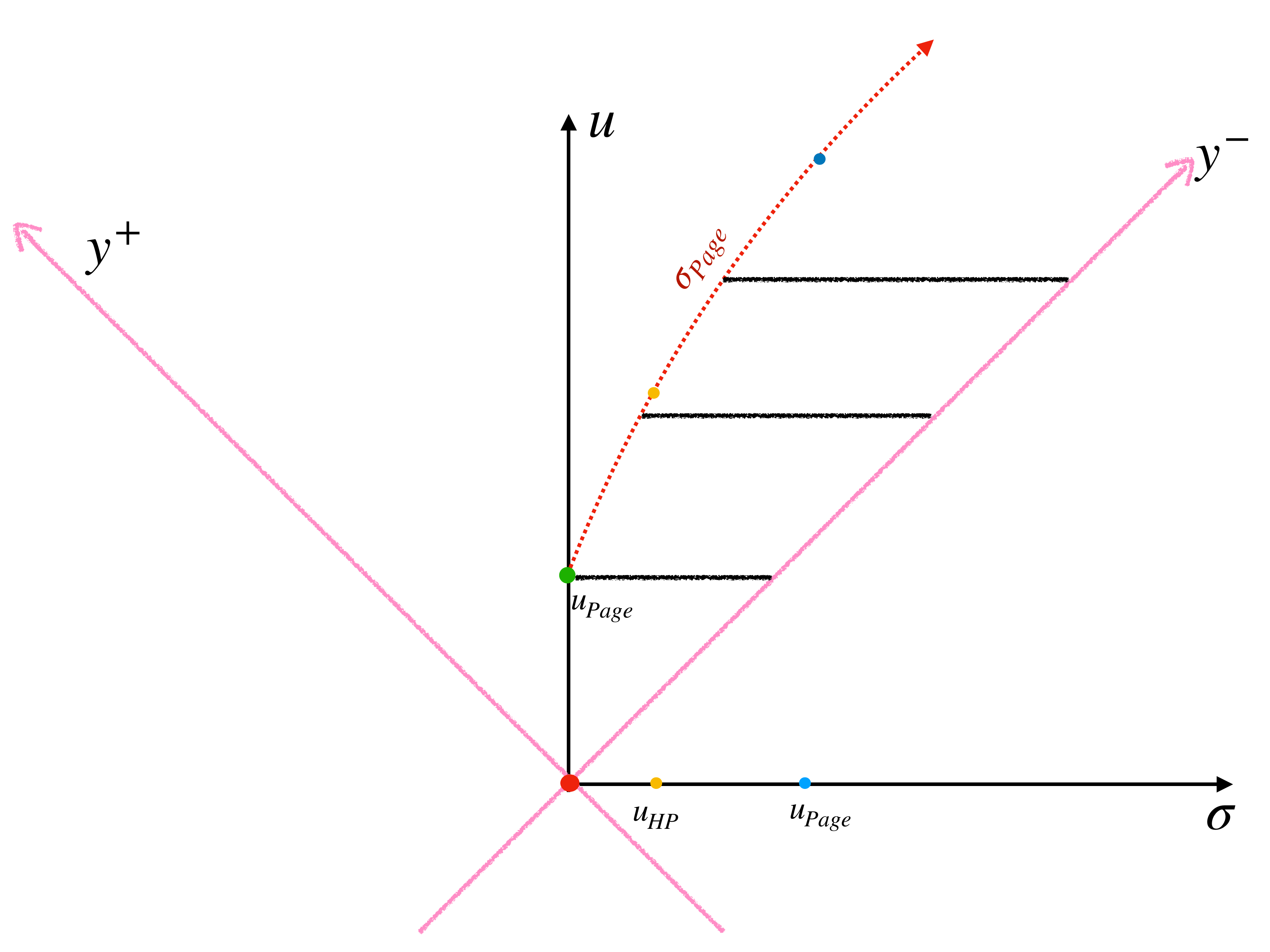}
	\caption{Smallest connected intervals that, together with \qml, are able to reconstruct a part of the black hole interior. The left endpoint $\sigma_\mt{Page}$ follows the path illustrated in figure~\ref{fig:bath_point}, while the right endpoint is anchored very close to the shockwave, as described by eq.~\reef{eq:smalldist}.}
	\label{fig:sigma2}
\end{figure}

We can now consider what happens when the point $x_2$ crosses the shock, and see if we can remove any more of the bath interval. This would amount to removing some of the early radiation after the evaporation began. In terms of the calculation, the difference now is that $x_2^+>0$ and therefore $\bar{z} = \(\frac{12 \pi}{c}E_S\)^{-2} \frac{i}{x^+}$ so that the expressions for the generalized entropies of the two channels in eqs.~\eqref{eq:twochannels} and~\eqref{eq:twochannels-other} is now
\beqs
\label{eq:twochannels2}
S^\textrm{gen}_{\QES''} + S_{1-2} &= \frac{c}{6} \log \(\frac{2 }{ \epsilon^2} \frac{ (x_1^+-x_2^+)(y_2^--y_1^-)}{\sqrt{f'(y_1^+)f'(y_2^+)}}\)+ \frac{\phi(x_{\QES''})}{4G_N}\,,\\
S^\textrm{gen}_{\QES-1}+S_2 & = \frac{c}{6} \log \(\frac{ 24 \pi E_S }{ c\epsilon^2}
\frac{ y_2^- x_2^+ (y_1^--y_\QES^-)  (x_\QES^+-x_1^+) \sqrt{f'(y_\QES^-)}}{(x_\QES^+-x_\QES^-)\sqrt{f'(y_1^+)f'(y_2^+)}} \) + \frac{\phi(x_\QES)}{4G_N}\,.
\eeqs
%and 
%\beqs
%\frac{4G_N}{\bar{\phi}_r} \Delta S^\textrm{gen} =& \(2\pi (T_0-T_1) + k\pi T_1(y_1^+-u_{HP})+\frac{k}{2}\log 2e\)
%\\&+ 2k \log \( \frac{c}{12 \pi E_S} \frac{(x_\QES^+-x_\QES^-)(x_1^+-x_2^+)(y_2^--y_1^-)}{y_2^-x_2^+(y_1^--y_\QES^-) %(x_\QES^+-x_1^+)\sqrt{f'(y_\QES^-)}}\)
%+ {\cal O}(k^2)\,.
%\eeqs
Using the position of the extremal surface in eq.~\eqref{QES_linear}, the approximation in eq.~\eqref{logapprox} 
%\beqs
%\frac{4G_N}{\bar{\phi}_r} \Delta S^\textrm{gen} =& \(2\pi (T_0-T_1) + k\pi T_1(3y_1^++u_{HP})+\frac{k}{2}\log 2e\)\\
%&+ 2k \log \( \frac{c}{24 \pi E_S} \frac{(x_1^+-x_2^+)(y_2^--y_1^-)}{y_2^-x_2^+(y_1^--y_1^++u_{HP})}\) + {\cal O}(k)\,,
%\eeqs
and plugging the positions of the endpoints $y_1$ and $y_2$ the difference in the entropies of the two channels $\Delta S^\textrm{gen}$ is
\beq\label{eq:deltaS}
\frac{4G_N}{\bar{\phi}_r} \Delta S^\textrm{gen} =\, 2k \log \( \(\frac{c}{12 \pi E_S}\)^2 \frac{ 8 \pi T_1}{3\sqrt{\pi T_1t_\infty}}    \frac{T_0}{T_1-T_0} \frac{(x_1^+-x_2^+)(\sigma_2-\sigma_\mt{Page})}{(u_\mt{Page}+\sigma_2)x_2^+(2\sigma_\mt{Page}+u_P^0)}\) + {\cal O}(k^2)\,.\\
\eeq
The term in the denominator $\(\frac{12 \pi E_S}{c}\)^2(T_1-T_0)(u_P^0+2\sigma_\mt{Page})\sim E_S^4 k/c^2T_1^3$ is very large and needs to be canceled by the separation $y_2^+$ of the point $y_2$ from the shock. Taking the ansatz $y_2^+ =  d \,\eta$ with 
\beq\label{housecat}
\eta =  \(\frac{c}{12 \pi E_S}\)^2 \frac{ 8 \pi T_1}{3(u_P^0+2\sigma_\mt{Page}) \sqrt{\pi T_1t_\infty}}    \frac{T_0}{T_1-T_0} \lesssim  \(\frac{c T_1}{6E_S}\)^4 \frac{\pi T_1}{k} \ll1 \,,
\eeq
we find that $y_2^-= 2 u_{\rm Page} - d\, \eta$ and $x_2^+ = y_2^+ +{\cal O}((y_2^+)^3)= d \eta + {\cal O}(\eta^3)$. Solving for $\Delta S^{\rm gen} = 0$ then gives
\beq\label{eq:dapprox}
d = x_1^+\frac{u_\mt{Page}-\sigma_\mt{Page}}{2u_\mt{Page}} - x_1^+ \frac{u_\mt{Page}-\sigma_\mt{Page}}{2u_\mt{Page}} \frac{u_\mt{Page}(x_1^++2u_{\mt{Page}})+\sigma_\mt{Page}(x_1^+-2u_\mt{Page})}{4u_{\mt{Page}}^2} \eta + {\cal O}(\eta^2)\,.
\eeq
Hence we find that the right endpoint must indeed anchored very close to the shock wave (at
$y^+_\mt{shock}=0$). That is,
\beq\label{eq:smalldist}
y_2^+ = x_1\, \frac{u_\mt{Page}-\sigma_\mt{Page}}{2\,u_\mt{Page}}\,\eta \sim  \(\frac{c T_1}{6E_S}\)^4 \frac{\pi T_1}{k} \ll t_\infty\,.
\eeq 
figure~\ref{fig:sigma2} shows the smallest connected intervals that are able to reconstruct a portion of the black hole interior.

%% file: sections/HolyInfo.tex
% !TEX root = ../BH_v04.tex

\subsubsection{Redundancy of the encoding} \label{five}

In examining the holographic entanglement and the corresponding entanglement wedge for \qml+bath, we found that the information needed to reconstruct interior of the black hole is encoded in a region in the bath extending from $\yp\simeq u-u_\mt{Page}$ to $\ys=u$ on a given time slice $u$ in the bath.\footnote{Recall that $y^\pm\equiv u\mp\y$, so that increasing positive $\y$ corresponds to moving further into the bath.} However, as may be expected for holography
\cite{Almheiri:2014lwa,Pastawski:2015qua,Harlow:2016vwg}, we will see that this encoding is redundant, here and in the next subsection. In this subsection, we examine the question of removing a smaller interval from the shortest connected bath interval that can still recover the black hole interior.\footnote{If one is favorably inclined to puns, one might call this process ``lyft''ing, since we are on our way to \"uberholography.} While in the following two subsections we will be working with the early-time protocol in mind for concreteness, the results in this subsection are qualitatively similar if we started from the shortest connected intervals in the late-time protocol of section~\ref{twoprime}, and in fact the main conclusion of subsection~\ref{uber} in eqs.~\eqref{eq:trickOrTreat} and~\eqref{eq:deq} is quantitatively the same.

\begin{figure}[t]
  \centering
  \includegraphics[width=0.7\textwidth]{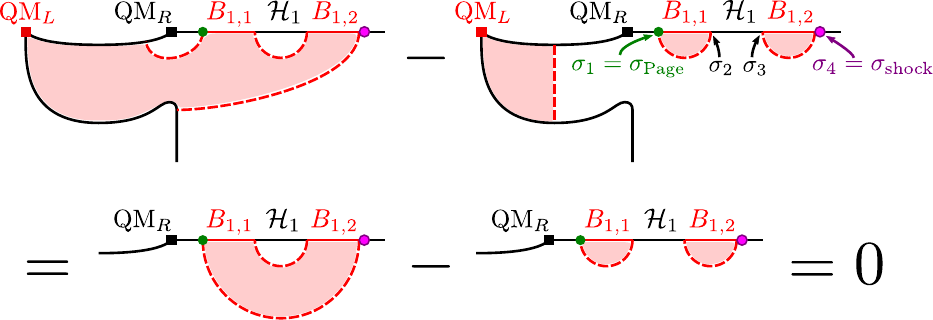}
  \caption{Excising the largest possible hole $\Hcal_1$ from the smallest possible interval $\Bcal_0=B_{1,1}\cup \Hcal_1\cup B_{1,2}$ of the bath such that recoverability of the black hole interior is preserved. Minimizing of $\Bcal_0$ \ie setting $\y_1=\yp$, allows us to equate the difference in generalized entropies of the first line with the differences in von Neumann entropies in the second line; maximization of $\Hcal_1$ is determined by the equality of latter branches.}
  \label{fig:mocha}
\end{figure}

Let us denote the bath interval described above as $\Bcal_0=\[\yp,\ys\]$. Now we ask how large a hole $\Hcal_1$  can we remove from $\Bcal_0$ while still preserving recoverability of the black hole interior? The desired configuration of HRT surfaces is sketched in the top left illustration of figure~\ref{fig:mocha}.  We are now left with two disjoint intervals in the bath $B_{1,1}=[\y_1=\yp,\y_2]$ and $B_{1,2}=[\y_3,\y_4=\ys]$, which combined with \qml\ are still able to reconstruct the black hole interior. To determine the allowed size and position of the hole, \ie to determine the allowed values of $\y_2$ and $\y_3$, we must compare the contributions of the different HRT surfaces. For example, the desired configuration (in the top left of figure~\ref{fig:mocha}) is given by
\beq\label{rabbit1}
S_{\qes-1,2-3,4} = S_{\qes-1}+S_{2-3} + S_{4}\,,
\eeq
where we have indicated the contributions of the separate components of the HRT surface on the right. For example, $S_4$ is the contribution of the geodesic connecting $y_4$ to the ETW brane, while $S_{\qes-1}$ corresponds to the generalized entropy which includes the length of the geodesic connecting $\y_1$ to the QES and also the dilaton contribution at the latter point. Now the competing configuration which limits the size of the hole is shown in the top right illustration of figure~\ref{fig:mocha}, and the corresponding holographic entropy is given by
\beq\label{rabbit2}
S_{\qes',1-2,3-4} = S_{\qes'} + S_{1-2} + S_{3-4}\,.
\eeq
In this case, QES$'$ indicates that the quantum extremal surface is distinct from that appearing in eq.~\reef{rabbit1}. In fact, in this configuration, QES$'$ corresponds to the bifurcation surface of the original black hole on the Planck brane.

A priori it may seem that comparing the entropies in eqs.~\reef{rabbit1} and \reef{rabbit2} will require some numerical analysis, however, the present comparison is simplified because we have chosen $\y_1=\yp$. This point marks the precise transition between two competing sets of HRT surfaces, as illustrated in figure~\ref{fig:3pts}. Hence at this precise point, we have
\beq\label{rift}
S_{\qes-1}+S_4 = S_{\qes'}+S_{1-4}\,.
\eeq
Substituting this expression into eq.~\reef{rabbit1} and taking the difference then yields
\beq\label{latter}
S_{\qes-1,2-3,4}-S_{\qes',1-2,3-4}= S_{1-4} +S_{2-3}- S_{1-2}-S_{3-4}\,,
\eeq
as illustrated by the bottom illustration of figure~\ref{fig:mocha}.
Note that the latter \reef{latter} is controlled entirely by the positions of the points in the bath, which are fixed, \ie the transition between the two branches in the top of figure~\ref{fig:mocha} is completely independent of the physics on the Planck brane, \ie of QES and QES$'$.\footnote{However, if instead, $\y_1$ was placed closer to the end of the bath (\ie closer to \qmr), then eq.~\reef{rift} would no longer hold and comparing eqs.~\reef{rabbit1} and \reef{rabbit2} would no longer be as simple.}

Hence in eq.~\reef{latter}, we are simply comparing the lengths of the corresponding HRT surfaces. This comparison can be made in terms of the $z$ coordinates, where the transition occurs at
\beq
\frac{|z_2-z_3|^2}{|z_4-z_3|^2} \frac{|z_4-z_1|^2}{|z_2-z_1|^2}=1\,,
\label{eq:nap}
\eeq
or in the $y^\pm$ coordinates, where
\beq
\frac{y_3^--y_2^-}{y_4^--y_3^-}\, \frac{f(y_2^+)-f(y_3^+)}{f(y_3^+)-f(y_4^+)}\, \frac{y_4^--y_1^-}{y_2^--y_1^-}\,
\frac{f(y_1^+)-f(y_4^+)}{f(y_1^+)-f(y_2^+)}= 1\,.
\label{eq:zzz}
\eeq

\begin{figure}[t]
	\centering
 \includegraphics[width=0.45 \textwidth]{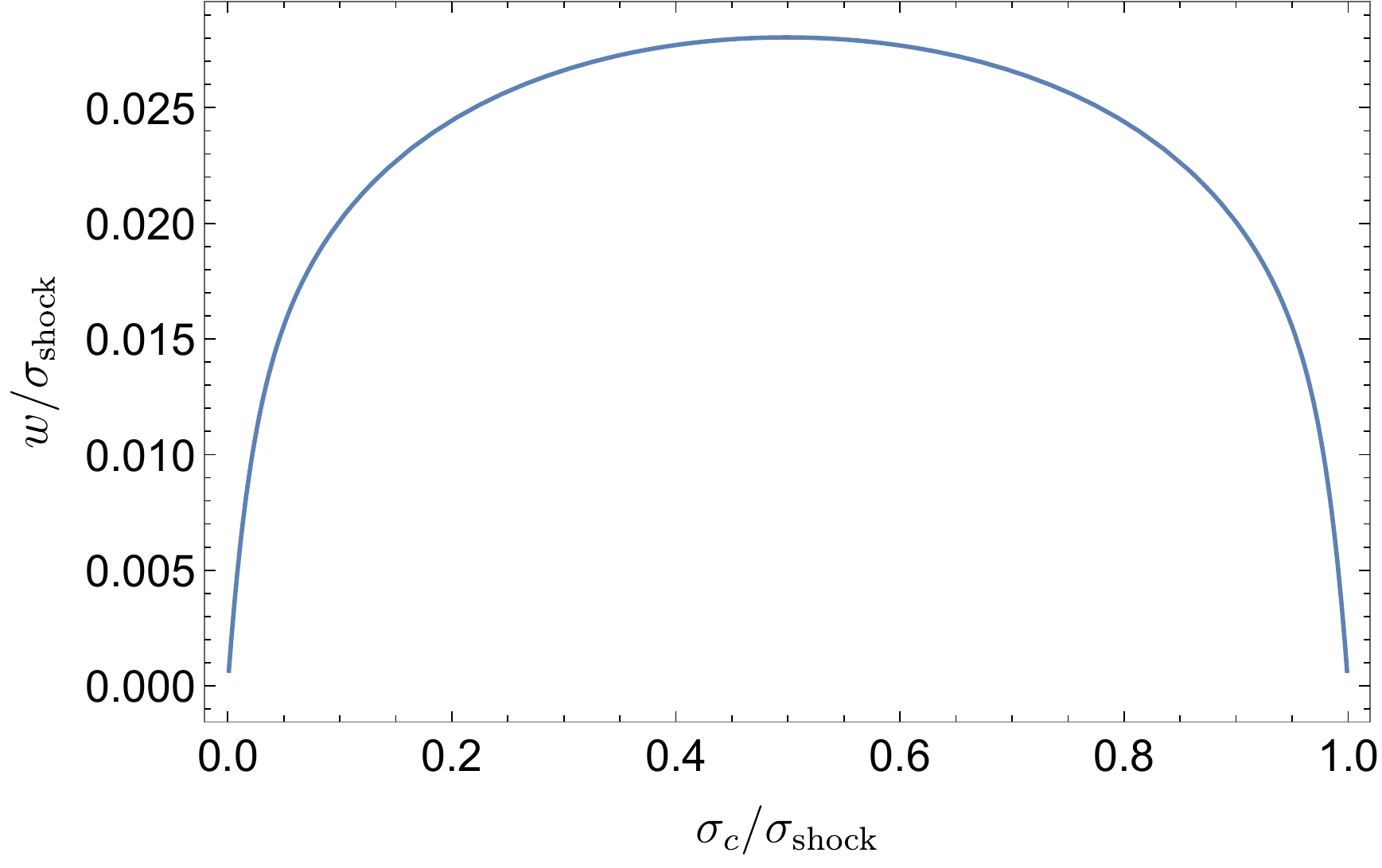}
 \includegraphics[width=0.45 \textwidth]{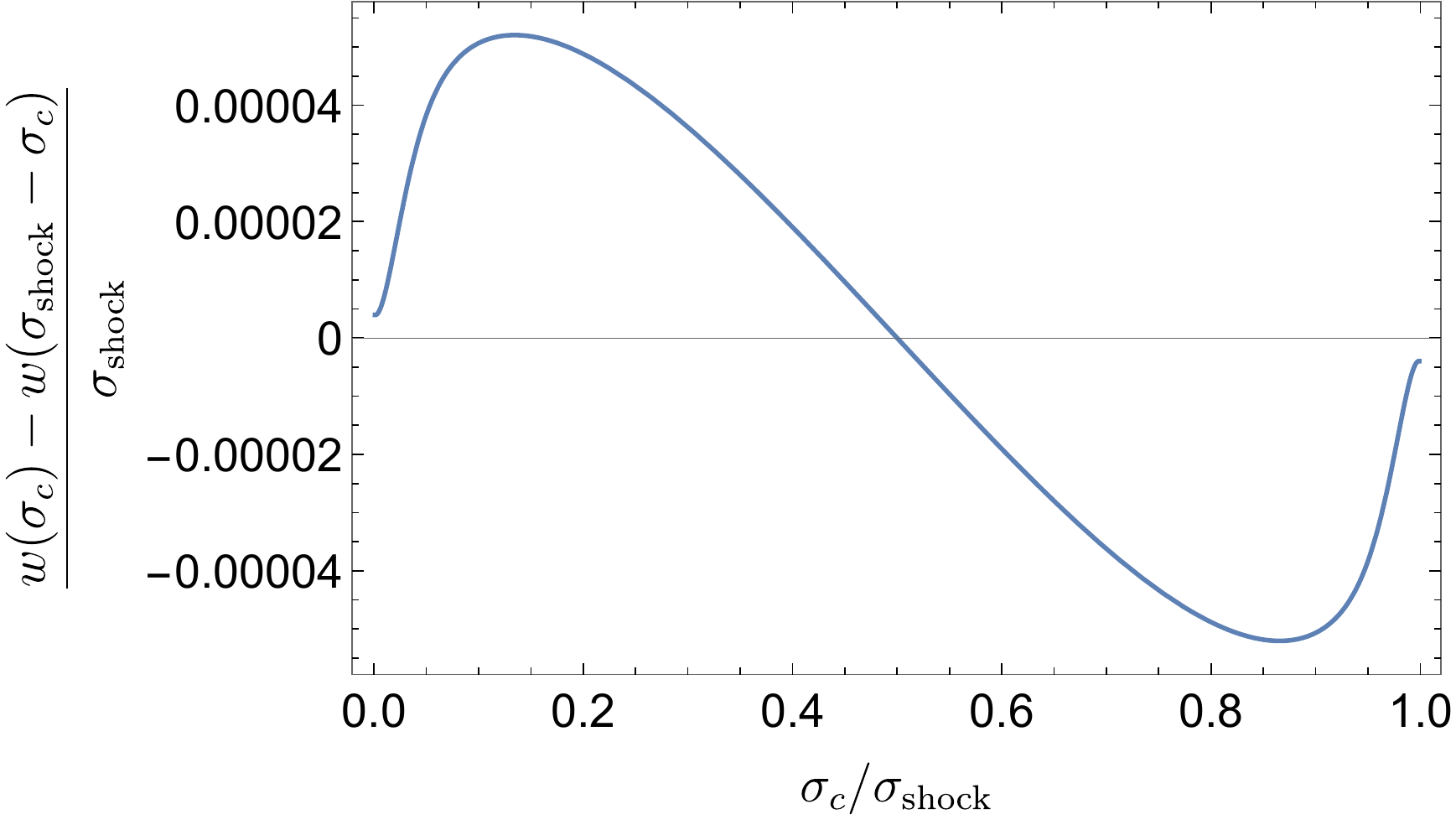}
 \caption{To the left, maximum width $w$ of the hole $\Hcal_1$ removed from the bath region $\Bcal_0$ as a function of the center of the interval $\y_c$. To the right, asymmetry in the maximum width about $\sigma_c=\sigma_{\rm shock}/2$. Here we consider the time slice $u= u_\mt{Page}$ so that $\Bcal_0=[\yp=0,\ys=u_\mt{Page}]$.  }\label{fig:5pts-u0}
\end{figure}

Now, of course, the width of our hole $\Hcal_1$, \ie $|\y_3-\y_2|$, depends on how it is positioned within the original interval $\Bcal_0=\[\yp,\ys\]$. As an example, in figure~\ref{fig:5pts-u0},
we consider $\Bcal_0$ with $\yp= 0$, \ie $u = u_\mt{Page}$,\footnote{Note that  there is no real loss of generality with this choice. Moving to a later time slice simply shifts the parameters to $u' =u_\mt{Page} + \Delta u$, $\yp'\simeq\Delta u$ and $\ys'=u_\mt{Page}+\Delta u$, which corresponds to just shifting $y_{1,2}^-$ by a constant while leaving $y_{1,2}^+$ unchanged. However, we observe that eq.~\eqref{eq:zzz} is invariant under a constant shifts in $y^-$ and so our analysis here would be unchanged.} and explore the maximum width of the interval that can be removed as a function of the center of the interval. In the figure, we see that the optimal choice, \ie the largest hole, is when we position the hole at the center of $\Bcal_0$. In the figure, we see that in this optimal configuration, we can remove approximately 10\% of the region $\Bcal_0$. The width of the hole shrinks rapidly as $\y_c$ approaches either $\yp$ or $\ys$ -- see further comments below. We can interpret this shrinking as indicating that the information in both the early Hawking radiation (near the shock) and the later radiation (near $\yp$) are extremely important in reconstructing the black hole interior.

The resulting plot in the left panel of figure~\ref{fig:5pts-u0} is almost symmetric about the midpoint.  The small asymmetry (shown in the right panel) is due to the nonlinearities of the mapping $f(y_i^+)$. Interestingly, this asymmetry is eliminated if we use the small $k$ approximation:\footnote{For the parameters in table~\ref{tab:baseline}, the difference between the full $f(u)$ and this approximation is less that an fraction of a percent, \ie $|f(u)-f_\mt{approx}(u)|/|f(u)|\lesssim
0.0015\%$.} $f(u) \simeq \frac{1}{\pi T_\infty} \tanh \( \pi T_\infty u \)$ where $T_\infty = \frac{1}{\pi t_\infty} =  {I_{1}\left[\frac{2 \pi T_{1}}{k}\right]}/{I_{0}\left[\frac{2 \pi T_{1}}{k}\right]}$. With this approximation, the identity $\tanh(x)-\tanh(y) = {\rm sech}(x) \,{\rm sech}(y) \sinh(x-y)$ can be used to simplify eq.~\eqref{eq:zzz} as
\beq
\frac{y_3^--y_2^-}{y_4^--y_3^-}\, \frac{\sinh(y_2^+-y_3^+)}{\sinh(y_3^+-y_4^+)} \,\frac{y_4^--y_1^-}{y_2^--y_1^-}\,
\frac{\sinh(y_1^+-y_4^+)}{\sinh(y_1^+-y_2^+)} = 1\,.
\eeq
Further, for the example shown in figure~\ref{fig:5pts-u0},\footnote{Again, the general result corresponds to shifting all the points to the left by  $\Delta u= u-u_\mt{Page}$.}
we then substitute $y_1^\pm = u_\mt{Page}$, $y_2^\pm=u_\mt{Page} \mp (\y_c-w/2)$, $y_3^\pm=u_\mt{Page} \mp (\y_c+w/2)$ and $(y_4^+,y_4^-)=(0, 2u_\mt{Page})$, which yields
\beq\label{simpA}
\frac{w}{u_\mt{Page}  -\y_c-w/2}\, \frac{\sinh w}{\sinh(u_\mt{Page}  -\y_c-w/2)}\, \frac{u_\mt{Page} }{\y_c-w/2}
\frac{\sinh u_\mt{Page} }{\sinh(\y_c-w/2)} = 1\,.
\eeq
Clearly, the resulting expression is invariant under $\y_c \to u_\mt{Page} -\y_c$, \ie the corresponding plot is exactly symmetric about the midpoint $\sigma_c=u_\mt{Page}/2$.
Hence in this approximation, the importance of the information in both the early and later Hawking radiation is equally weighted for the reconstruction of the black hole interior.

%We plot the difference between the using the approximation in figure~\ref{fig:5pts-approx}.
%

In closing this section, we note that the initial and final slopes of the curve in the left panel of figure~\ref{fig:5pts-u0} are universal for holographic CFTs. This is because the question of how large a hole can be exciseded near the endpoint of an interval without triggering a phase transition is one which probes the UV entanglement structure. To see this, let us, without loss of generality, take in the RHS of eq.~\eqref{latter} the endpoints, $\y_2$ and $\y_3$, of the hole to be very close to the endpoint $\y_1=\yp$. Maximizing the size of the hole to the verge of triggering the transition between the two branches amounts to setting the RHS of eq.~\eqref{latter} to zero. In the limit of the hole tending towards the point $\y_1$, we have $S_{1-4}=S_{3-4}$; moreover, the dependence of $S_{3-4}$ on point $\y_3$ is extremely weak relative to the dependence of $S_{1-2}$ and $S_{2-3}$ on the location and size of the hole. Thus, we find that $S_{1-2}\sim S_{2-3}$ for maximally-sized holes close to  $\y_1$. Since these latter entropies probe short distances, this relation gives the same constraint on points $\y_{1,2,3}$ as in the vacuum case, \ie $|\y_1-\y_2|\sim |\y_2-\y_3|$. This corresponds to slopes of $\pm {2}/{3}$ at the endpoints of figure~\ref{fig:5pts-u0}, \ie near $\y_1$, we have $w\simeq \frac23(\y_c-\y_1)$ while near $\y_4$, $w\simeq \frac23(\y_4-\y_c)$. These results might be contrasted with the largest holes that can be removed from $\Bcal_0$ in these limits, \ie $w< 2(\y_c-\y_1)$ and $w< 2(\y_4-\y_c)$. This comparison gives a quantitative measure that the $w$ is indeed shrinking {\it rapidly} near the endpoints of $\Bcal_0$, as commented above.

\subsubsection{\"Uberholography} \label{uber}

\input{sections/uberholo}

%% file: sections/uberholo.tex
% !TEX root = ../BH_v04.tex
Having considered removing a single hole from the bath region $\Bcal_0=\[\yp,\ys\]$, it is natural to generalize our analysis to arbitrarily many holes. Specifically, one may ask: what is the smallest total length of disconnected regions in $\Bcal_0$ needed, in conjunction with QM$_L$, to reconstruct the interior of the black hole? In fact, by an iterative process where, at each step, a hole is punched into each connected region in this bath region, this total length can be reduced arbitrarily close to zero. This procedure was designated `\"uberholography', where a bulk region is encoded in a subset of the boundary with lower (fractal) dimension than the dimension of the boundary~\cite{Pastawski:2016qrs}.

\begin{figure}[t]
  \centering
  \begin{subfigure}[t]{\textwidth}
\centering
    \includegraphics[width=0.8\textwidth]{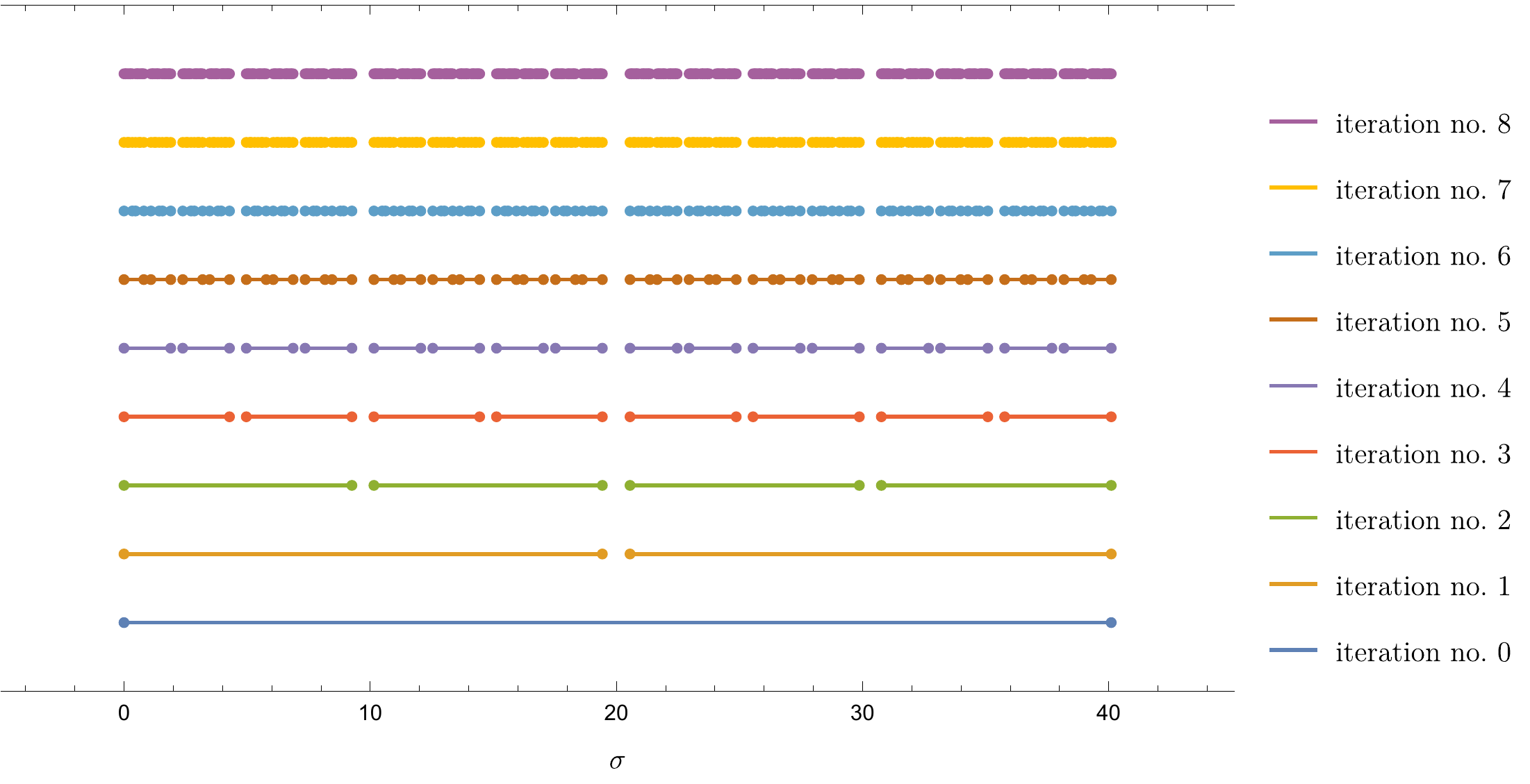}
    \caption{Interval of the bath needed to reconstruct black hole interior, iteratively hole-punched.}
    \label{fig:latte}
  \end{subfigure}
  \par\bigskip
  \begin{subfigure}[t]{0.45\textwidth}
    \includegraphics[width=\textwidth]{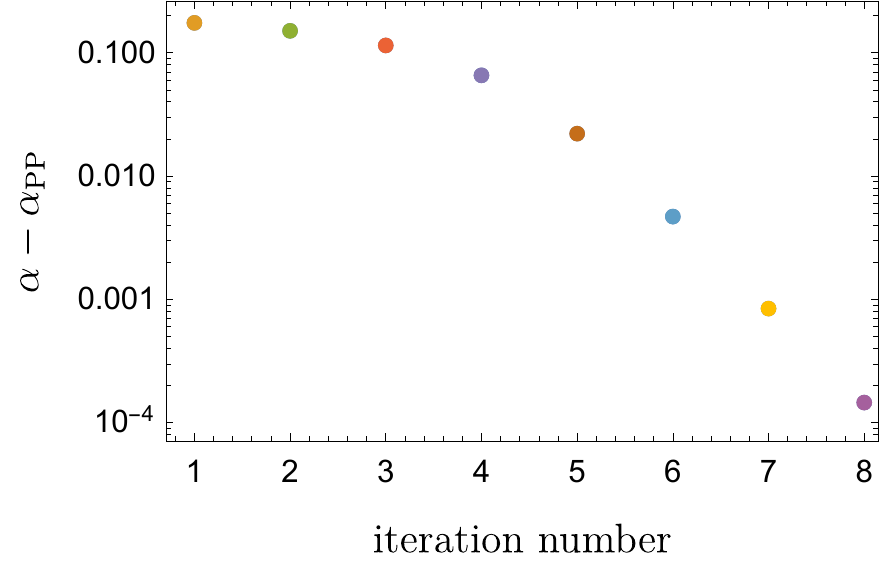}
    \caption{Parameter $\alpha$, defined in eq.~\eqref{eq:sleepy}, which, in the infinite iteration limit, gives the fractal dimension of the bath region needed to recover the black hole interior.}
    \label{fig:overtime}
  \end{subfigure}
  \hfill
  \begin{subfigure}[t]{0.45\textwidth}
    \includegraphics[width=\textwidth]{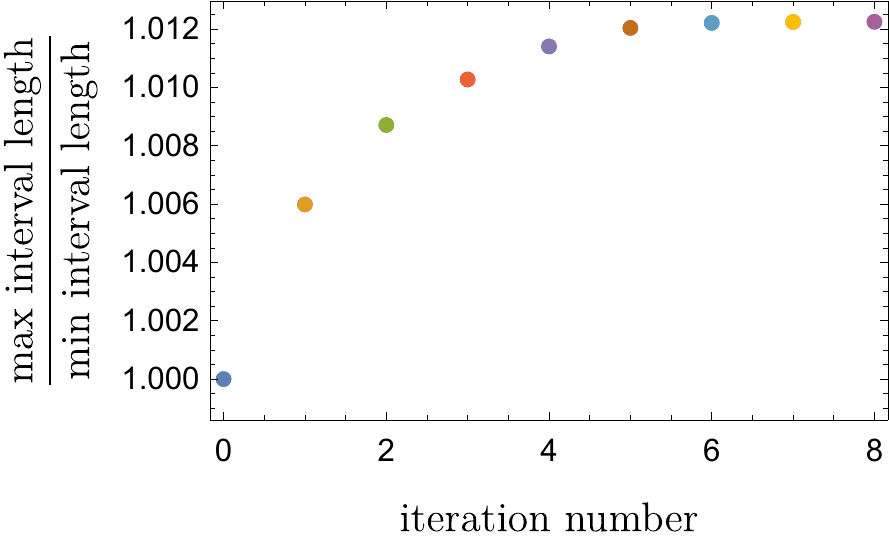}
    \caption{Ratio between the maximal and minimal lengths of connected intervals after each iteration.}
    \label{fig:yawn}
  \end{subfigure}
  \caption{Iterative process of punching maximally-sized holes into the interval of the bath needed (together with QM$_L$) to reconstruct the black hole interior. Here, the original interval of the bath under consideration stretches from the AdS-bath boundary to the shock on the time slice corresponding to the Page time on the boundary.}
  \label{fig:overtimePay}
\end{figure}

% \begin{figure}[t]
%   \centering
%   \begin{subfigure}[t]{\textwidth}
%     \includegraphics[width=\textwidth]{figures/intervals__baseline__u_uPage__y1Offset_-1}
%     \caption{Interval of the bath needed to reconstruct black hole interior, iteratively hole-punched.}
%     \label{fig:bacardi}
%   \end{subfigure}
%   \par\bigskip
%   \begin{subfigure}[t]{0.45\textwidth}
%     \includegraphics[width=\textwidth]{figures/alpha__baseline__u_uPage__y1Offset_-1}
%     \caption{Parameter $\alpha$, defined in eq.~\eqref{eq:sleepy}, whose limit for infinite iterations gives the Minkowski dimension of the bath region needed to recover the black hole interior.}
%     \label{fig:absolut}
%   \end{subfigure}
%   \hfill
%   \begin{subfigure}[t]{0.45\textwidth}
%     \includegraphics[width=\textwidth]{figures/lengthRatio__baseline__u_uPage__y1Offset_-1}
%     \caption{Ratio between the maximal and minimal lengths of connected intervals after each iteration.}
%     \label{fig:zirkova}
%   \end{subfigure}
%   \caption{Iterative process of punching maximally-sized holes into the interval of the bath needed (together with the $L$ quantum system) to reconstruct the black hole interior. Here, the original interval of the bath under consideration stretches from a point at $\frac{y^+-y^-}{2}=-1$ to the shock on the time slice corresponding to the Page time at $\frac{y^+-y^-}{2}=-1$.}
%   \label{fig:jackDaniels}
% \end{figure}

We illustrate this process in figure~\ref{fig:latte}. We begin, as in section~\ref{sec:cocacola}, with the smallest interval $\Bcal_0$ on a constant time slice of the bath such that the black hole interior can be recovered from QM$_L$ and $\Bcal_0$. For concreteness, we have positioned the first endpoint $\min(\Bcal_0)=\sigma_\Page$ at the AdS-bath boundary in figure~\ref{fig:overtimePay} --- we find qualitatively similar results when this endpoint is chosen inside the bath. In the first round of the iterative process, we punch a maximally-sized hole $\Hcal_1$ into the initial interval $\Bcal_0$ while preserving recoverability of the black hole interior, as discussed in section~\ref{five}. What remains is the union $\Bcal_1=\Bcal_0\setminus \Hcal_1=B_{1,1}\cup B_{1,2}$ of two intervals $B_{1,1},B_{1,2}$. Before proceeding to the inductive step, we emphasize again that the task of maximizing $\Hcal_1$ can be reduced into a simple problem that involves comparing channels of the Von Neumann entropy of the disconnected region $\Bcal_1$, as written in eq.~\eqref{latter} and illustrated in the first equality of figure~\ref{fig:mocha}. A similar reduction can be made in all further iterative steps of the hole-punching procedure, so that we need only consider Von Neumann entropy channels of the surviving region $\Bcal_n$ in the bath.\footnote{Indeed, the problem would be identical to the vacuum case considered in~\cite{Pastawski:2016qrs} save for the conformal transformation taking $z$ to $x,y$ coordinates.}

Due to the maximization of the hole $\Hcal_1$, the two channels shown in the last line of figure~\ref{fig:mocha} give the same entropy. For the inductive step, it is simplest to consider the second channel shown. Since, in this channel, the entanglement wedges for $B_{1,1}$ and $B_{1,2}$ are disconnected, we may separately consider punching maximally-sized holes in $B_{1,1}$ and $B_{1,2}$. Thus, the process described in the previous paragraph can be repeated, now with $B_{1,1}$ or $B_{1,2}$ taking the place of $\Bcal_0$. Indeed, this procedure may be performed iteratively: given a disconnected region $\Bcal_n=B_{n,1}\cup\cdots\cup B_{n,2^n}$ composed of intervals $B_{n,m}$, we may punch a maximally-sized hole $H_{n+1,m}$ into each $B_{n,m}$ while maintaining recoverability of the the black hole interior; the result is a smaller region $\Bcal_{n+1}=\Bcal_n\setminus \Hcal_{n+1}$, where $\Hcal_{n+1}=H_{n+1,1}\cup\cdots\cup H_{n+1,2^n}$.

At each step, we may define the quantities
\begin{align}
  r_n =& \frac{|\Bcal_n|}{|\Bcal_{n-1}|},
  &
  \alpha_n =& \frac{\log 2}{\log \frac{2}{r_n}}
  \label{eq:sleepy}
\end{align}
describing the rate at which the total length $|\Bcal_n|$ of the region in the bath shrinks over iterations. In figure~\ref{fig:overtime}, we plot $\alpha_n$, showing that it approaches the constant value
\begin{align}
  \alpha_\infty
  = {\alpha_\PP}
  \equiv \frac{\log 2}{\log(\sqrt{2}+1)}
  \approx 0.786
  \label{eq:trickOrTreat}
\end{align}
obtained for the CFT vacuum in~\cite{Pastawski:2016qrs}. Thus, we find that the region $\Bcal_\infty$ of the bath needed, with QM$_L$, to recover the black hole interior exhibits uberholography --- it has zero total length. Moreover, as we shall show momentarily, $\alpha_\infty$ gives the fractal dimension $d(\Bcal_\infty)$ of $\Bcal_\infty$. Hence, we see that $\Bcal_\infty$ has the same fractal dimension $\alpha_\infty=\alpha_\PP$ as for uberholography in the vacuum case. The universality of $\alpha_\PP$ may be explained by the fact that the UV entanglement excised by uberholography is determined predominantly by the vacuum entanglement structure. Explicitly, for our case, despite the conformal transformation from eq.~\eqref{eq:nap} to eq.~\eqref{eq:zzz}, for small interval sizes, eq.~\eqref{eq:zzz} still reads as though it were comparing vacuum entropy channels:
\begin{align}
%   \begin{split}
%   (y_3^- - y_2^-)(y_2^+ - y_3^+) (y_4^- - y_1^-)(y_1^+ - y_4^+)
%   =& (y_4^- - y_3^-)(y_3^+ - y_4^+) (y_2^- - y_1^-)(y_1^+ - y_2^+) \\
%   &+ \Ocal\left((\text{distance between points})^5\right).
% \end{split}
\begin{split}
\frac{|y_2-y_3|^2|y_1-y_4|^2}{|y_3-y_4|^2|y_1-y_2|^2}+ \Ocal\left(f''\cdot(\text{distance between points})\right)
=& 1.
\end{split}
\end{align}

It is straight-forward to show that $\alpha_\infty$ gives the dimension of $\Bcal_\infty$ by making use of the fact that the ratio $\frac{\max_m |B_{n,m}|}{\min_m |B_{n,m}|}$ of maximal and minimal lengths of the consituents of $\Bcal_n$ approaches a constant in the infinite iteration limit $n\to\infty$, as verified in figure~\ref{fig:yawn}. Recall that the (Minkowski) dimension of the set $\Bcal_\infty$ is defined to be
\begin{align}
  d(\Bcal_\infty)
  \equiv& \lim_{\epsilon \to 0} \frac{\log N(\epsilon)}{\log(1/\epsilon)},
  \label{eq:bedtime}
\end{align}
where $N(\epsilon)$ is the minimal number of $\epsilon$-diameter balls (in this case, $\epsilon$-length intervals) needed to cover $\Bcal_\infty$. For any small $\epsilon$, it is possible to find the first iteration $n=n^+(\epsilon)$ such that $\max_m |B_{n,m}|\le\epsilon$ and also the last iteration $n=n^-(\epsilon)$ such that $\epsilon\le\min_m |B_{n,m}|$. Since $\max_m |B_{n,m}|$ and $\min_m |B_{n,m}|$ differ only by a constant factor in the $n\to\infty$ limit, it follows that
\begin{align}
n^\pm \sim \frac{\log(\epsilon)}{\log(r_{n^\pm}/2)}
\label{eq:allnighter}
\end{align}
where $r_n/2$ gives the factor by which the average length of single intervals shrinks over the $n$th iteration. By monotonicity in $N(\epsilon)$, we also have
\begin{align}
  2^{n^-}\le N\left(\min_m |B_{n^-,m}|\right) \le N(\epsilon) \le N\left(\max_m |B_{n^+,m}|\right)\le 2^{n^+}.
  \label{eq:tired}
\end{align}
Using eqs.~\eqref{eq:allnighter} and~\eqref{eq:tired}, we have from eq.~\eqref{eq:bedtime} and the definition~\eqref{eq:trickOrTreat} of $\alpha_\infty$,
\begin{align}\label{eq:deq}
  d(\Bcal_\infty)
  = \alpha_\infty
\end{align}
as claimed. Eqs.~\eqref{eq:trickOrTreat} and~\eqref{eq:deq} are the main results of this subsection. Lastly, we emphasize once again that despite the fact that we have started from the shortest connected intervals of the early-time protocol, the results are the same if we start from the shortest connected intervals of the late-time protocol of section~\ref{twoprime}.

%% file: sections/Later.tex
% !TeX spellcheck = en_US
% !TEX root = ../BH_v04.tex

\subsection{Late-time protocol: forgetting the early-time radiation}\label{twoprime}

In section~\ref{two}, we asked the question of how much of the bath is required to reconstruct the interior of the black hole in combination with \qml while focusing on the Hawking radiation emitted at early times.  A different approach is to ask how much of the early-time radiation can we ignore but still keep the ability to reconstruct the interior of the black hole. Concretely, we can anchor $\sigma_1 = 0$ for times later than $u_\mt{Page}$ and see how small $\sigma_2$ can be while still keeping the recoverability of the black hole interior. The two competing channels are the same as the ones in the early-time protocol, and are illustrated in figure~\ref{fig:3pts}. The difference is that the left endpoint of the bath interval is now anchored at the AdS-bath junction, \ie $\sigma_1=0$, and the right endpoint is no longer anchored at the shock,  \ie $\sigma_2 < \sigma_{\rm shock}$.
 \begin{figure}[t]
 	\centering\includegraphics[width=4.80in]{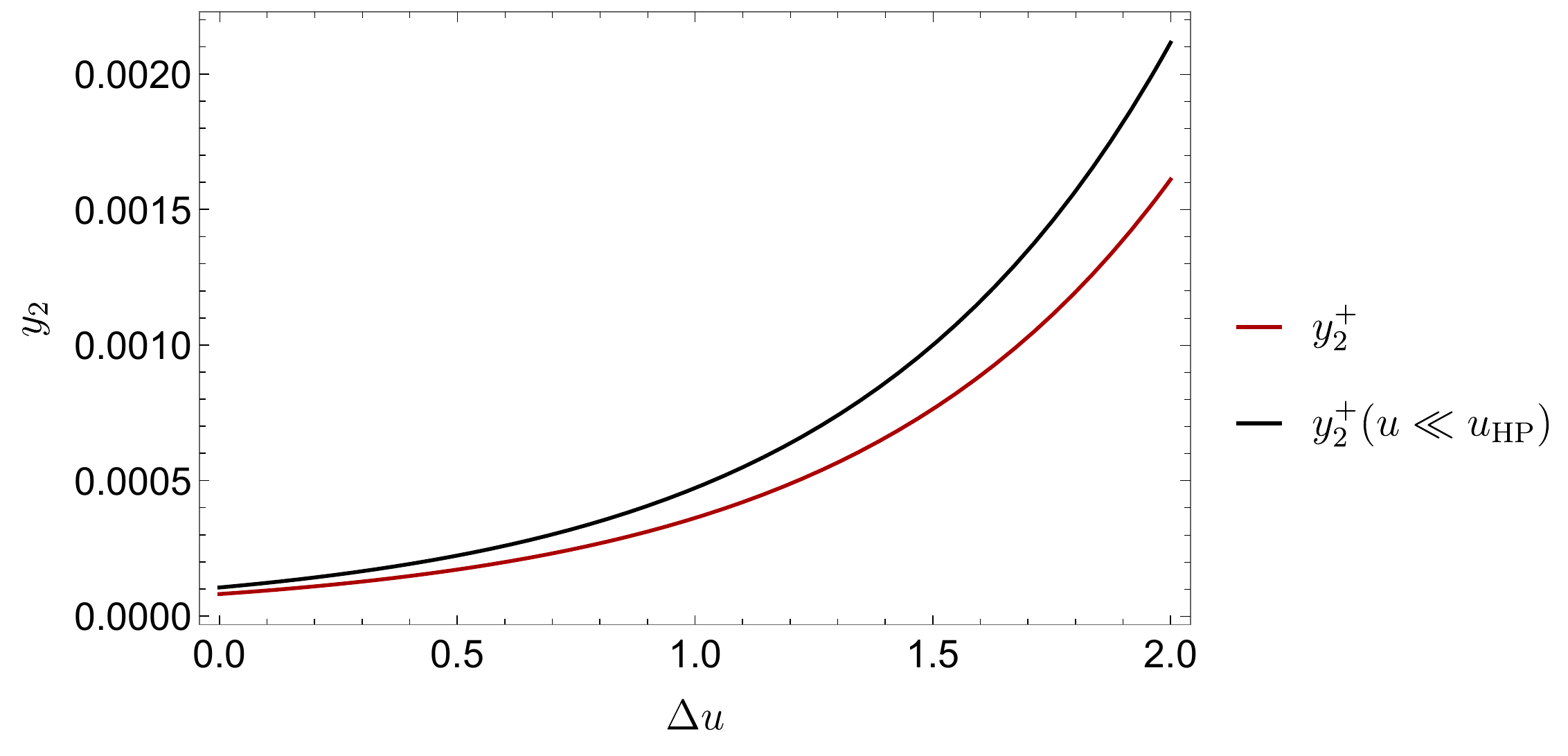}
 	\caption{The time evolution of $y^+_2$ with dependence on $\Delta u= u- u_{\mt{Page}}$. The red line is derived from the direct numerical calculation, while the blue line represents eq.~\eqref{y2p_early}. }\label{fig:y2p_early}
 \end{figure}

As in eq.~\reef{barcode}, we  need to consider the equivalence condition, 
\begin{equation}
%\frac{4 \GN}{\bar{\phi}_r}\( 
 S^\textrm{gen}_{\QES''} + S_{1-2}  - S^\textrm{gen}_{\QES-1}-S_2
 % \) 
=0
\end{equation}
with the new endpoints,
which is equivalent to 
\begin{equation}\label{newdeltaS}
\begin{split}
& 2k \log \( \frac{ 12 \pi E_S }{ c} \frac{f(y^+_2)y^-_2}{\(t-f(y^+_2) \)\sigma_2} \)  \\
& = 2\pi T_0 - \frac{\phi(x^\pm_{\QES})}{\bar{\phi}_r} 
 + 2k \log \(  \frac{(x_\QES^+-x_\QES^-)}{(y_1^--y_\QES^-) (x_\QES^+-x_1^+)\sqrt{f'(y_\QES^-)}}\) \,,
 \end{split}
\end{equation}
where the dilaton is derived in eq.~\eqref{dilatexpasion} and the bulk entropy on the right hand side is as in eq.~\eqref{bulkentropy}. The above equation can not be solved analytically in general, and so we examine different for different regimes of $\Delta u\equiv u - u_{\Page}$.

\begin{figure}[t]
	\centering
	\includegraphics[width=0.45 \textwidth]{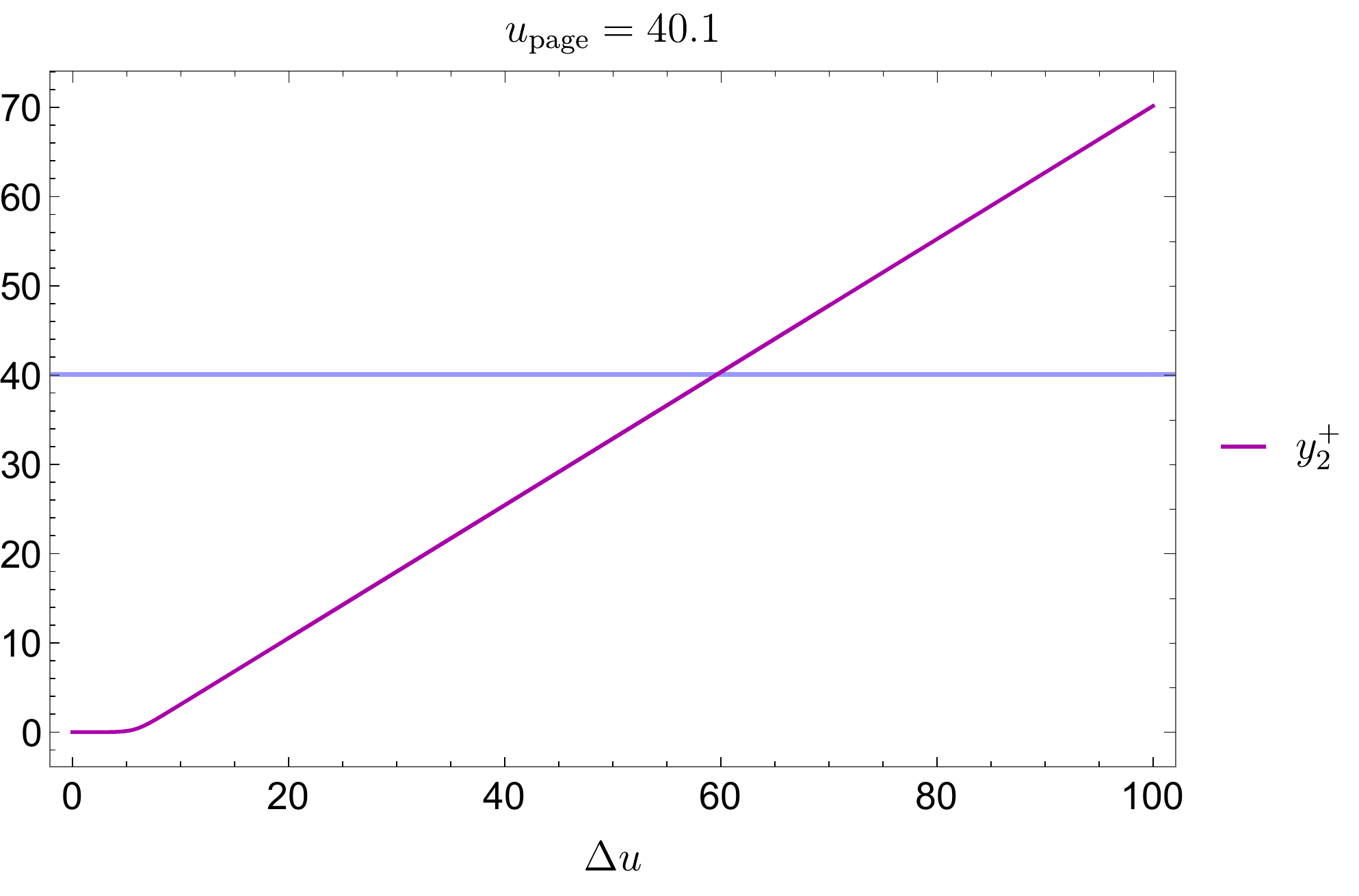}
	\includegraphics[width=0.54 \textwidth]{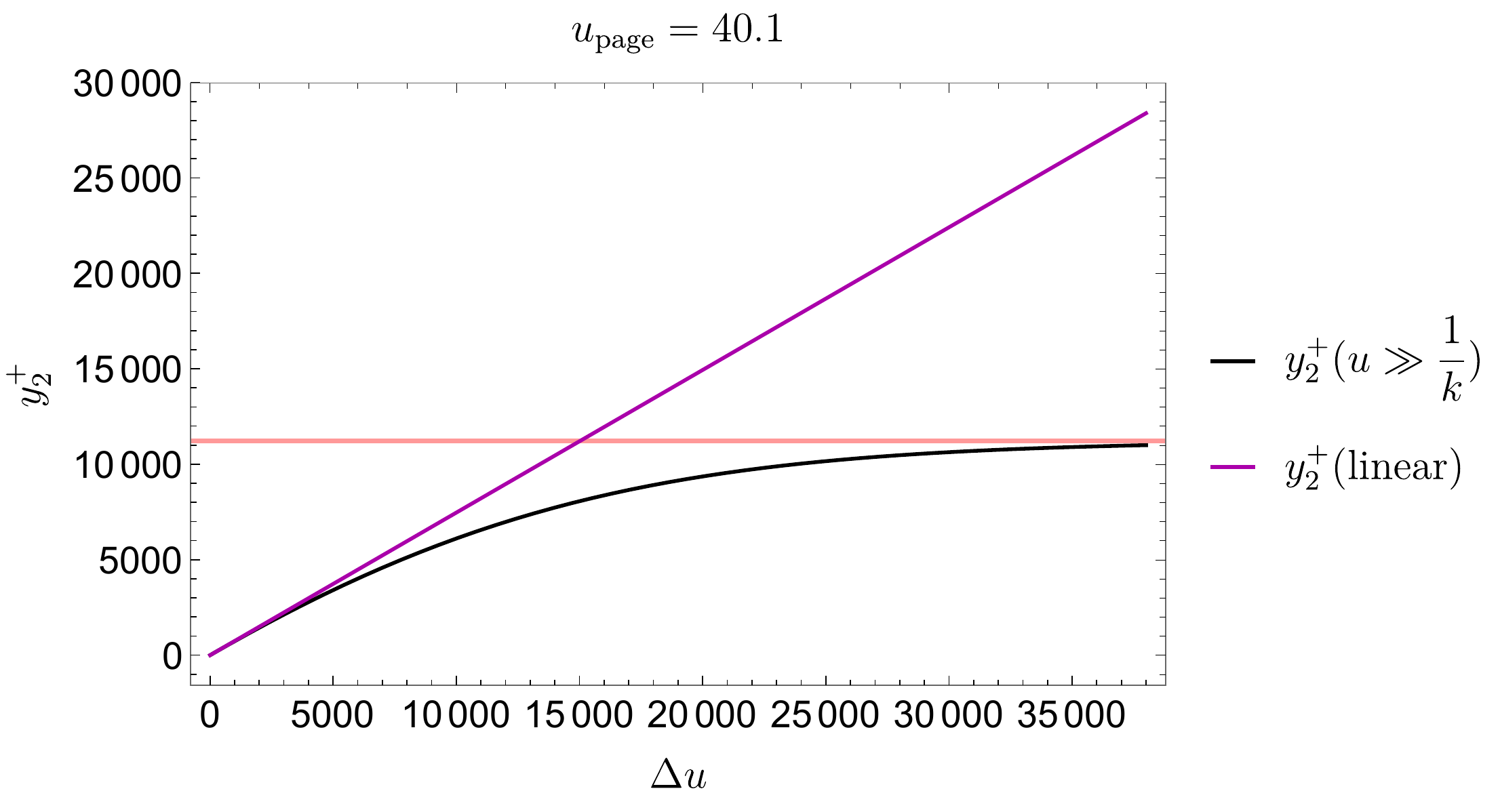}
	\caption{The time evolution of $y^+_2$ with dependence on $\Delta u= u- u_{\mt{Page}}$. Left: The numerical results from the full linear generalized entropy. The horizontal line indicates the $y^+_2=u_{\mt{Page}}$.  Right: Black curve shows the results with exponential dilaton term. The horizontal line represents the limit of $y^+_2$ defined in~\eqref{null_shift}.}\label{fig:y2p}
\end{figure}

When $\Delta u$ is smaller than the Hayden-Preskill scrambling time $u_\mt{HP}$, the distance of the right endpoint of the bath interval to the shock $y_2^+$ is still very small. This is shown in the plateau region in the beginning of figure~\ref{fig:y2p}. For $y_2^+ \ll t_\infty$, we can use\footnote{ This approximation only works for small $y_2^+$. In previous sections, we dealt with times $u$ of the order of the Page time or larger, and then~\eqref{logapprox} is a much better approximation.}
 \begin{equation}\label{smalluapprox}
 \begin{split}
 x_2^+ = f(y^+_2) &\approx   t_{\infty} \tanh \( \frac{y^+_2}{ t_{\infty} } \) \,, \\
 \log \(  \frac{f(y^+_2)y^-_2}{\(t-f(y^+_2) \)\sigma_2} \)  &\approx    \log \frac{y^+_2}{t_{\infty}}  +\log 2 +\frac{y^+_2}{2u}  +\frac{y^+_2}{t_{\infty} } \approx \log \frac{y^+_2}{t_{\infty}}  +\log 2\,. \\
 \end{split}
 \end{equation}
 Solving for $\Delta S=0$ then leads to the solution 
 \begin{equation}\label{y2p_early}
 \begin{split}
 y^+_2 (u) \approx \frac{c t_\infty}{2^{23/4} \pi E_{\mt{S}} u_{\mt{HP}}} \exp\bigg[& \frac{1}{4}+\frac{\pi (T_0 -T_1)}{k}  + \frac{\pi T_1}{2} \(3 u + u_{\mt{HP}} \) \\
&+\frac{k}{8} \(  -\frac{1}{\pi T_1} +(3-2\pi T_1 u_{\mt{HP}})(u-u_{\mt{HP}}) \)\bigg]\,,\\
 \end{split}
 \end{equation}
 for $u \lesssim u_\mt{HP}$. As expected, we find an exponential increase of $y^+_2(u)$ for early times. The comparison with numerical results are shown in figure~\ref{fig:y2p_early}.

\begin{figure}[t]
	\centering
	\includegraphics[width=\textwidth]{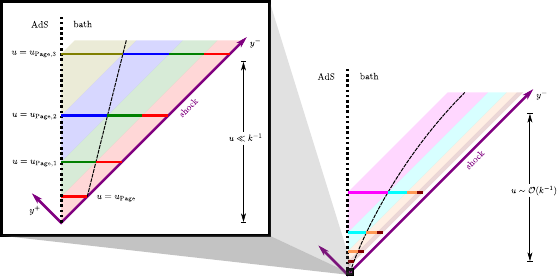}
	\caption{The dotted line in the right figure illustrates the evolution of $y^{\pm}_2$ with respect of time $u$. The left figure is the linear region with the approximation of $y^+_2$ described by eq.~\eqref{linear_y2p}. }
	\label{fig:mushy}
\end{figure}

We now move on to later times, when $\Delta u$ is of the order of the Page time, but still less than ${\cal O}(k^{-1} \log k)$. The above approximation of eq.~\eqref{smalluapprox} will break down. For times with $\Delta u$ comparable to the Page time we find numerically that the separation increases linearly with $\Delta u$, as can be seen in figure~\ref{fig:y2p}. We now proceed to show this linear behavior analytically. 
Using the results of section~\ref{sec:QES} in eqs.\eqref{dilatexpasion} and~\eqref{bulkentropy}, the only new term we need to consider are
\begin{equation}\label{newapprox}
\begin{split}
\log \(  \frac{f(y^+_2)y^-_2}{\(t-f(y^+_2) \)\sigma_2} \)  &\approx   \log \( \frac{u+\sigma_2}{2 \sigma_2} \)   + \frac{4\pi T_1}{k} \(  1- e^{-\frac{k}{2} (u-\sigma_2)}   \)  \,,\\
\log \(  \sqrt{f'(u)}\) &\approx \log 2 +\frac{k}{8\pi T_1} -\frac{ku}{4} -\frac{2\pi T_1}{k}(1-e^{-\frac{ku}{2}}) \,,
\end{split}
\end{equation}
where we have taken the approximation $f(y^-_+) \approx t_{\infty}$ for $u-\sigma \gg t_\infty$, which is satisfied in the region with linear behavior. We also note that the $\log \(\frac{u+\sigma_2}{2 \sigma_2} \)$ is a small contribution because of the log function and $\sigma$ also increase with $u$. Furthermore, if we take the small $ku$ expansion again and keep the liner terms, this approximations leads us to the following solution
\begin{equation}
\begin{split}
\sigma_2 \( u\) &= \frac{T_1-T_0}{2T_1k}   + \frac{1}{4} \(  u -u_{\mt{HP}}\) + \frac{1}{2\pi T_1}\log \(\frac{16 E_S \pi u_\mt{HP}(u+\bar{\sigma}_2)}{(2e)^{1/4}c \bar{\sigma}_2}\) - \frac{k}{8} u_\mt{HP}^2 + \mathcal{O}(k) \,,
\end{split}
\end{equation}
where $\bar{\sigma}_2= \frac{T_1-T_0}{2T_1k}   + \frac{1}{4} \(  u -u_{\mt{HP}}\)$ is the leading order term of $\sigma_2(u)$.\footnote{The $u$ dependence inside the log is very small, since for $\Delta u$ much larger than $u_\mt{Page}$ we have $\log \frac{(u+\sigma_2)}{\sigma} \approx \log 5 $.} It is straightforward to add higher $k$ corrections to this approximation, but we only need the first order terms to show that $\sigma_2$ depends almost linearly in $u$ for $\Delta u$ of the order of the Page time and up to ${\cal O}(k^{-1})$. Thus, in this regime, we find a linear
evolution for the distance of the endpoint of the bath interval to the shock:
\begin{equation}\label{linear_y2p}
 y_2^+(u) = u -\sigma_2 (u) \simeq \frac{3 }{4} (u -u_{\Page}) \,,
\end{equation}
where the slope is fixed to be $\frac{3}{4}$ at leading order, and we have ignored the correction from order of $\mathcal{O}(k)$.\footnote{The approximation is in $\partial_{u} y^+_2(u) \approx \frac{1}{4} e^{-\frac{k}{2}(y^-_\QES-y^+_2)} \partial_u y^-_\QES +\frac{1}{2} e^{-\frac{k}{2}(u-y^+_2)}$, reducing to $\frac{3}{4}$ when $u$ is order $u_{\Page}$ and to $0$ for $ku \gg 1$.}
The linear behavior is illustrated in  figure~\ref{fig:mushy}.

For very late times of $\mathcal{O}(k^{-1})$, the small $ku$ approximation in \eg eq.~\eqref{Slinear} breaks down. This is due to the breakdown of the dilaton approximation in eq.~\eqref{dilatexpasion}. The correct expression for times of ${\cal O}(k^{-1})$ is
\begin{equation}\label{null_y2p}
\phi \approx \bar{\phi}_r \( 2 \pi T_1 e^{-\frac{k}{2}y^-_{\mt{\QES}}} -\frac{k}{2} \log 2e \) \,.
\end{equation}
Correspondingly, the linear decrease of generalized entropy is replaced by a much slower exponential decrease. Using the improved dilaton contribution in eq.~\eqref{null_y2p}, as well as the approximation in eq.~\eqref{newapprox}, we can solve eq.~\eqref{newdeltaS} numerically and plot the results in figure~\ref{fig:y2p}.  Focusing on the large $u$ limit, we can give an approximation for the surface $y^+_2$
at very late time with $u \gg k^{-1}$ 
\begin{equation}
 y_2^+ \approx  \frac{2}{k} \log \(  \frac{16 \pi T_1}{4\pi T_1 (2T_1-T_0) +4k \log \( \frac{16E_{\mt{S} }\pi u_{\mt{HP}}}{c}\)  - k  (1+4\pi T_1 u_{\mt{HP}}-\log 8 )-2\pi T_1 k^2 u }    \) \,.
\end{equation}
This surface is becoming null for very large $u$. In an approximation that holds up to late times of order $u \sim \mathcal{O}(k^{-1} \log\frac{T_1}{k})$,\footnote{Note that we can not simply take $u$ to infinity to derive this leading order behaviour because the semi-classical model will break down in the late-late-time regime with $u \gg k \log\frac{T_1}{k}$. Here we assume $u \sim y^-_{\QES}$ approaches the very late-time limit. However, this formula \reef{null_shift} does not hold for
$u\to\infty$. } the asymptotic behaviour is 
\begin{equation}\label{null_shift}
 y_{2}^+ \simeq  \frac{2}{k} \log\( \frac{4 T_1}{2T_1- T_0} \) + \mathcal{O}(1) \,.
\end{equation}
We observe that the order-one correction above is also a constant, however, $u$-dependent terms appear on the right-hand side at order $k$.

Before closing here, let us also comment on the late-time protocol applied to the quench-scrambling phase transition. The calculation to find the behavior of the right endpoint with $u>u_{\mt{QS}}$ is similar to the one for the Page transition carried earlier in this section. However, the result is that as we increase the time up to the Page time, the distance of the right point  to the shock $y^+_2(u)$ starts from a very small value \ie $\ys-\y_2\sim  \(\frac{c T_1}{6E_S}\)^2 t_{\infty} \ll t_\infty$
and then \emph{decreases} exponentially for $u_{\mt{QS}} < u < u_{\Page}$. That is, the left boundary very quickly approaches the null curve defined by the shock, \ie $\y_2\simeq u$.
This contrasting  behavior originates from the increase of bulk entropy in scrambling phase, \ie the linear term in eqs.~\eqref{Sgen_before_k} and~\eqref{Sgen_before_k2}.

\subsubsection{Redundancy and Efficiency of Encoding}

With the protocol introduced above,  we found that we can reconstruct the black hole interior with the bath interval $\widetilde\Bcal_1=[\y_1=0,\y_2 = \yb(u)]$, where $\yb$ is the minimum value of $\y_2$ defined by  eq.~\eqref{linear_y2p}, \ie
\beq \label{slackr2}
\yb=(1-\q)\, u + \q\,\uP \,, \quad \gamma = \frac{3}{4}
\eeq
where $\gamma$ receives corrections at order $k$ which only become relevant at times of order $k^{-1}$, and which slowly change the slope to zero at very late times of order $k^{-1}\log{\frac{T_1}{k}}$. Therfore $\yb$ 
defines a time-like boundary for the endpoints of these minimal intervals, as shown in figure~\ref{fig:mushy}. Assuming the information flows at the speed of light,\footnote{As indicated by the evolution of $\yp$ in section~\ref{sec:pepsi}.}  this result points to a redundancy of the encoding of the black hole interior. That is, the black hole interior is encoded in the Hawking radiation emitted over many finite time intervals, but at times much later than the Page time $\uP$.  In general, if we begin to collect the radiation at an arbitrary time $u_\mt{initial}>\uP$ and we can reconstruct the black hole interior with radiation collected  (at $\y_1=0$) in the time interval $[u_\mt{initial},u_\mt{final}]$ with
\beq\label{slackr44}
u_\mt{final} = \frac{u_\mt{initial}}\q+\uP\,.
\eeq
This time $u_\mt{final}$ is determined by the intersection of the null ray entering the bath at $u_\mt{initial}$ with the curve $\yb$, such that all of the information flowing into the bath in the above time interval is captured in the interval $[0, \yb(u_\mt{initial})]$ on this final time slice.

As a concrete example, we can discard all of the Hawking radiation emitted before $\uP$, but we are still able to reconstruct the black hole interior by collecting the radiation emitted in $u\in [\uP,\upp{1}]$ where $\upp{1}-\uP=\uP/\q$. Further, this process can be repeated again, \ie we discard the radiation before $\upp{1}$ but the black hole interior is recovered if we collect the subsequent radiation up to a time $\upp{2}$. Repeating the process repeatedly, one finds that
\beq\label{slackr4}
\upp{n}-\upp{n-1}=\frac{\uP}{\q^{n}}\,.
\eeq
Since $\q<1$,  these intervals are becoming longer and longer. This suggests that while the information about the black hole interior is still encoded in the radiation collected at later times, the density of this information becomes less dense at much later times. That is, the encoding of the information is becoming less efficient at later times -- see further comments in section~\ref{sec:discuss}.  

These results depend on the simple linear growth of $\yb$ in eq.~\reef{slackr2}. However, we also showed above that this behaviour breaks down at late times, with this boundary  approaching a null curve \reef{null_shift} at very late times  -- see figure~\ref{fig:mushy}. This means that the size of the successive intervals, \ie $\upp{n}-\upp{n-1}$, would grow even more quickly than the geometric behaviour shown in eq.~\reef{slackr4}. With the final asymptotic expansion of $\yb$ following a null curve, we would conclude that for times beyond 
\beq\label{slackrmax}
u_\mt{max}\simeq  \frac{2}{k} \log\( \frac{4 T_1}{2T_1- T_0} \)\,,
\eeq
we could never collect enough information to reconstruct the black hole interior. This conclusion should be tempered by the fact that our semi-classical understanding of the {\aims} model will break down at times of order $u \gtrsim k^{-1} \log\frac{T_1}{k}$. Combining eq.~\reef{slackrmax} with the expressions for $\upp{n}$ following from eq.~\reef{slackr2},\footnote{Explicitly, one finds that $\upp{n}= \frac{1-\gamma^{n+1}}{\q^n\,(1-\gamma)}\, u_{\mt{Page}}$.}
suggests a finite redundancy of the encoding of the black hole interior in the Hawking radiation with
\beq\label{Nmax}
n_\mt{max}\simeq\frac{\log\(2\uP/k\)}{\log\q}\,.
\eeq
More precisely, the black hole information is encoded in a finite number of distinct time intervals roughly given by eq.~\reef{Nmax}.

Of course, there is nothing special about these intervals $[\upp{n+1},\upp{n}]$. As indicated in eq.~\reef{slackr44}, we can  reconstruct the black hole interior with radiation collected in general time intervals $[u_\mt{initial},u_\mt{final}]$, beginning at any arbitrary $u_\mt{initial}>\uP$.
Further, on the time slice $u=u_\mt{final}$, we could remove intermediate segments between $\y_1=0$ and $\y_2=u_\mt{final} - u_\mt{initial}$ as in section~\ref{five} or even implement the \"uberholography process as in section~\ref{uber}. Of course, this indicates that the reconstruction of the black hole interior does not require all of the radiation in the time interval $[u_\mt{initial},u_\mt{final}]$. Rather, the \"uberholography process suggests collecting the radiation on some fractal subset of this time interval. All of these considerations certainly point to a remarkable redundancy in time for the encoding in the Hawking information of information about the black hole interior. It would be interesting to understand if and how this pattern of redundancies is manifest in other models of black hole evaporation.

%% file: sections/Discussion.tex
% !TEX root = ../BH_v04.tex

In this paper, we examined the flow of information in black hole evaporation as described by the {\aims} model~\cite{Almheiri:2019psf,Almheiri:2019hni}. This model involves two systems: JT gravity coupled to a two-dimensional holographic CFT, and an infinite bath, comprised of the same holographic CFT on a half-line. The former is prepared as an eternal black hole, which is dual to a thermofield-double state entangling {\qml} and \qmr, while the bath is prepared in its vacuum state. These two systems are connected by a quantum quench, and the subsequent evolution of the entanglement entropy of \qml+bath subsystem exhibits three phases: the quench phase, in which the QES on the Planck brane is fixed at the bifurcation surface of the initial black hole; the scrambling phase, in which the QES moves slowly away from this bifurcation surface; and the late-time phase, in which QES is just behind the event horizon of the evaporating black hole.

In the example of the eternal AdS$_2$ black hole with reflecting boundary conditions at the asymptotic boundary, the QES for {\qml} (or \qmr) alone will be the bifurcation surface. Hence the information in this subsystem can be used to reconstruct the exterior region on the left (or right) side of the black hole. That is,  the entanglement wedge for \qml\ is the entire region outside of the left event horizon, as shown in the left plot in figure~\ref{fig:entwedges}. Considering the information flow after the quench, since the position of the QES for the \qml+bath subsystem is fixed in the initial quench phase, the Hawking radiation is carrying negligible information into the bath. That is, any information about the black hole interior would only be at order one in the large $c$ expansion of the holographic CFT.\footnote{In the analysis of~\cite{Almheiri:2019psf} for a general CFT, the QES already begins to move away from the bifurcation surface during the quench phase. Of course, there is also a smooth cross-over between the quench and scrambling phases in their model.}

The onset of the scrambling phase marks the time when the Hawking radiation begins to contain information about the interior. In the scrambling phase, the information flow is detected by the QES, and is order $c$, but Hawking radiation absorbed by the bath only carries enough information for {\qml}+bath to reconstruct a small additional region behind the horizon of the left side (and to the past of the shockwave), as illustrated in the middle plot in figure~\ref{fig:entwedges}. However, once the black hole has passed the Page transition and entered into the late-time phase, the QES jumps to be behind the right event horizon (and to the future of the shockwave), and so the bath has acquired enough information for \qml +bath to reconstruct a much larger portion of the black hole interior (see the right plot in figure~\ref{fig:entwedges}).

Let us comment on the HRT surfaces and the encoding of the black hole interior in the late-time phase (see figure~\ref{faze}). We note that in this regime, the black hole interior provides a classic example of the quantum error correcting encoding that is characteristic of holography~\cite{Pastawski:2015qua,Almheiri:2014lwa}. We are considering three subsystems of the boundary, \qml, {\qmr} and the bath. In this configuration the information about the black hole interior cannot be recovered from any one of these subsystems; however, combining any two of them allows us to reconstruct the interior information. In our discussion, the focus was on the combination \qml+bath, but a quick examination of the HRT surfaces in figure~\ref{faze} shows that it is also included in the entanglement wedges of either \qml+{\qmr} or \qmr+bath.%\footnote{\jh{new:} \rcm{I would drop this footnote} We have assumed here that $S_0$ is large enough that we have not entered the regime where the entanglement wedge of the bath includes the quantum extremal islands of~\cite{Almheiri:2019hni}. If this were the case, then the entanglement wedge of \qml+{\qmr} would exclude this region of the black hole interior.}

\begin{figure}[t]
  \centering
  \includegraphics[width=0.8\textwidth]{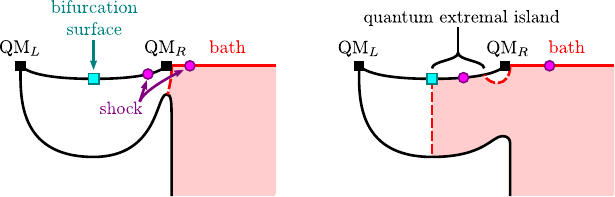}
  \caption{Quench (left) and late-time (right) phases for the entropy of the bath.}
  \label{fig:noMoreFiguresPlz}
\end{figure}

However, the above discussion is not complete. Eventually, on a time scale much larger than those considered here, the bath on its own will make a Page transition. Initially, the bath is in the analog of the quench phase with the HRT surface sketched in the left panel of figure~\ref{fig:noMoreFiguresPlz}. It then makes a transition to a late-time phase with the HRT surfaces sketched in the right panel, where a quantum extremal island~\cite{Almheiri:2019hni} has formed. Here we implicitly assume a large intrinsic gravitational entropy for the JT model, \ie we are assuming that $S_0=\phi_0/(4\GN )\gg1$ in eq.~\eqref{Sgen}.\footnote{A standard assumption is that $\phi_0\gg\phi_r/\epsilon$ in the spacetime regions of interest~\cite{Maldacena:2016upp} -- see eq.~\reef{Sgen}.} This contribution to the generalized entropy adds a heavy penalty for HRT surfaces which end on the Planck brane, and so it would delay the onset of the late-time phase and the appearance of the quantum extremal island. Note that the transitions in the main text (for the entropy to \qml+bath), one is always comparing branches where a single HRT geodesic ends on the Planck brane and so $S_0$ did not play a role. Further, one can argue that if $S_0\gtrsim \Delta S$ (the change in the black hole entropy generated by the shock wave, \ie in going from $T_0$ to $T_1$), then the branch corresponding to the scrambling phase never dominates and so the Page transition corresponds to going directly from the quench branch to the late-time branch. Of course, in the latter phase with the quantum extremal island, the bath by itself now encodes sufficient information to reconstruct a portion of the black hole interior. The fact that this other Page transition takes place much later suggests that early-time scrambling is important for the reconstruction of the black hole interior, as suggested in~\cite{Penington:2019npb}.
It would be interesting to repeat the detailed analysis that we have performed in this paper considering just the bath on its own.

% In the original thought experiment, the scrambling phase is physically interesting, as it makes concrete some aspects of the Hayden-Preskill thought experiment~\cite{HayPre07}. This phase only appears after a sharp phase transition, and immediately afterwards a small amount of the interior is in the entanglement wedge of \qml+bath. This observation matches well with the heuristic picture of a black hole as a highly efficient scrambler of quantum information. The approximately linear rate of increase in the exposure of the black hole interior represents the (limited but) increasing ability of an outside observer to decode the black hole interior even before the Page time. Once the Page time is reached, there is a sudden and massive increase in the entanglement wedge of \qml+bath, again in accord with expectations from the Hayden-Preskill thought experiment. \rcm{I didn't understand how??}

As the \qml+bath system continues to evolve beyond the Page time, the wedge region grows relatively slowly as the bath continues to absorb more Hawking radiation. That is, the information carried by the radiation coming after the Page transition is less important for the reconstruction of the black hole interior. Eventually, one expects the entanglement wedge of the \qml+bath subsystem to extend to the right boundary of the AdS$_2$ geometry at $t_\infty$ (where the dilaton vanishes), but we can not trust the model to these very late times. However, a more appropriate comment might be to say that the information is less densely encoded in the late-time radiation -- see further comments below.

 In this late-time phase, we found in sections~\ref{two} and~\ref{sec:cocacola} that the information needed to reconstruct the black hole interior propagates at nearly the speed of light into the bath. That is, (a large portion of) the black hole interior could be reconstructed using the Hawking radiation captured on the time slice $u=u_\mt{Page}$ between $\y_1=0$ and $\y_2=\uP$, together with \qml. However, on a later time $u>\uP$, we could reproduce essentially the same reconstruction using the Hawking radiation captured between $\y_1=\yp\simeq u-\uP$ and $\y_2=\ys=u$ instead.\footnote{Of course, it is reasonable to expect that no information about the black hole interior is encoded in the bath beyond the position of the shockwave, since this portion of the bath is not in causal contact with the quench point.}  Of course, this is consistent with the information being carried into the bath by massless right-moving quasi-particles in the two-dimensional CFT \cite{Leichenauer:2015xra}. Similar behaviour was also recently observed in \cite{Rozali:2019day}. Of course, as shown in eq.~\reef{slackr1} (see also figure~\ref{fig:bath_point}), there are corrections to $\yp$ in the small $k$ expansion. However, the corrected (timelike) boundary still rapidly approaches a (slightly) shifted null ray. In section~\ref{sec:cocacola}, we also showed that the early Hawking radiation is extremely important in the above reconstruction protocol.
That is, the right boundary $\y_2$ of the bath region must be extremely close to the shockwave, \ie $\ys-\y_2\sim  \(\frac{c T_1}{6E_S}\)^4 \frac{\pi T_1}{k} \ll t_\infty$ as in eq.~\eqref{eq:smalldist}.

The importance of the early and late time Hawking radiation in this protocol was examined more closely in section~\ref{five}, where we considered removing an intermediate interval from the bath region -- see figure~\ref{fig:3pts}. As shown in figure~\ref{fig:5pts-u0}, the size of the intermediate interval is maximal when it is at the center and quickly decreases as this interval approaches either the shockwave or the boundary $\yp$. This is indicative of a clear separation of the radiation into early and late pieces. Of course, the process of systematically removing intermediate intervals from the bath region can be continued, cutting out smaller and smaller subregions, as discussed in section~\ref{uber}. Repeating this process ad infinitum, following~\cite{Pastawski:2016qrs}, we produce a fractal structure which, in combination with \qml, contains enough information to reconstruct the interior of the black hole from which the interior of the black hole can still be reconstructed. It is interesting that the (Minkowski) dimension characterizing this fractal matches that found for the CFT vacuum in~\cite{Pastawski:2016qrs}. This match arises because the very small intervals only probe the correlations of the CFT deep in the UV, and these must match in both settings.

In section~\ref{twoprime}, we considered a different reconstruction procedure that focused on the later radiation by anchoring the bath interval at $\y_1=0$. We found that the minimal size $\y_2 = \yb$ for which the information in \qml+bath still allowed us to reconstruct a large portion of the black hole interior follows time-like boundary, as shown in figure~\ref{fig:mushy}.  Using eq.~\eqref{slackr2}, we found a redundancy with the information about the black hole interior being encoded in the Hawking radiation emitted in the time intervals $[\upp{n+1},\upp{n}]$ after the Page time $\uP$. 

Of course, this redundancy is consistent with the Hayden-Preskill thought experiment~\cite{HayPre07}. The latter indicates that if a few qubits are dropped into an old black hole, the information can be recovered after the scrambling time by combining (essentially) the same number of qubits from the subsequent radiation with (all of) the early Hawking radiation. However, the radiated qubits need not be those radiated immediately after the scrambling time, but rather can be collected from the subsequent radiation at any time -- see also \cite{Yoshida:2018ybz,Yoshida:2019qqw}. From this perspective, the initial eternal black hole at temperature $T_0$ plays the role of the old black hole and early radiation, \ie \qmr\ is the old black hole while \qml\ plays the role of the early radiation. The black hole is `rejuvenated' by dropping in the shock wave and the information can be recovered after $\uP$, which then plays the role of the scrambling time in this discussion. However, as noted above, the information need not be collected immediately after the Page time but in any sufficiently large interval after $\uP$. This analogy might be made more precise by regarding the shock wave as a `heavy diary', as discussed in \cite{Penington:2019npb} -- see also \cite{Yoshida:2017non}.

Of course, as indicated by eq.~\reef{slackr4}, or more generally by eq.~\reef{slackr44}, the length of the time interval needed to collect sufficient information grows at later times. We suggested that this indicates the encoding is becoming less dense or less efficient at later times. However, the temperature of the black hole is (slowly) falling, and so one might wonder if the reduction in the flux of Hawking radiation accounts for this effect. However, the flux flowing into the bath (at $\y_1=0$) is given by $T_{y^+y^+}(u) \sim T_1^2 e^{-ku}$, as shown in eq. \eqref{Schwarzian}. Hence this reduction only becomes noticeable on time scales of order $u\sim1/k$. A simple calculation shows that an interval $[0,\sigma_2]$ needed to capture a fixed amount of Hawking radiation, as counted by energy or number of quanta (\ie $E/T_\mt{eff}$), barely exhibits any growth at early times, \ie  in the regime where eq.~\reef{slackr4} is valid.\footnote{Our conclusion assumes $(T_1-T_0)/T_1 \ll 1$ and uses $\uP\sim(T_1-T_0)/(k\,T_1)$ from eq.~\eqref{uPage}.} Hence  the reduction of Hawking radiation over time does not  explain the growth of $\sigmab$, and the natural explanation is once again that the redundant encoding of information simply becomes less efficient over time. However, we should note that the different time intervals are not reconstructing precisely the same interior region. Rather the latter also grows with time, and so this way partially account for the growth in $\yb$.

We also note that the reduction of the Hawking flux, \ie $T_{y^+y^+}(u) \sim T_1^2 e^{-ku}$, is a central factor in the nonlinear behaviour in the growth of $\yb$ found at time scales of order $u\sim 1/k$, as shown in figure~\ref{fig:mushy}. More directly in our calculations, the reduction in the corresponding gravitational entropy \reef{null_y2p} on the QES produces this effect. As a result,
$\yb(u)$ approaches a null ray, as shown in eq.~\reef{null_shift}, in this nonlinear regime.  We then infer that the information in the Hawking radiation is too depleted beyond $u_\mt{max}$ -- see eq.~\reef{slackrmax} -- to collect enough quanta to reconstruct the black hole interior. Of course, our semi-classical understanding of the {\aims} model breaks down at times of order $u \gtrsim k \log\frac{T_1}{k}$, and so nonperturbative effects may still allow for such a reconstruction.

In wrapping up this discussion, we reiterate that there is a remarkable redundancy in the encoding of the black hole interior in the Hawking radiation. In section~\ref{uber}, we explicitly showed that the interior information was still available after numerous subintervals were gouged out of the initial parcel of radiation emitted between the quench and $\uP$, to the point where it was reduced to a fractal structure.  The reconstruction was also possible with the radiation collected (at $\y_1=0$) in the interval $[u_\mt{initial},u_\mt{final}]$, beginning at any arbitrary $u_\mt{initial}>\uP$ and with $u_\mt{final}$ given by eq.~\reef{slackr44}. Again, the \"uberholography approach could again be applied to perforate any such interval with holes.  It would, of course, be interesting to understand if this pattern of redundancies appears in other models of black hole evaporation.

In closing, we observe that our analysis in section~\ref{sec:excise} focused on the Page transition between the scrambling and late-time transitions. However, this discussion can easily be extended to the first transition between the quench and scrambling phases, corresponding to the onset of scrambling, and the results are more or less the same. One important difference is that the trajectory for the $\sigma_{\mt{QS}}$ analog of $\sigma_\mt{Page}$ in section~\ref{sec:pepsi} is null for all times, unlike the trajectory of $\sigma_\mt{Page}$ which asymptotes towards a null path as is shown in figure~\ref{fig:bath_point}. As was noted towards the end of section~\ref{twoprime}, the position of the $\sigma_{\mt{Turn}}$ in the quench-to scrambling phase transition shows different behaviour to the $\sigma_{\mt{Turn}}$ of the Page transition. In particular, as we increase the time from $u_\mt{QS}$ up to the Page time $u_\mt{Page}$, the distance of the right point  to the shock $y^+_2(u)$ starts from a very small value \ie $\ys-\y_2\sim  \(\frac{c T_1}{6E_S}\)^2 t_{\infty} \ll t_\infty$
and then \emph{decreases} exponentially. It was noted that the contrasting  behavior originates from the increase of bulk entropy in the scrambling phase, \ie the linear term in eqs.~\eqref{Sgen_before_k} and~\eqref{Sgen_before_k2}.

Furthermore, in our discussion, for simplicity we set the boundary entropy to zero, \ie $\log g=0$ in eq.~\reef{Sbulk_hol}. This choice does not affect the Page transition in any way, as we have said. The reason is that neither of the two competing geodesics terminates on the end-of-the-world brane in this case. However, the first (quench-to-scrambling) transition will be shifted if we choose $\log g\ne 0$. On the scrambling phase branch, bulk geodesic connects a boundary point in the bath to the QES on the Planck brane. However, in the quench phase, the HRT surfaces are comprised of two geodesics terminating on the ETW brane. Therefore, the corresponding generalized entropy would be increased by a term $4\log g$. If we consider figure~\ref{fig:Sgen01}, then the transition time would move to an even earlier time (assuming that $\log g>0$).